%% file: paper.tex
\documentclass[sigconf]{acmart}

\usepackage{algorithmic}
\usepackage{xcolor}
\usepackage{pgf}
\usepackage{svg}

\usepackage[percent]{overpic}
\usepackage{tikz}
\usepackage{subcaption}
\usepackage{graphicx}
\usepackage{multirow}
\usepackage{subcaption}
\usepackage{rotating}
\usepackage{mwe} %
\usepackage{array}
\usepackage{enumitem}
\setitemize{noitemsep,topsep=0pt,parsep=0pt,partopsep=0pt}
\usepackage{amsmath}
\usepackage{commath}
\usepackage[noabbrev]{cleveref}

\setcopyright{none}
\copyrightyear{2024}
\acmYear{2024}
\acmDOI{}
\acmConference[]{}{}{}
\acmPrice{}
\acmISBN{}

\usepackage{booktabs} %

\usepackage[ruled]{algorithm2e} %

\SetAlFnt{\small}
\SetAlCapFnt{\small}
\SetAlCapNameFnt{\small}
\SetAlCapHSkip{0pt}

\begin{document}

\title{Neural Two-Level Monte Carlo Real-Time Rendering}

\author{Mikhail Dereviannykh}
\orcid{1234-5678-9012}
\affiliation{%
  \institution{Karlsruhe Institute Of Technology}
  \country{Germany}
}
\author{Dmitrii Klepikov}
\orcid{0000-0001-9577-7047}
\affiliation{%
  \institution{Karlsruhe Institute Of Technology}
  \country{Germany}
}

\author{Johannes Hanika}
\orcid{1234-5678-9012}
\affiliation{%
  \institution{Karlsruhe Institute Of Technology}
  \country{Germany}
}

\author{Carsten Dachsbacher}
\orcid{1234-5678-9012}
\affiliation{%
  \institution{Karlsruhe Institute Of Technology}
  \country{Germany}
}
    
\renewcommand{\shortauthors}{Dereviannykh et al.}

\begin{abstract}
We introduce an efficient Two-Level Monte Carlo (subset of Multi-Level Monte Carlo, MLMC) estimator for real-time rendering of scenes with global illumination. Using MLMC we split the shading integral into two parts: the radiance cache integral and the residual error integral that compensates for the bias of the first one.
For the first part, we developed the \emph{Neural Incident Radiance Cache} (NIRC) leveraging the power of fully-fused tiny neural networks \cite{NRC} as a building block, which is trained on the fly. The cache is designed to provide a fast and reasonable approximation of the incident radiance: an evaluation takes 2-25$\times$ less compute time than a path tracing sample. This enables us to estimate the radiance cache integral with a high number of samples and by this achieve faster convergence. For the residual error integral, we compute the difference between the NIRC predictions and the unbiased path tracing simulation.\\
Our method makes no assumptions about the geometry, materials, or lighting of a scene and has only few intuitive hyper-parameters. We provide a comprehensive comparative analysis in different experimental scenarios. Since the algorithm is trained in an on-line fashion, it demonstrates significant noise level reduction even for dynamic scenes and can easily be combined with other importance sampling schemes and noise reduction techniques.
\end{abstract}



\begin{teaserfigure}
\setlength{\abovecaptionskip}{1pt}
  \begin{tabular*}{\linewidth}{%
  @{}
  p{.677\linewidth}
  @{\hspace*{.01\linewidth}}
  p{.313\linewidth}
  @{}}
    \includegraphics[clip,trim=0 65 0 0,width=\linewidth]{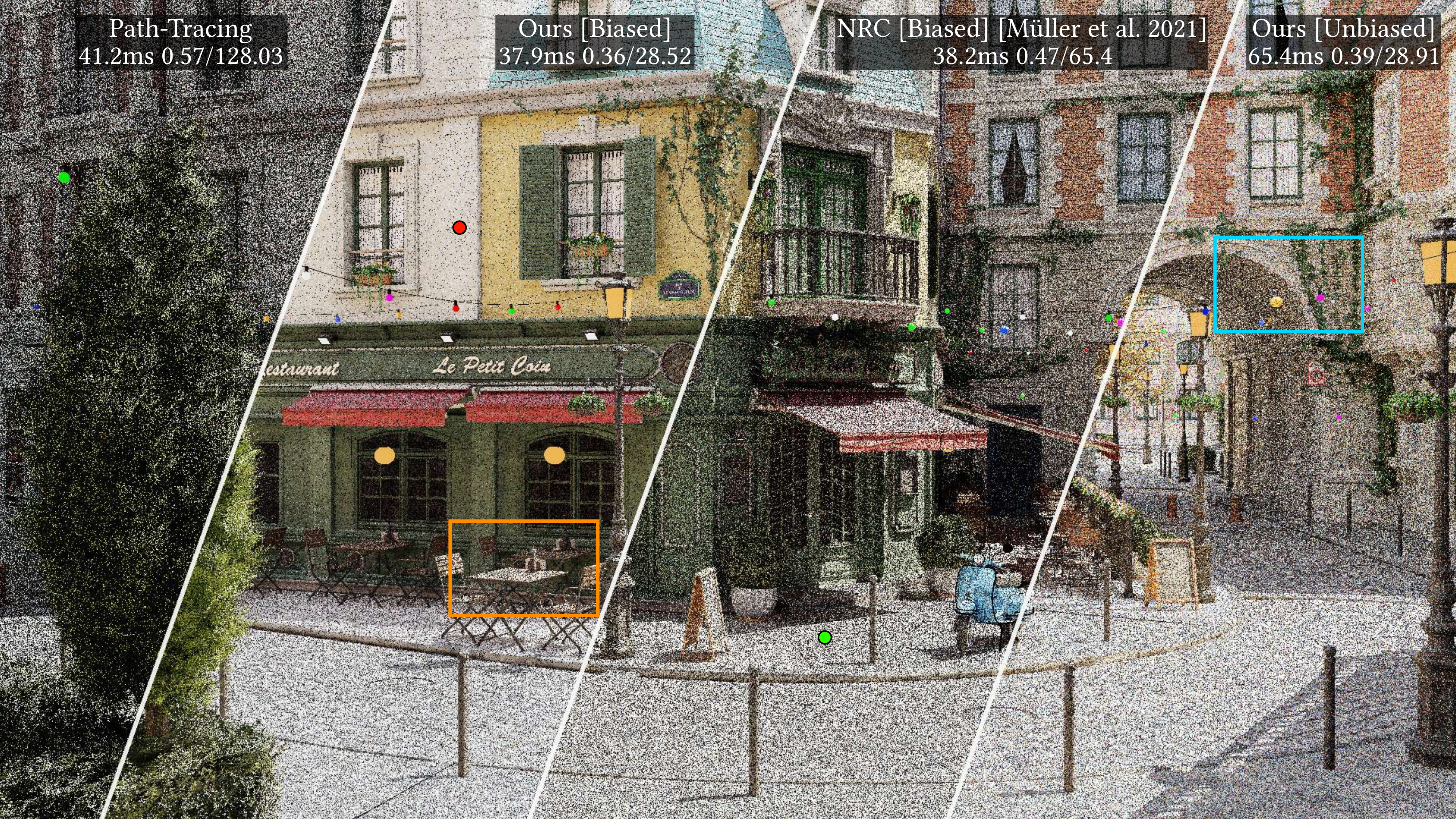} &
    \begin{tabular*}{\linewidth}[b]{@{}p{.495\linewidth}@{\hspace*{.01\linewidth}}p{.495\linewidth}@{}}
        \includegraphics[width=\linewidth]{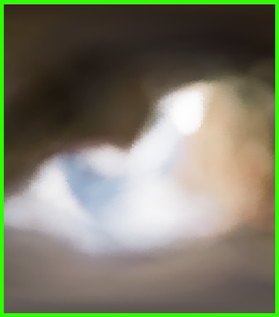}&
        \includegraphics[width=\linewidth]{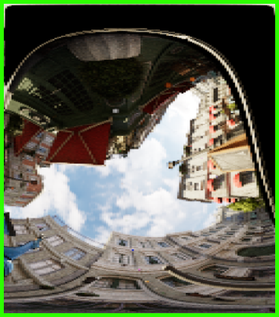}\\
        \includegraphics[width=\linewidth]{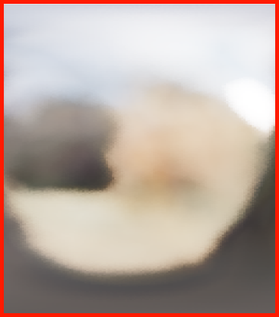}&
        \includegraphics[width=\linewidth]{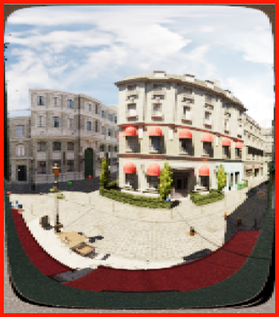}
    \end{tabular*}\\
    \begin{tabular}[b]{%
    @{}
    p{.192\linewidth}@{\hspace*{.01\linewidth}}
    p{.192\linewidth}@{\hspace*{.01\linewidth}}
    p{.192\linewidth}@{\hspace*{.01\linewidth}}
    p{.192\linewidth}@{\hspace*{.01\linewidth}}
    p{.192\linewidth}}
        \includegraphics[width=\linewidth]{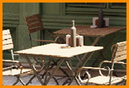}&
        \includegraphics[width=\linewidth]{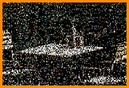}&
        \includegraphics[width=\linewidth]{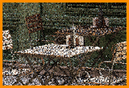}&
        \includegraphics[width=\linewidth]{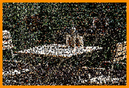}&
        \includegraphics[width=\linewidth]{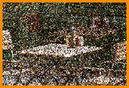} \\
        \includegraphics[width=\linewidth]{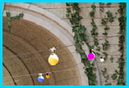} &
        \includegraphics[width=\linewidth]{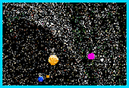}&
        \includegraphics[width=\linewidth]{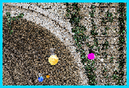}&
        \includegraphics[width=\linewidth]{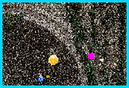}&
        \includegraphics[width=\linewidth]{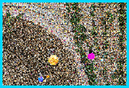} \\ 
        \centering Reference & 
        \centering Path Tracing & 
        \centering Ours [Biased] & 
        \centering NRC [Biased] & 
        \centering Ours [Unbiased]
      \end{tabular}&
    \includegraphics[width=1.0\linewidth]{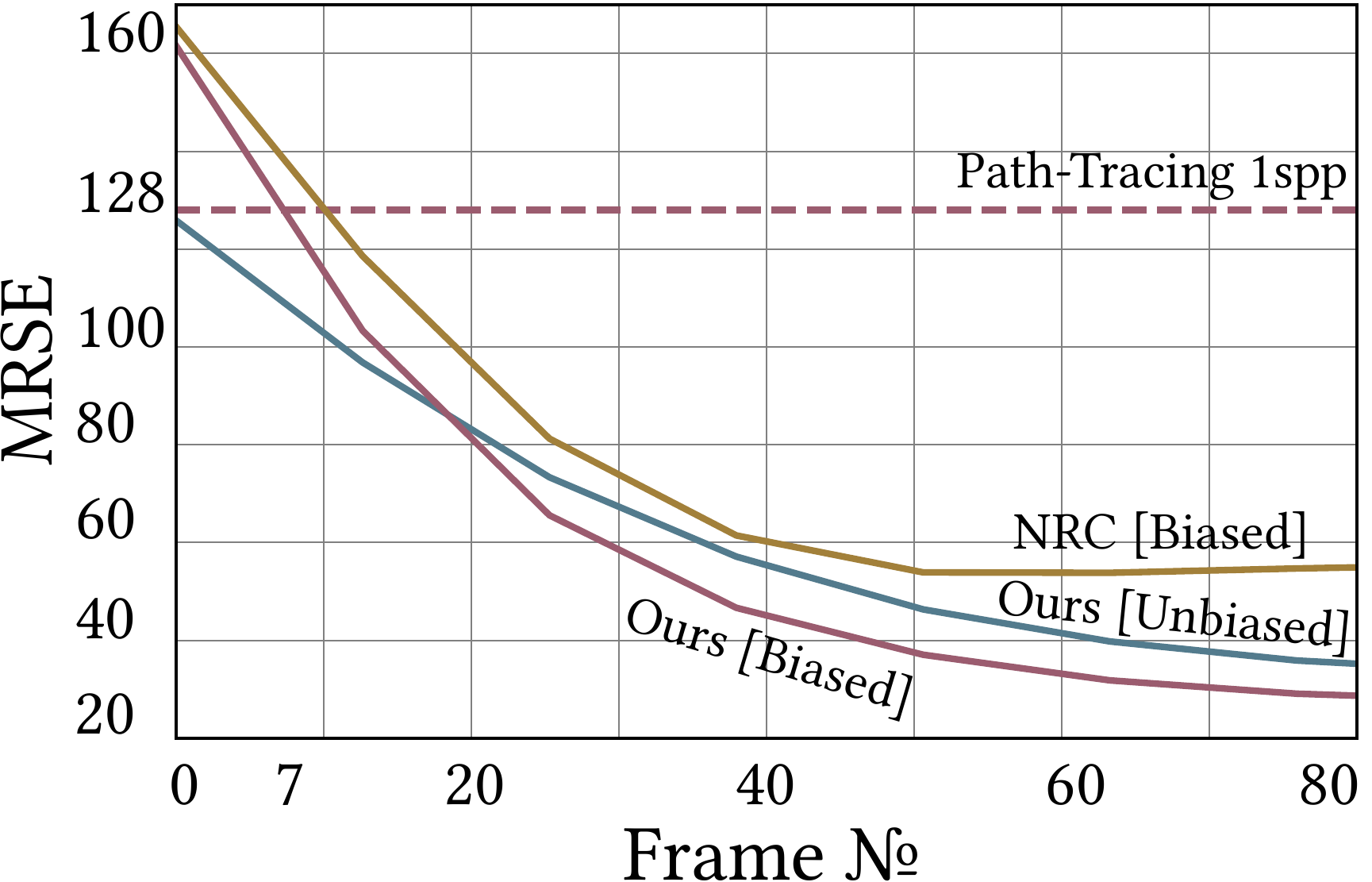}
    \end{tabular*}
  \label{fig:teaser}
  \caption{The Bistro Exterior scene (RTX 3080 at 1080p) rendered using one sample per pixel to compare our \emph{Neural Incident Radiance Cache} (NIRC) to path tracing as well as the Neural Radiance Cache (NRC) \cite{NRC}. 
  As does NRC, we replace tracing long light transport paths with a cache lookup. Our cache, however,
  stores \emph{incident} radiance, avoiding a ray cast before lookup.
  In addition, combining \emph{Two-Level Monte Carlo} (a subset of Multi-Level
  Monte Carlo, MLMC) with our, NIRC enables us to estimate the cache error and
  thus to remove bias. For each render, we show computation time, the
  perceptual image difference according to \reflectbox{F}LIP \cite{FLIP}, and
  the \emph{Mean Relative Squared Error} (MRSE). On the right, we show NIRC
  cache visualizations and a convergence plot.}
  \setlength{\abovecaptionskip}{10pt}
\end{teaserfigure}

    \maketitle
\newcolumntype{P}[1]{>{\centering\arraybackslash}p{#1}}

\section{Introduction}

Rendering global illumination effects remains challenging in interactive and real-time applications, even with modern hardware-accelerated ray tracing. 
Monte Carlo methods, which are used to compute a solution to the  \emph{Rendering Equation} \cite{Kajiya1986} are prone to noise due to their stochastic nature, amplified by complex materials and occlusions in a scene.
Techniques such as importance sampling, and more recently learning-based
approaches have demonstrated effectiveness in addressing these issues,
mitigating the noise problem significantly using radiance
caching~\cite{Křivánek2009,ward1988,giradiancecache},
photon mapping~\cite{photonmapping,photonmapping2}, adaptive sampling~\cite{mitchell1987}, vertex connection and merging \cite{vertexmerging} and path guiding~\cite{guidingcourse}.

Our work also depends on caching:
we use several path tracing samples to initialize a cached representation of the incident radiance field in the scene.
This cache can then be evaluated to yield a fast approximation of the actual light field. \\
In our approach, we introduce a novel caching method using fully-fused tiny
neural networks, inspired by the \emph{Neural Radiance Cache} (NRC)~\cite{NRC},
and multiresolution hash encoding~\cite{InstantNGP}. The important
novelty of our method is that we employ \emph{Neural Incident Radiance Caches} (NIRCs)
which are specifically designed to cache incident radiance (as opposed to
outgoing radiance with NRCs). This allows us to query the cache for incident
radiance at a shading point post-material evaluation to efficiently approximate
the shading integral without additional ray tracing.
Our combination with a \emph{Two-Level Monte Carlo} scheme enables us to compensate for the resulting bias. For biased rendering, we introduce a new Balanced Termination Heuristic (BTH) to enhance path termination efficiency, leveraging the strengths of NIRCs while addressing their limitations with glossy details. The BTH allows us to use our cache instantly at the primary bounce since it predicts incident radiance directly, unlike the NRC \emph{Spread Angle Heuristic} (SPH), which is unsuitable for stopping at the primarily visible surface.

Along with that, we assessed the differences between \emph{Control Variates} (CV) and Two-Level Monte Carlo methods by comparing analytical models, such as \emph{Spherical Harmonics} (SH) and mixtures of \emph{von Mises-Fisher} (vMF) lobes, commonly used in the earlier research \cite{pantaleoni2020online}\cite{OnlineEMPG}, alongside \emph{Neural Control Variates} (NCV) \cite{NCV}. While NCV is specifically designed to satisfy Control Variates requirements by providing analytical integral computation, our results show that NIRC in the Two-Level Monte Carlo framework can achieve lower variance of the residual error estimator, requiring significantly fewer training frames and offering higher generalization.

Furthermore, the NIRC is designed to predict incident radiance for batches of 10 to 50 rays, amortizing the computation of the spatial feature for requests for the same shading point.
This can lead to significant speedups compared to NRC.
We also introduce an alternative encoding of the input to the multi-layer perceptron (MLP)~\cite{MLP}, leading to a much more precise
angular representation of the radiance field which is required for sufficiently accurate approximation of the shading integral.
To train the cache, we derive a loss function which includes the two-level Monte Carlo estimator; further, we also evaluate a simplified loss based on the relative L2 difference.

Additionally, we developed an error-based path termination algorithm to ensure a fair equal-bias comparison between the NIRC and the NRC. Our cache analysis shows that the NIRC can achieve up to 130.5 times fewer indirect radiance bounces, performing indirect ray bounces for only 0.1-1\% of pixels while maintaining similar bias levels compared to the NRC. Using the SPH, NIRC achieves a Relative Squared Bias up to 4.12$\times$ lower than NRC, though it increases noise. If we accept higher bias, BTH reduces light bounces by up to 3.54 times and improves mean relative square error by up to 6.67 times, enhancing real-time rendering performance and accuracy.

Finally, we investigate the specific use case of environment map lighting.
Instead of approximating the radiance \cite{nenv2023}, the neural cache is used to
predict the visibility term, which is
similar to a concurrent work \cite{fujieda2023neural}. However, there's a
notable difference in the encoding methods: while Fujieda et al. employ a
grid-based method for encoding directions, our method utilizes Spherical
Harmonics for this purpose. We observe that the \emph{Neural Visibility Cache}
(NVC) achieves significantly better reconstruction results compared to the NIRC.
This improvement directly contributes to lower variance in the unbiased
two-level estimator, enhancing the overall performance for 
environment map lighting.

\noindent To sum up, our contributions are:
\setlength{\leftmargini}{1.2em}
\begin{itemize}
\item the introduction of the \emph{Neural Incident Radiance Cache} (NIRC)
\item a \emph{Multi-Level Monte Carlo} (MLMC) estimator to remove bias
\item new evidences suggesting that the MLMC framework in combination with the NIRC can achieve superior variance reduction in less training time over CV
\item a comprehensive analysis comparing NRC and NIRC
\item a set of performance optimizations enabled by the NIRC concept
\item a specialized solution for environment map lighting.
\end{itemize}

\section{Background}

\paragraph*{Light transport simulation}

The rendering equation~\cite{Kajiya1986} describes how light interacts with a scene. The radiance $L_{o}(x, \omega_{o})$ leaving a point $x$ in direction $\omega_{o}$ is:
\begin{align}
L_{o}(x, \omega_{o}) =  L_{e}(x, \omega_{o}) + \int_{H^{+}} L_{i}(x, \omega_{i})f_{r}(x,\omega_{i},\omega_{o})
\cos\theta_{i} d\omega_{i}, \label{eq:RE}
\end{align}
where $L_{e}(x, \omega_{o})$ is the radiance emitted at $x$ in direction $\omega_{o}$. The reflected light is computed by integrating over all incident light directions $H^{+}$. $L_{i}(x, \omega_{i})$ is the  incident light, $f_{r}(x,\omega_{i},\omega_{o})$ is  the bidirectional reflectance distribution function, and $\cos\theta_{i}$ is the foreshortening.
A typical Monte Carlo estimator for $L_{o}(x, \omega_{o})$ is:
\begin{align}
\hat{L}_{o}(x, \omega_{o}) \approx  L_{e}(x, \omega_{o}) + \frac{1}{N} \sum_{i=1}^{N} \frac{L_{i}(x, \omega_{i})f_{r}(x,\omega_{i},\omega_{o})\cos\theta_{i}}{p(\omega_{i})},
\label{eq:MC_RE}
\end{align}
where $N$ is the number of sampled light paths, and $p(\omega_{i})$ is the probability density function used to sample incident directions  $\omega_{i}$. 

\paragraph*{Multi-Level Monte Carlo}

\emph{Multi-Level Monte Carlo} (MLMC) methods~\cite{MLMC1,MLMC_STAR} 
have been applied in various fields, but have not yet found widespread application in rendering. 
In this paper, we introduce a two-level Monte Carlo scheme (a subset of MLMC)
to turn the integration based on our NIRCs into an unbiased estimator. We begin with the standard Monte Carlo estimator for an integral $F$ over a domain $\mathcal{D}$:
\begin{equation} \label{eq:3}
F \approx \frac{1}{N} \sum_{i}^{N} \frac{f(X_{i})}{p(X_{i})},
\end{equation}
where $\{X_{i}\}$ are samples drawn from a probability distribution $p(x)$. 

Our two-level MC consists of 1) a cache-based estimator \(F_{c}\) with an approximating function \(f_{c}(x, \mathbf{w})\) and optimizable parameters 
\(\mathbf{w}\):
\begin{align}
F_{c} \approx \frac{1}{N_{c}} \sum_{i}^{N_{c}} \frac{f_{c}(X_{i}, \mathbf{w})}{p_{c}(X_{i})}, \label{eq:Fc}
\end{align}
and 2) a residual error estimator \(F_{r}\):
\begin{align}
F_{r} \approx \frac{1}{N_{r}} \sum_{i}^{N_{r}}\frac{f(Y_{i}) - f_{c}(Y_{i}, \mathbf{w})}{p_{r}(Y_{i})}. \label{eq:Fr}
\end{align}
We obtain the full two-level estimator \(F_{tl}\) as
\begin{align}
F_{tl} \approx F_{c} + F_{r}, \label{eq:Fml}
\end{align}
where \(X_{i} \sim p_c(X_{i})\) and \(Y_{i} \sim p_r(Y_{i})\). 
Note that we can choose the number of samples for each estimator, \(N_{c}\) and \(N_{r}\), individually. 
If the optimizable function \(f_{c}\) approximates the original function \(f\) well and is cheap to evaluate we can allocate more computation to the estimator \(F_{c}\) using more samples \(N_{c}\) per frame. The basis for \(f_{c}\) is the Neural Incident Radiance Cache (NIRC), which will be introduced and defined in \Cref{sec:nirc}.
\begin{figure}
    \centering
    \includegraphics[width=.5\textwidth]{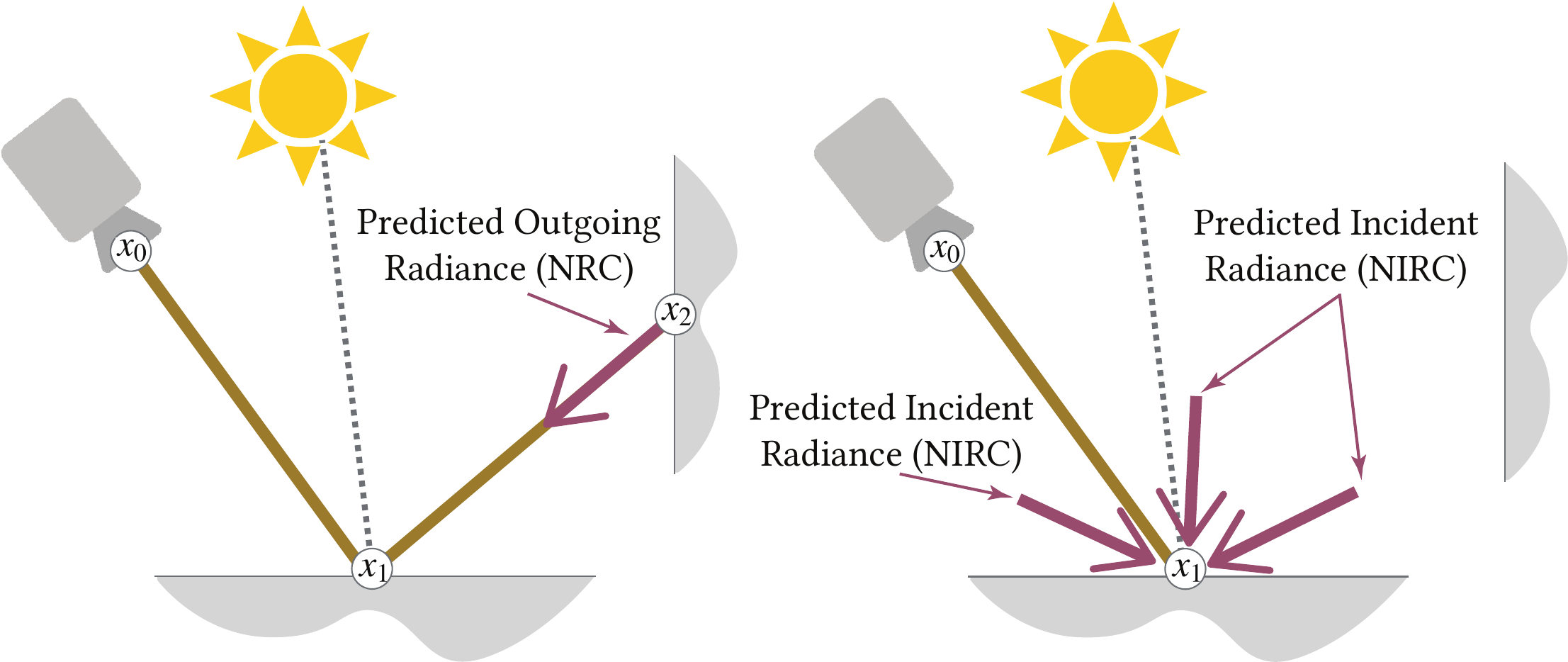}

    \caption{This figure illustrates the application of the \emph{Neural Radiance Cache} (NRC, left) \cite{NRC} and our \emph{Neural Incident Radiance Cache} (NIRC, right) in the biased path tracing. In NRC, a path is traced from the camera \(x_{0}\) to surface point \(x_{1}\), where it scatters and then terminates at \(x_{2}\) in the cache. In contrast, NIRC stops tracing already at \(x_{1}\) and estimates radiance using Monte Carlo, as described in \Cref{eq:L_c}, by sampling scattering rays based on BRDF and querying the NIRC to predict incident radiance. In practice, the exact termination point is determined by different heuristics for both models (see \Cref{sec:heuristics}).}
    \label{fig:paths}
\end{figure}

\section{Related Work}
\paragraph*{Radiance Caching} Since the seminal work of Ward et al.~\cite{ward1988}, numerous advances have been made in caching techniques. 
Important developments include the widely employed irradiance volumes~\cite{IrradianceVolume} and the introduction of radiance caching~\cite{Křivánek2005} using spherical harmonics for directional domain representation. These methods have seen significant enhancements for both offline~\cite{Dubouchet2017, Marco2018, Zhao2019} and real-time rendering~\cite{DDGI}, and  new approaches to processing and storing lighting information~\cite{Vardis2014Radiance, Silvennoinen2017, Rehfeld2014, Binder2018, pantaleoni2020online, Scherzer}. The popularity of neural networks leads to the Neural Radiance Cache (NRC)~\cite{NRC} which offers a method for dynamic scenes, learning during rendering and leveraging low-cost computation over memory access by utilizing fully-fused neural networks. A combination of NRC with multiresolution hash encodings significantly accelerates training convergence while improving the quality of radiance signal reconstruction~\cite{InstantNGP}. Moreover,   pre-trained neural networks have been shown to support dynamic parameters such as mesh positions or light parameters~\cite{NPRT,ActiveExploration} and still predict a reasonable approximation of global illumination. 
In contrast to the aforementioned works, we focus on efficient caching of incident radiance, a step that significantly improves rendering performance and lets us use the cache for an efficient unbiased Monte Carlo estimator for real-time rendering.

\paragraph*{Neural Methods} Neural Methods comprise a variety of techniques: Xie et al.~\cite{NeuralSTAR} provide a detailed analysis of such techniques with an emphasis on \emph{Neural Radiance Fields} (NeRFs)~\cite{NERF}.
Neural Radiosity~\cite{NeuralRadiosity} has parallels with traditional radiosity techniques but computes a solution to the rendering equation by applying neural networks to minimize the residuals.  Additionally, deep neural networks can be used for guiding path tracing by generating scattering directions~\cite{NIS, NeuralPhotonGuiding, NeuralSubsurfaceRGL} and by this enhancing the efficiency of Monte Carlo integration in light transport simulation.

\paragraph*{Control Variates (CV)} A control variate $g(x)$, with a known expected value $E[g(x)]$, can be used to estimate the integral of a function $f(x)$ with:
\begin{equation} \label{eq:cv}
E[f(x)] \approx E[g(x)] + \frac{1}{N} \sum_{i=1}^{N} \frac{f(X_i) - g(X_i)}{p(X_{i})},
\end{equation}
where $X_i$ are points sampled according to a probability distribution function \(p\). 
Control Variates show a certain similarity to Multi-Level Monte Carlo, however, they differ in that the expected value $E[g(x)]$ is computed analytically. This is also an inherent limitation: CVs require a mathematical framework for analytical integration over the domain. %
In our case, the expected value is derived from the cache-based integral in a MLMC framework.  

\paragraph*{Neural Control Variates (NCVs)  ~\cite{NCV}} NCVs use Normalizing Flows to model Control Variates. These models are primarily chosen for their capability to ensure that the final integral of the function equals to 1 multiplied by a prediction of another neural network.
Despite their theoretical appeal, experimental results on the performance of Normalizing Flow models~\cite{NCV, NIS, NeuralAppearance} have demonstrated notable computational demands for both training and inference. Moreover, NCVs require to maintain a second neural network which predicts the integral coefficient. These resource requirements often exceed what is practical for real-time or interactive rendering.

One motivation for our work was the potential advantage of MLMC over (N)CV in reducing the variance of the residual error. This requires a computationally efficient mathematical basis for numerically estimating the integral, which we achieve with NIRCs.

\section{Neural Incident Radiance Cache}
\label{sec:nirc}
The computation of \(L_{i}(x, \omega_{i})\) is the most resource-intensive part
of \Cref{eq:RE}. This is due to the need to trace rays, find
intersection points, fetch surface material parameters and evaluate
\Cref{eq:RE} for a set of positions and directions, which often results in
noisy estimations. 
In this work, we address this problem by firstly splitting \(L_{i}(x, \omega_{i})\) term into indirect lighting part \(L_{ind}(x, \omega_{i})\) and direct \(L_{nee}(x, \omega_{i})\):
\begin{align}
    \label{eq:Lsplit}
    L_{i}(x, \omega_{i}) = L_{ind}(x, \omega_{i}) + L_{nee}(x, \omega_{i})
\end{align}
since we will use MIS combination with next event estimation for \(L_{nee}(x, \omega_{i})\). Secondly, we propose a cache that approximates \(L_{ind}(x, \omega_{i})\) via a neural network while computing the other terms normally:
\begin{align} \label{eq:11}
L_{ind} \approx f_{c}(x, \omega_{i}, \omega_{o}, \phi, \mathbf{w} ) =
n_{i}(x, \omega_{i},
\phi,\mathbf{w})f_{r}(x,\omega_{i},\omega_{o})\cos\theta_{i}.
\end{align}
Here \(n_{i}(x, \omega_{i}, \phi, \mathbf{w} )\) is a neural network with optimizable parameters (weights) \(\mathbf{w}\), conditioned on a surface position \(x\), incident vector \(\omega_{i}\), and a set of additional features \(\phi\) provided for facilitating the search of correlation between target values and input parameters. This function \(n_{i}\) serves as the \emph{Neural Incident Radiance Cache} (NIRC), approximating incoming lighting from all (hemi)-spherical directions onto a shading point.

Using \Cref{eq:11}, we are able to formulate an unbiased Multi-Level Monte Carlo estimator for computing the rendering equation:
\begin{align}
    \label{eq:L_mlmc}
      \hat{L}_{o}(x, \omega_{o}) &\approx  L_{e}(x, \omega_{o}) + \hat{L}_{c}(x, \omega_{o}) + \hat{L}_{r}(x, \omega_{o})\\
    \label{eq:L_c}
   \hat{L}_{c}(x, \omega_{o}) &\approx \frac{1}{N_{c}} \sum_{i=1}^{N_{c}} \frac{n_{i}(x, \omega_{i}, \phi,\mathbf{w})f_{r}(x,\omega_{i},\omega_{o})\cos\theta_{i}}{p(\omega_{i})} \\
    \label{eq:L_r}
\hat{L}_{r}(x, \omega_{o}) &\approx \frac{1}{N_{r}} \sum_{i=1}^{N_{r}}
\frac{(L_{i}(x, \omega_{i})-n_{i}(x, \omega_{i},
\phi,\mathbf{w}))f_{r}(x,\omega_{i},\omega_{o})\cos\theta_{i}}{p(\omega_{i})}
\end{align}

\subsection{Neural Network Architecture}
\setlength{\belowcaptionskip}{-10pt}
\begin{figure}
\def\svgwidth{\linewidth}
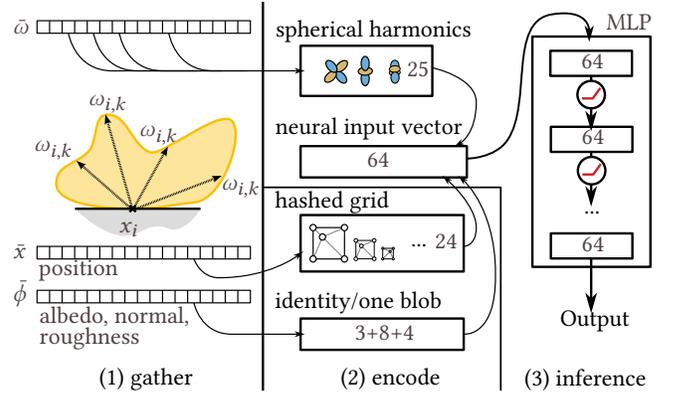
\caption{The stages of our neural inference pipeline. Initially, requests in the
path tracing pass are gathered (1), utilizing multiple buffers to store per-surface
parameters along with the incident direction vectors \(\omega_{i}\). Next, the
per-surface parameters are encoded (2) and aggregated.
Finally, the spherical harmonics coefficients for the
requested \(\omega_{i}\) are calculated and appended in shared memory as input for the
multi-layer perceptron inference (3).
Our optimized implementation never stores the neural input and output vectors in global memory.
}
\label{fig:architecture}
\end{figure}
\setlength{\belowcaptionskip}{0pt}

Inspired by previous work we leverage the power of fully-fused executed
multi-layer perceptions (MLP) \cite{NRC} to represent a radiance field.
MLP can efficiently approximate signals defined over a
five-dimensional manifold conditioned on surface a position (3D) and an outgoing direction (2D), even with real-time performance on GPUs.
However the expressive power of a tiny MLP is limited, and its convergence rate
is adversely affected if input parameters are forwarded without applying any
nonlinear transformation to their domain space. Therefore, we employ encodings
inspired by the following neural rendering literature \cite{InstantNGP, NRC,
NERF_REF}:

\begin{itemize}
    \item The shading point position \(x\) is used for querying a multiresolution hash encoding with trainable features. The resolution is adjusted based on the complexity of the scenes, but in the majority of our experiments, we used 12 levels with 2 features per level.
    \item To enhance the NIRC's precision in the spherical domain, we propose estimating spherical harmonics coefficients for the input parameter \(\omega_{i}\) \cite{NERF_REF}. Our experiments suggest that using 4 bands is a good compromise between increasing the neuron count and maintaining the network's directional representational quality.
    \item As proposed by \cite{NRC}, using the shading normal, albedo, and roughness as additional input parameters \(\phi\) for NIRC, without altering their corresponding encodings.
\end{itemize}

\setlength{\abovecaptionskip}{0pt}
\setlength{\belowcaptionskip}{0pt}
\begin{figure}[ht]
    \centering    
    \includeinkscape[width=0.48\textwidth]{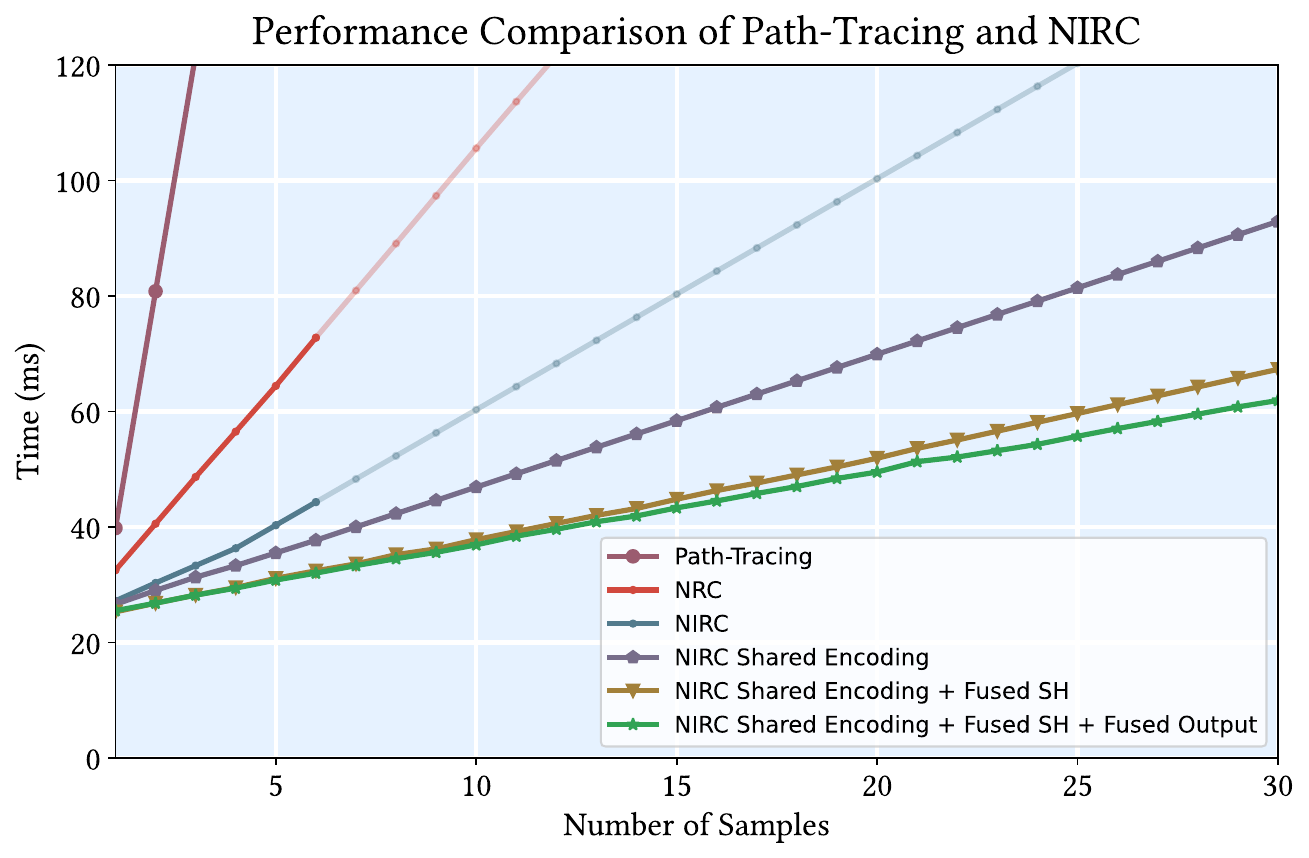}
    \caption{Scalability and ablation study with the Bistro Exterior scene
    rendered on an RTX 3080 at 1080p. We show incident lighting computed using
    conventional path tracing and our NIRC; the latter has been tested with
    various optimization strategies. The results demonstrate improved
    scalability of the NIRC method, particularly when it is optimized
    by exploiting shared surface encodings (1.74$\Uparrow$ improvement), fused
    spherical harmonics coefficients computation (2.62$\Uparrow$), and direct
    output forwarding from the shared memory using atomic variables
    (3.64$\Uparrow$). For the NRC and the naive NIRC implementations, we show
    extrapolated performance timings after the 6th frame because of memory
    limitations (light blue and light red).}
    \label{fig:performance_plot}
\end{figure}
\setlength{\abovecaptionskip}{10pt}
A notable challenge in integrating high-quality encodings before executing an
MLP is performance degradation. Trainable features, used in hash encoding for shading
point positions, require us to fetch numerous float variables from scattered memory locations.
However, it is possible to amortize memory accesses across all requests for the
same shading surface, given the way we use the cache (\Cref{eq:L_c}). This idea
is extended to other shared input parameters \(\phi\) and their
corresponding encoding outputs.
Furthermore, the computation of spherical harmonics coefficients for a
relatively high number of bands can create bandwidth problems,
moving 25 float variables into global memory per request.
To alleviate this, we advocate
for a fully-fused execution approach, not just for MLP inference, but also for
the spherical harmonics encoding layer. Our algorithm calculates the necessary
coefficients and immediately stores them in a section of shared memory
reserved for future inference as input neurons. Additionally, we found that
packing and compressing directional components maximizes system bandwidth.

Lastly, we suggest an optimization that avoids intermediate buffers used
in the original tiny-cuda-nn framework during output:
We write inference results into a final frame buffer directly from shared memory using atomic variables. This allows us to increase the number of
neural samples per pixel within the time budget.
We illustrate the final inference pipeline in \Cref{fig:architecture}.

By integrating these natural optimizations, which emerge from our cache design
choices and the manner in which we use it to estimate the Rendering Equation, we
achieve up to 3.64$\times$ speedup compared to the original neural inference
pipeline, depending on the sample count, as demonstrated in
\Cref{fig:performance_plot}. The performance cost of one neural
sample can be as low as 2.69\% of the corresponding path tracing simulation
in our experiments. This significant performance improvement allows us to
allocate more samples per pixel, reducing the variance of the stochastic
estimator $L_c$.
\subsection{Radiance Cache Optimization}
\setlength{\abovecaptionskip}{1pt}
\begin{figure}
    \centering
    \includegraphics[width=.5\textwidth]{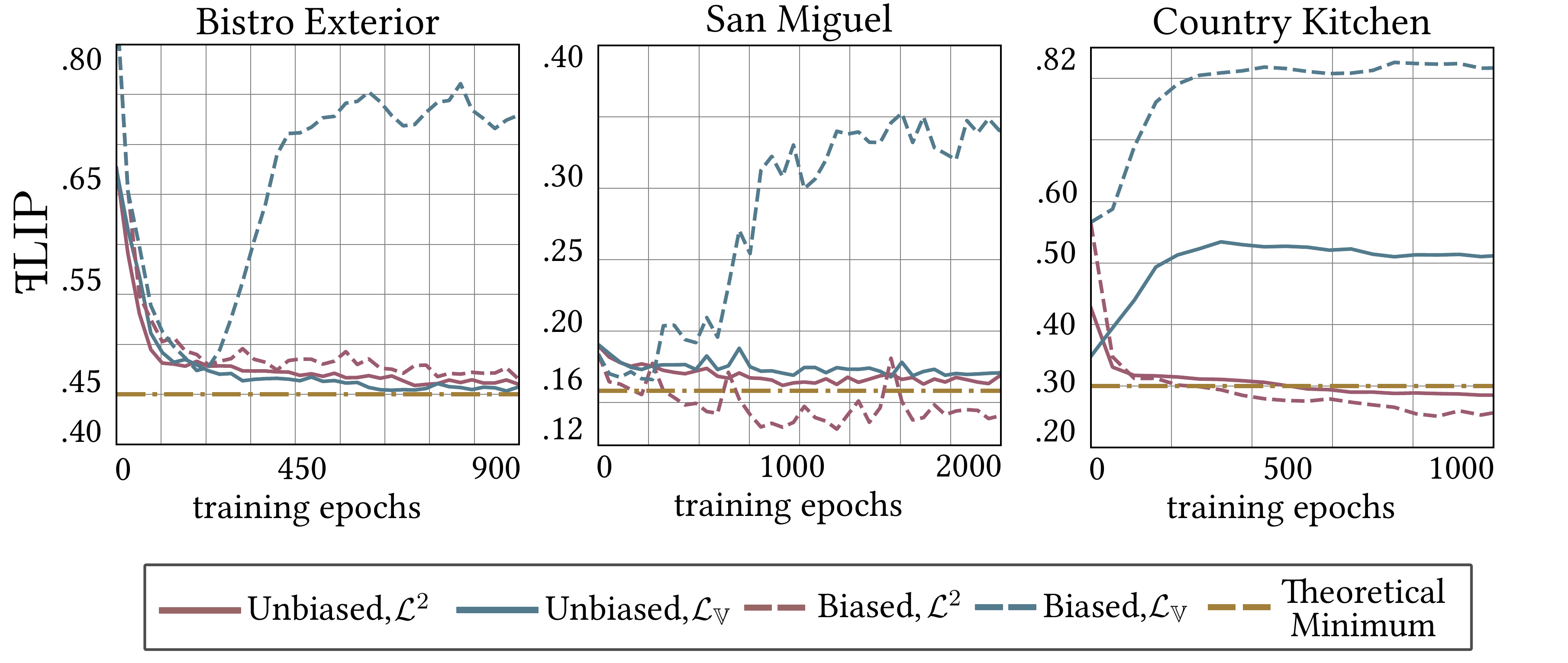}

    \caption{We explore the implications of employing $\mathcal{L}^{2}$ and
    variance-based $\mathcal{L}_{\mathbb{V}}$ loss functions for the neural
    cache optimization with our two-level MC estimator and parameters
    $N_{c} = 4$ and $N_{r} = 1$. We compute the \reflectbox{F}LIP error between
    our renders, utilizing Neural Incident Radiance Cache, and reference renders
    every training epoch. The results demonstrate a detrimental effect on the
    quality of renders of the biased estimator when the
    $\mathcal{L}_{\mathbb{V}}$ loss function is applied.
    This is because the variance-based loss does not necessarily minimize the residual $F_r$.
    The data also shows divergence during training, likely due to issues when
    estimating $F_r$ accurately,
    which is required for evaluating $\mathcal{L}_{\mathbb{V}}$.}
    \label{fig:optimization}
\end{figure}

To effectively optimize the trainable parameters \(\mathbf{w}\), we must
derive a suitable and practical loss function. Recall that Multi-Level Monte
Carlo (MLMC) methods aim to reduce the overall variance of the estimator. This
reduction is achievable when the variance of the residual estimator is
minimized. The residual estimator \(F_{r}\) \Cref{eq:Fr}, which can be defined
as \(F_{r} = \mathbb{E}\left[ \frac{f(X) - f_{c}(X,\mathbf{w})}{p_{r}(X)}
\right]\), represents the expectation of the normalized difference between the
function \(f\) and its controlled approximation \(f_{c}\). The variance of
\(F_{r}\) is given by
\begin{equation} \label{eq:8}
\mathbb{V}(\langle F_{r} \rangle) =  \int_{\mathcal{D}}\left(\frac{f(x) - f_{c}(x,\mathbf{w})}{p_{r}(x)} - F_{r}\right)^{2}p_{r}(x)\,dx.
\end{equation}
We can derive a numerical one-sample estimator for this variance of \(\langle
F_{r} \rangle\) which is then utilized as a loss function:
\begin{equation} \label{eq:var}
\langle\mathbb{V}(\langle F_{r} \rangle)\rangle \approx \left(\frac{f(x) - f_{c}(x,\mathbf{w})}{p_{r}(x)} - F_{r}\right)^{2}\frac{p_{r}(x)}{q_{r}(x)}.
\end{equation}
Here, \(q_{r}(x)\) denotes the probability density function of the sampling
points used for estimating the variance. We employ the same sampling methods for
estimating the variance in \Cref{eq:var} as well as for estimating the
residual error in \Cref{eq:Fr}, leading to $q_{r}=p_{r}$.
This results in the final variance-based loss function:
\begin{equation} \label{eq:final_loss}
\langle\mathcal{L}_{\mathbb{V}}(f, f_{c}, \mathbf{w})\rangle \approx \left(\frac{f(x) - f_{c}(x,\mathbf{w})}{p_{r}(x)} - F_{r}\right)^{2}.
\end{equation}
We observe that we essentially obtain a squared distance between the cache and ground truth function, offset by the residual error's expectation. This suggests the optimization problem has potentially many solutions for any possible shift value.

Empirical experiments shown in \Cref{fig:optimization} highlight the adverse impact of integrating the residual error's expectation into the optimization point, justifying our reliance on the basic squared difference loss function:
\begin{equation} \label{eq:basic_diff_loss}
\langle \mathcal{L}^{2}(f, f_{c}, \mathbf{w}) \rangle = \frac{(f(x)-f_{c}(x, \mathbf{w}))^{2}}{p_{r}(x)}
\end{equation}

Considering the high dynamic range of the data we divide the metric by the squared \(f(x)\) to ensure higher gradient weights for dark regions of a rendering scene as in \cite{RelativeLoss}:
\begin{equation} \label{eq:12}
\langle \mathcal{L}_{rel}^{2}(f, f_{c}, \mathbf{w}) \rangle = \frac{(f(x)-f_{c}(x, \mathbf{w}))^{2}}{p_{r}(x)(sg(f_{c}(x, \mathbf{w})^{2})+\epsilon)}
\end{equation}

Where \(\epsilon\) safeguards against division by zero and \(sg(z)\) denotes that gradient computations for operation \(z\) are not propagated.

\begin{figure*}[htbp]
\setlength{\abovecaptionskip}{0pt}
    \centering
    
\footnotesize
\renewcommand{\arraystretch}{0.5} %
\setlength\tabcolsep{3pt} %
\begin{tabular}{@{}
  p{.020\linewidth}
  @{\hspace*{.000\linewidth}}>{\centering\arraybackslash}p{.112\linewidth}
  @{\hspace*{.003\linewidth}}>{\centering\arraybackslash}p{.112\linewidth}
  @{\hspace*{.003\linewidth}}>{\centering\arraybackslash}p{.112\linewidth}
  @{\hspace*{.003\linewidth}}>{\centering\arraybackslash}p{.112\linewidth}
  @{\hspace*{.003\linewidth}}>{\centering\arraybackslash}p{.112\linewidth}
  @{\hspace*{.003\linewidth}}>{\centering\arraybackslash}p{.112\linewidth}
  @{\hspace*{.006\linewidth}}>{\centering\arraybackslash}p{.28\linewidth}}
& & \multicolumn{2}{c}{\textbf{\small First Level Monte Carlo}} & \multicolumn{3}{c}{\textbf{\small Control Variates}} & \\
    \addlinespace[5pt]
    & GT Integrand & NIRC & SS NIRC & NCV & SH & VMF & EMA $\mathbb{V}_{rel}(\langle F_{r} \rangle)$ vs Frame Number \\
    \addlinespace[1pt]
    & \textit{VRAM:} &  20MB + 50KB &  68.2MB + 55KB & 20MB + 2MB & \multicolumn{1}{c}{69.1MB}  & 71MB & \\
    \addlinespace[1pt]
    \multirow{2}{*}[0.5cm]{\rotatebox{90}{\small{Cornell Box}}} &
    \includegraphics[width=\linewidth]{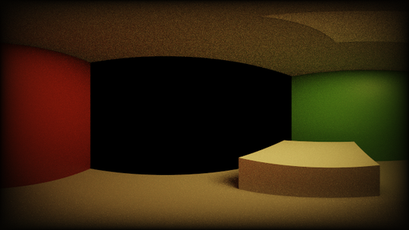} &
    \includegraphics[width=\linewidth]{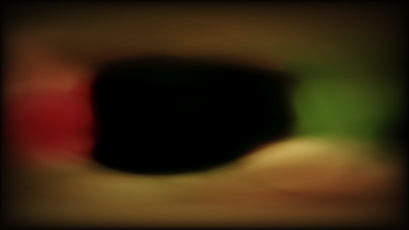} &
    \includegraphics[width=\linewidth]{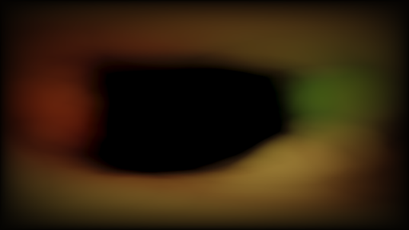} &
    \includegraphics[width=\linewidth]{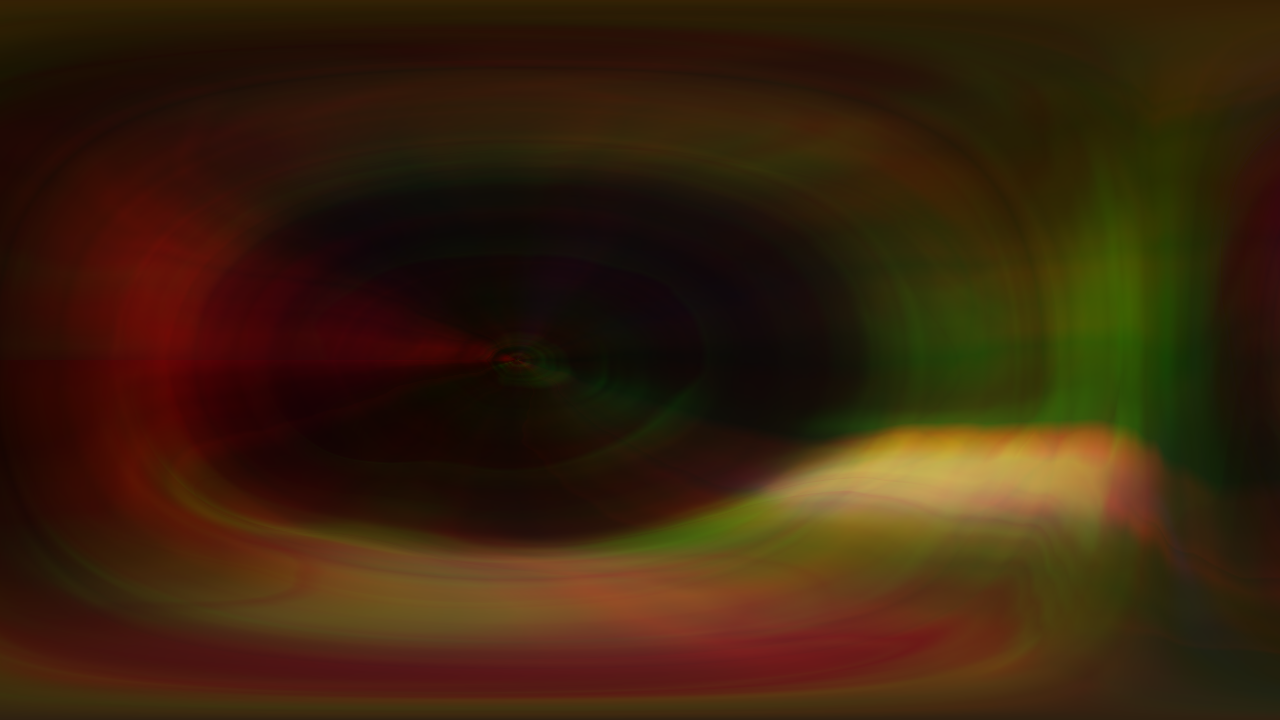} &
    \includegraphics[width=\linewidth]{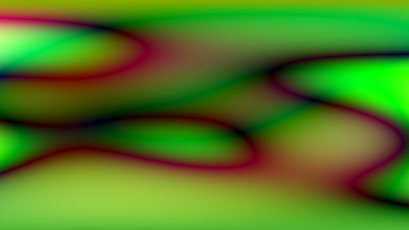} &
    \includegraphics[width=\linewidth]{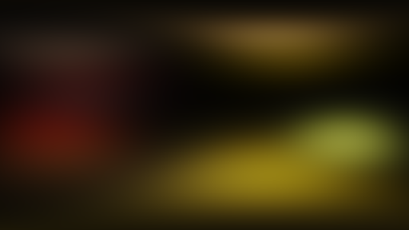} &
    \multirow{4}{*}[1cm]{\includeinkscape[width=\linewidth]{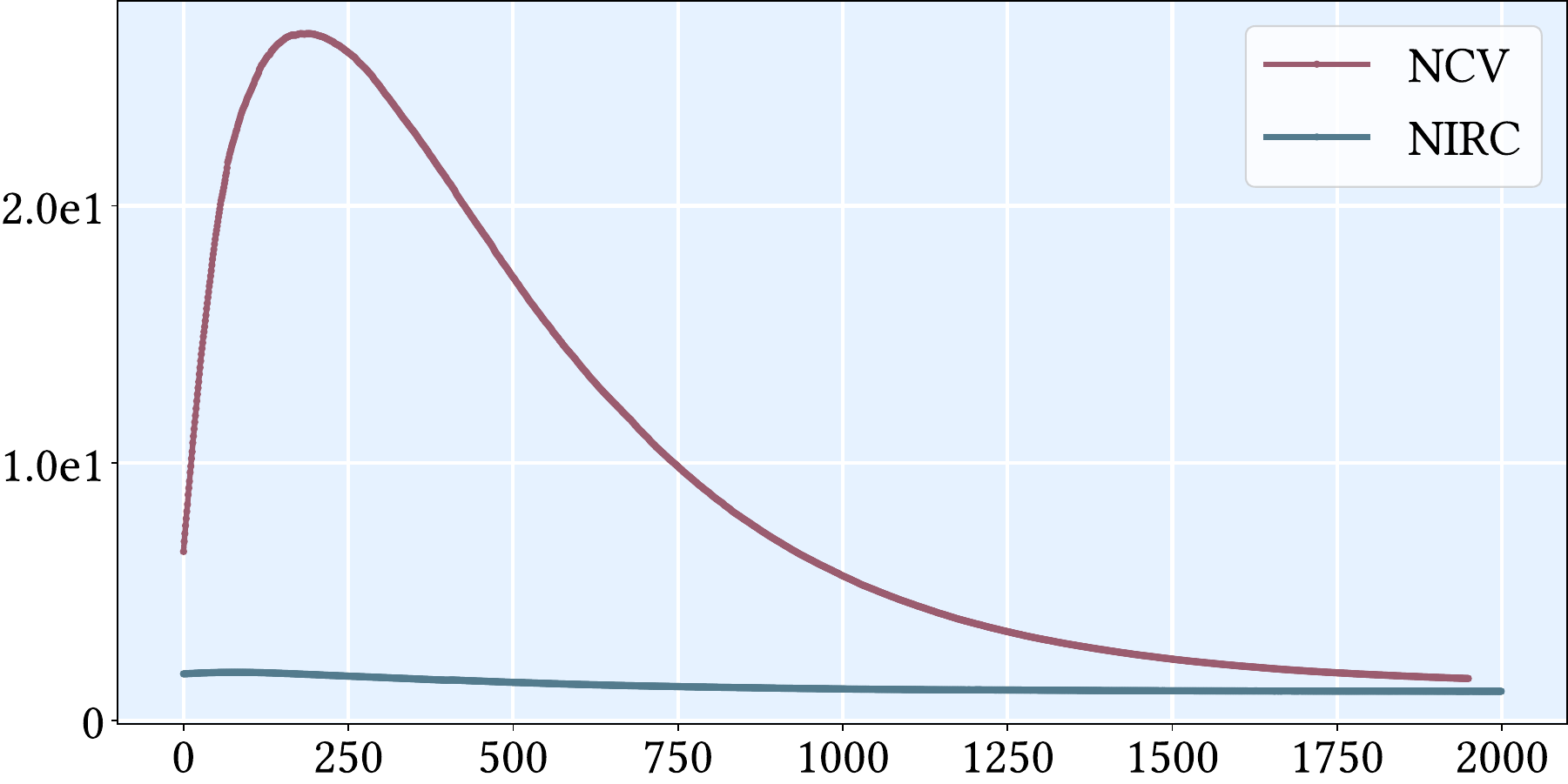}} \\
    &
    \includegraphics[width=\linewidth]{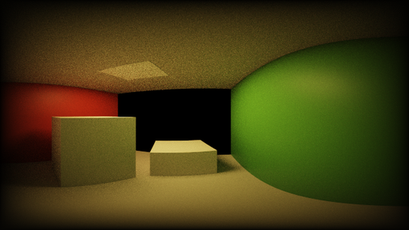} &
    \includegraphics[width=\linewidth]{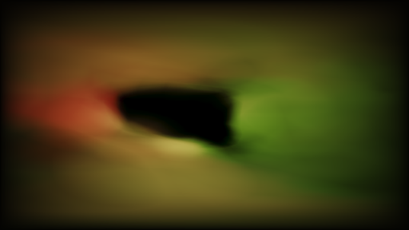} &
    \includegraphics[width=\linewidth]{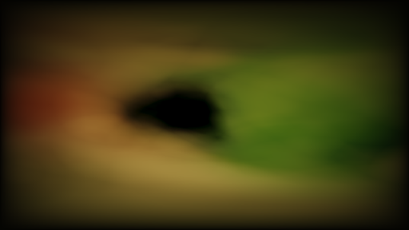} &
    \includegraphics[width=\linewidth]{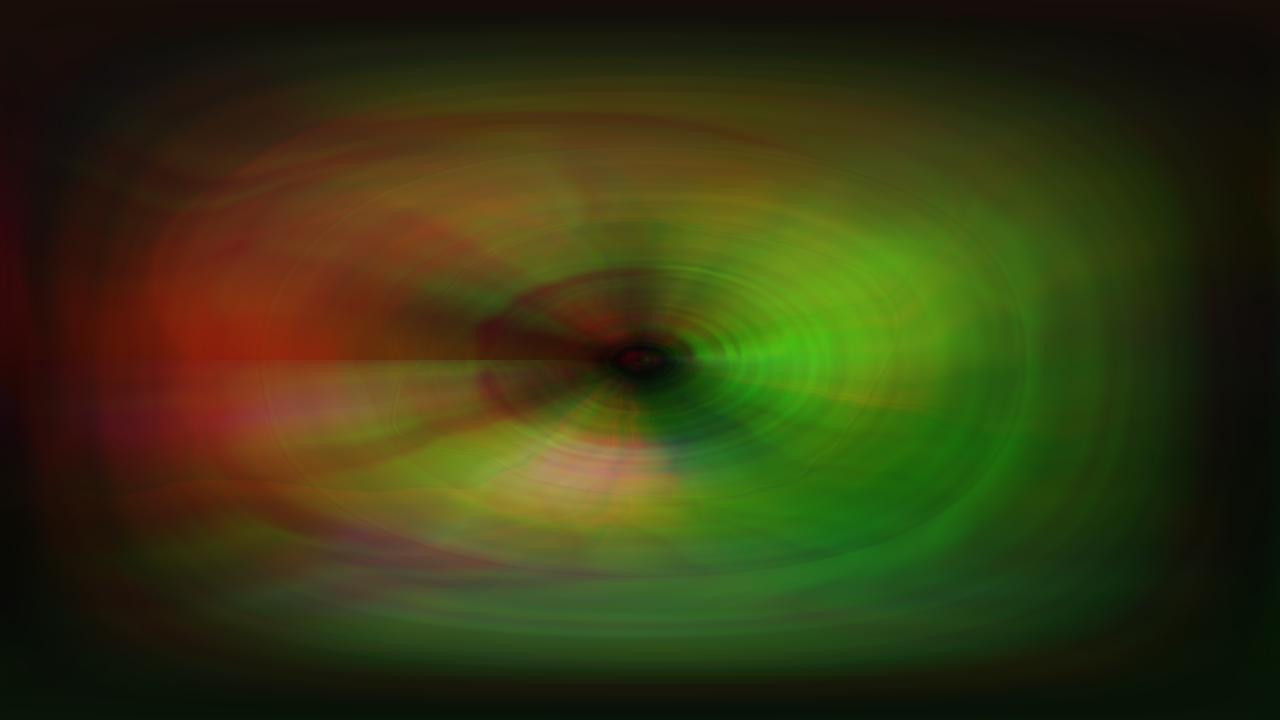} &
    \includegraphics[width=\linewidth]{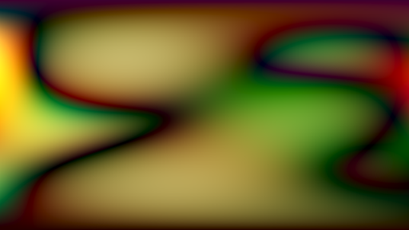} &
    \includegraphics[width=\linewidth]{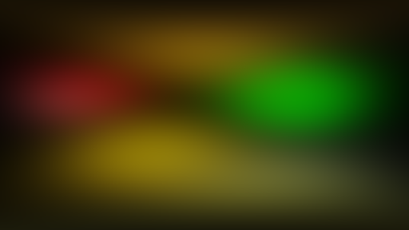} & \\
    & $\mathbb{V}_{rel}(\langle F_{r} \rangle)$: & \small \textbf{1.13} &  \small \textbf{1.43} & \small 1.35 & \small 2614   & \small 3.39 & \\
    &   &   &  & & & & \\
    \multirow{2}{*}[0.8cm]{\rotatebox{90}{\small{Bistro Exterior}}} &
    \includegraphics[width=\linewidth]{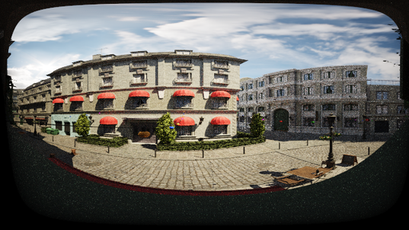} &
    \includegraphics[width=\linewidth]{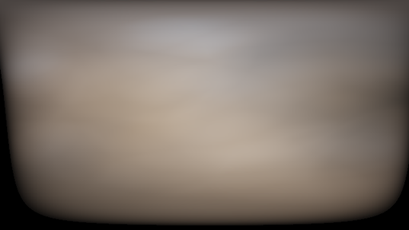} &
    \includegraphics[width=\linewidth]{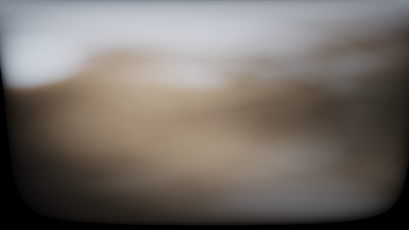} &
    \includegraphics[width=\linewidth]{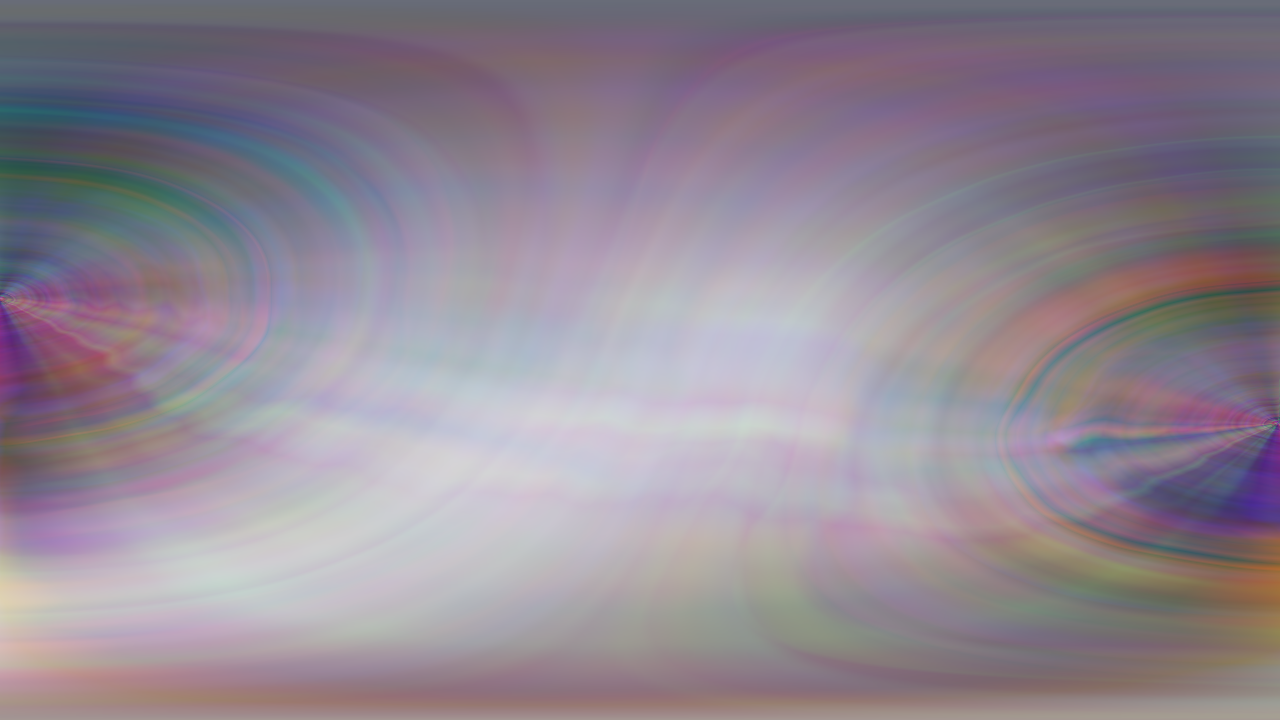} &
    \includegraphics[width=\linewidth]{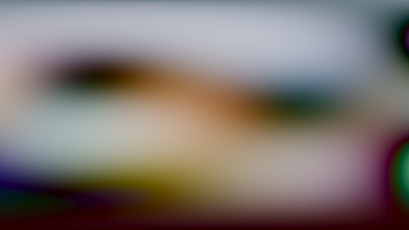} &
    \includegraphics[width=\linewidth]{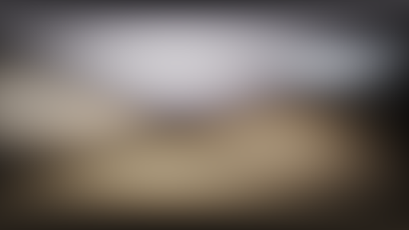} &
    \multirow{4}{*}[1cm]{\includeinkscape[width=\linewidth]{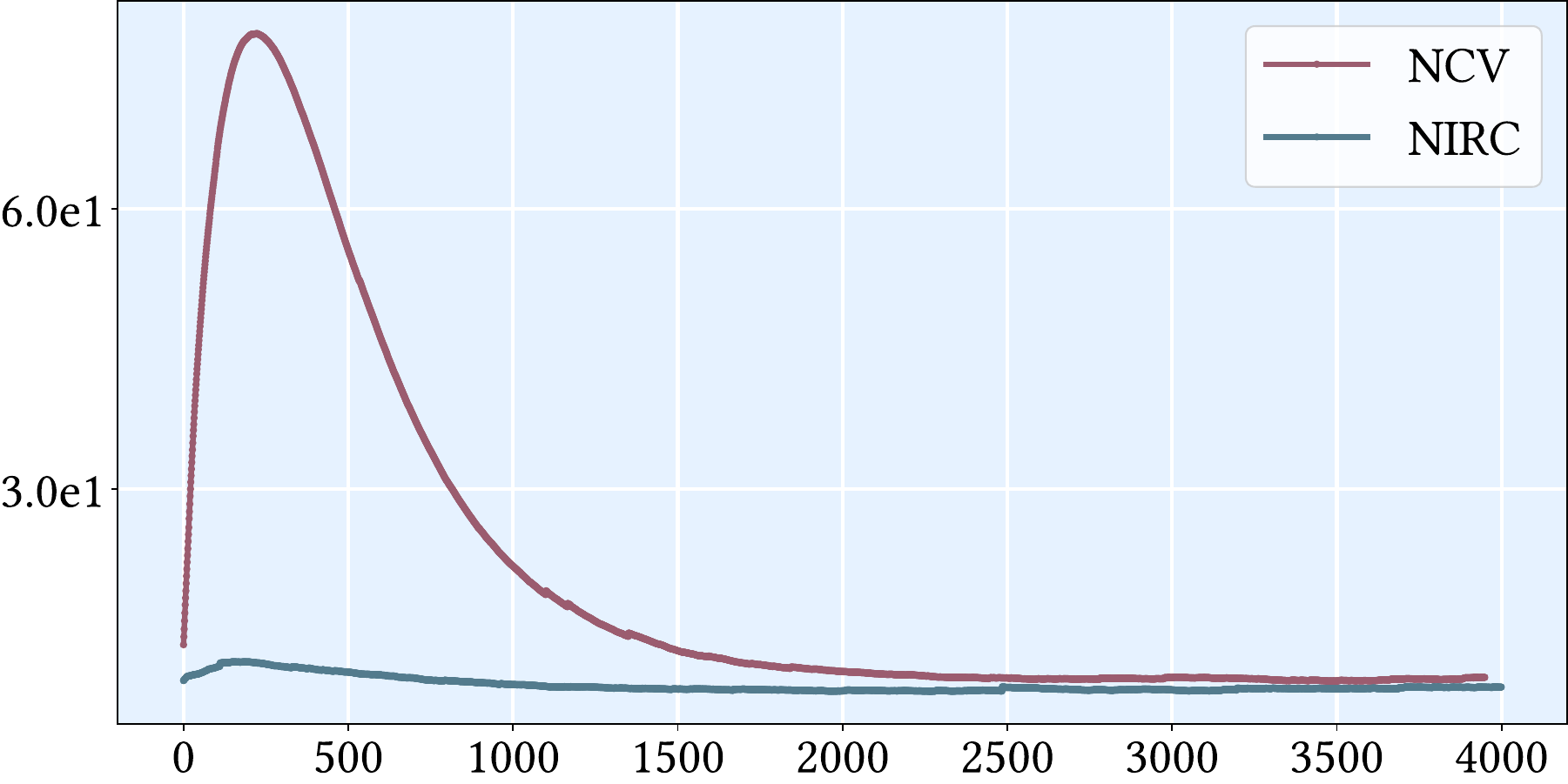}} \\
    &
    \includegraphics[width=\linewidth]{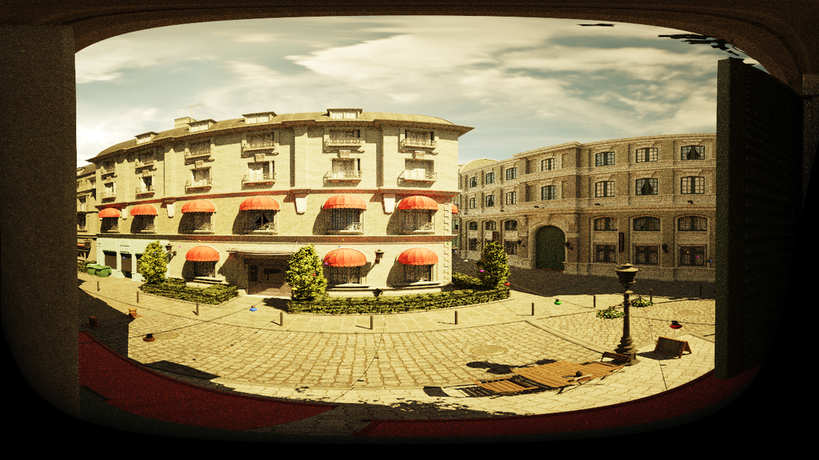} &
    \includegraphics[width=\linewidth]{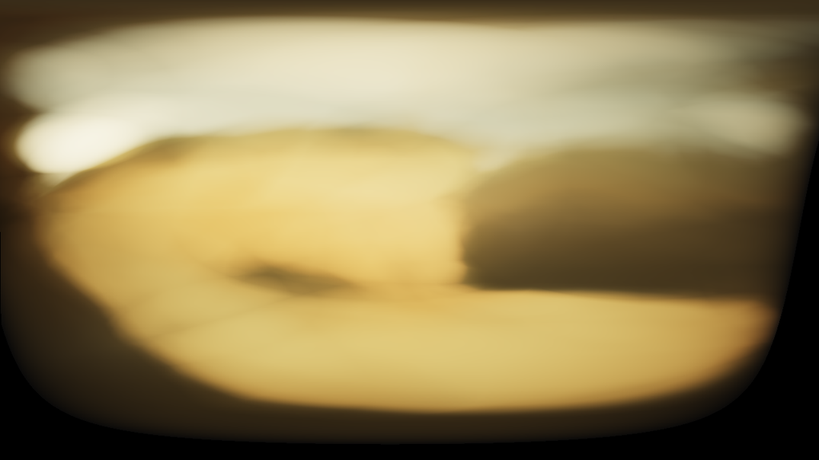} &
    \includegraphics[width=\linewidth]{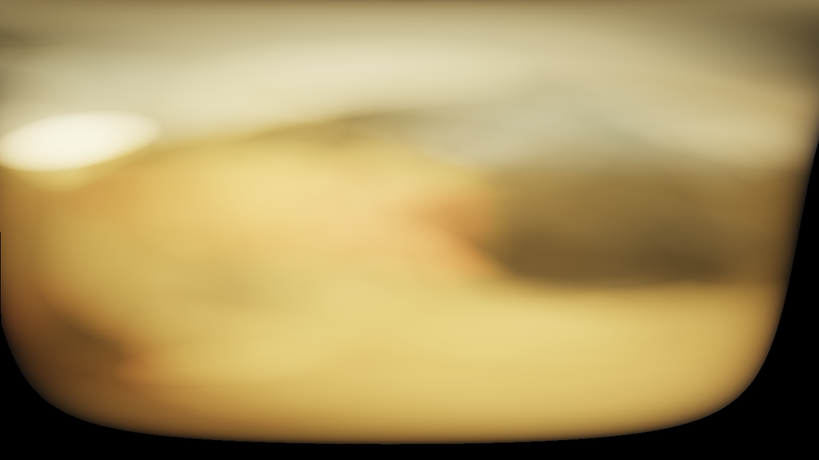} &
    \includegraphics[width=\linewidth]{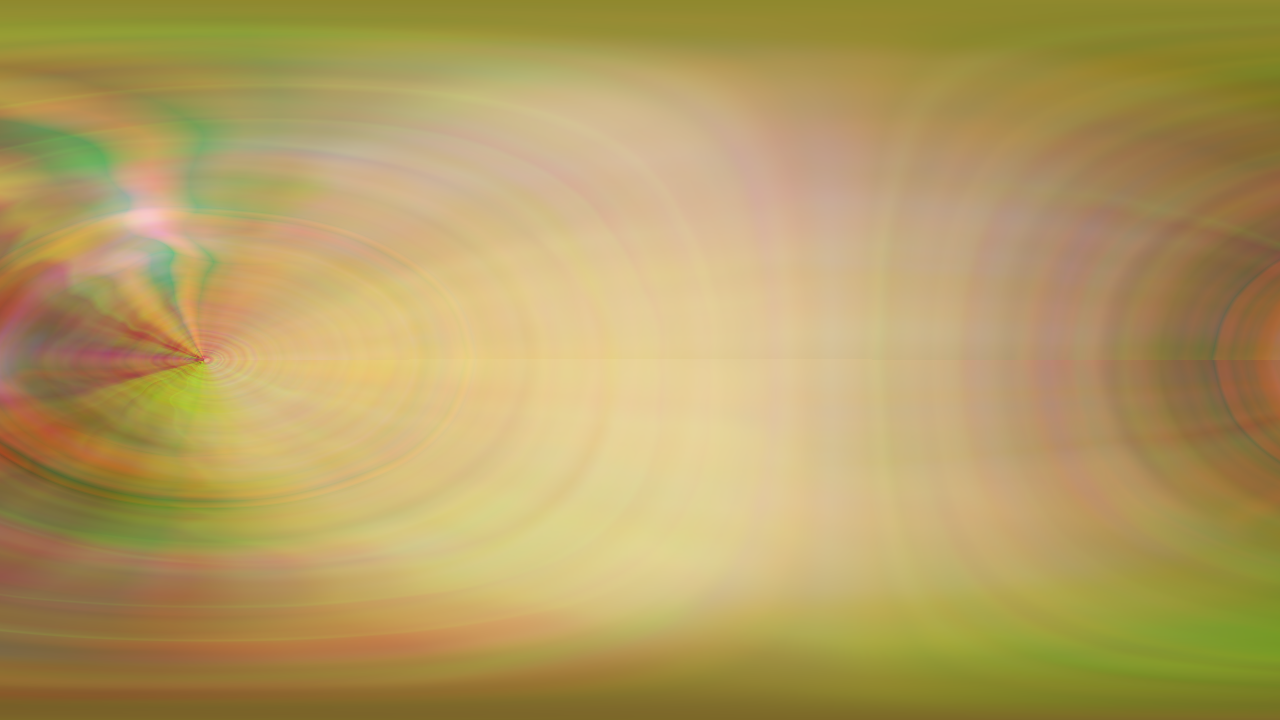} &
    \includegraphics[width=\linewidth]{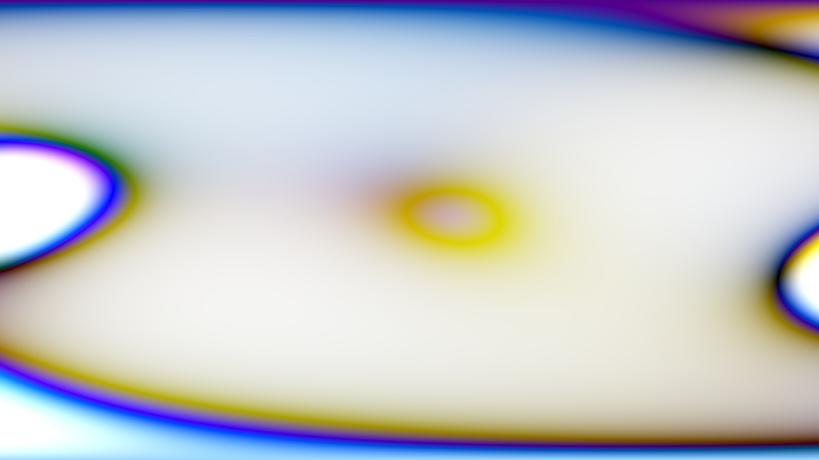} &
    \includegraphics[width=\linewidth]{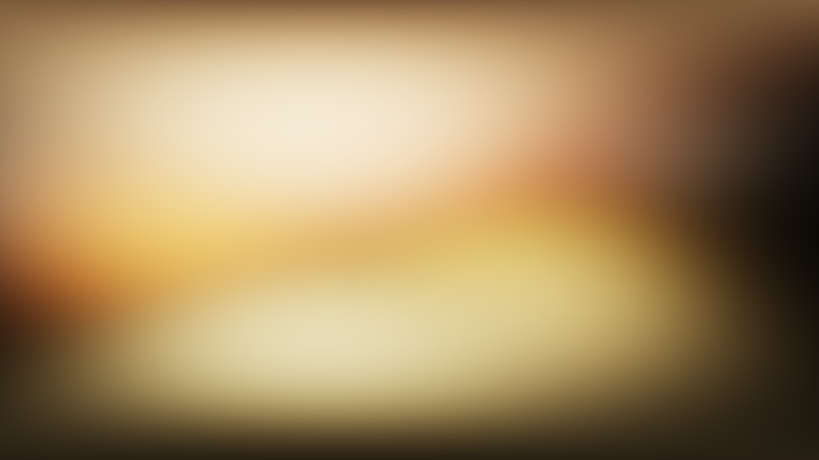} & \\
    & $\mathbb{V}_{rel}(\langle F_{r} \rangle)$: & \small \textbf{9.2} &  \small \textbf{8.8} & \small 10.12 & \small 828.3 & \small 30.0 & \\
    &   &   &  & & & & \\
    \multirow{2}{*}[0.7cm]{\rotatebox{90}{\small{Country Kitchen}}} &
    \includegraphics[width=\linewidth]{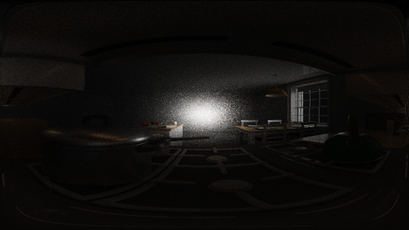} &
    \includegraphics[width=\linewidth]{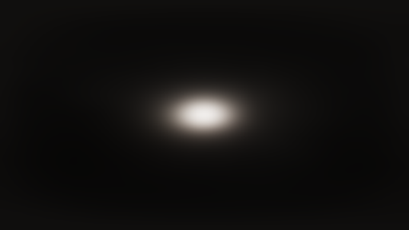} &
    \includegraphics[width=\linewidth]{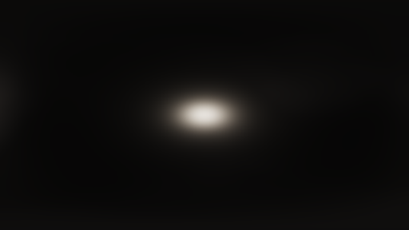} &
    \includegraphics[width=\linewidth]{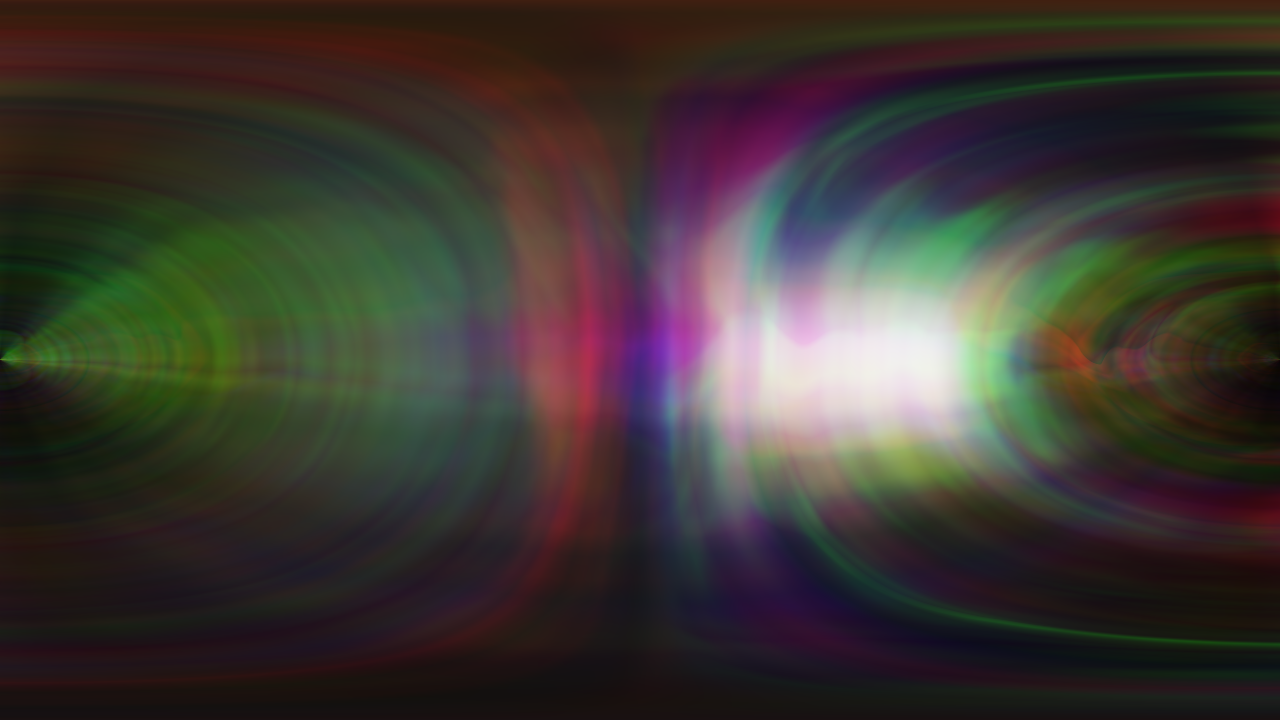} &
    \includegraphics[width=\linewidth]{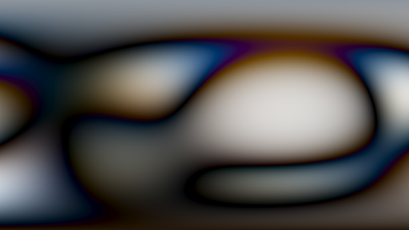} &
    \includegraphics[width=\linewidth]{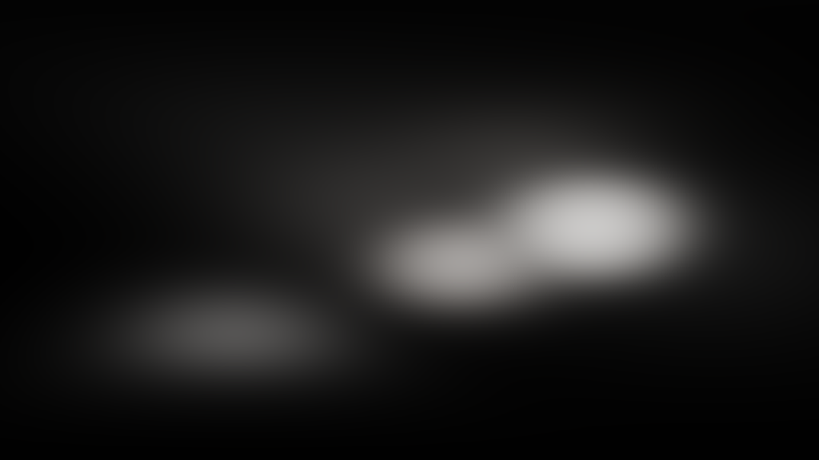} &
    \multirow{4}{*}[1cm]{\includeinkscape[width=\linewidth]{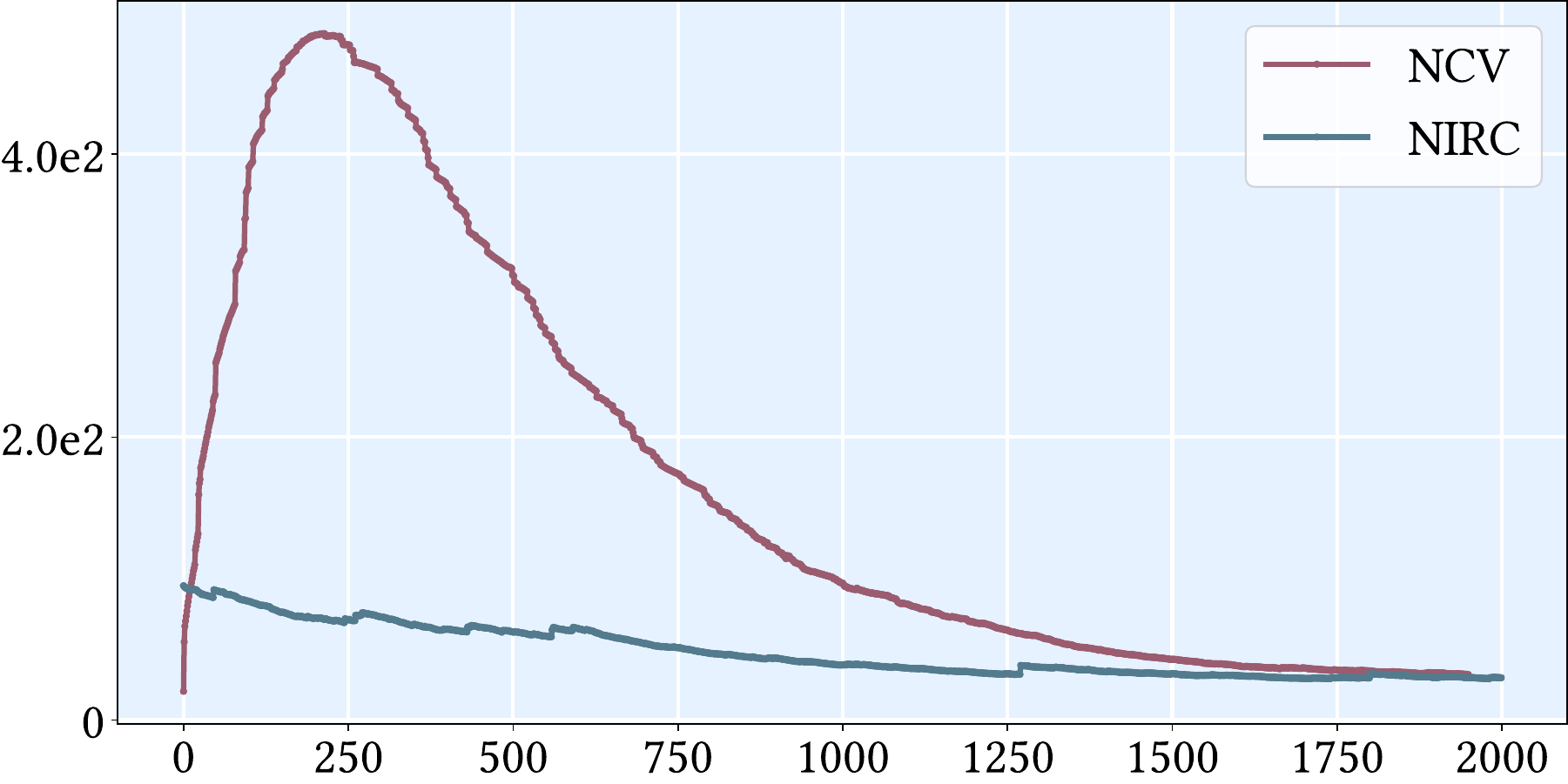}} \\
    &
    \includegraphics[width=\linewidth]{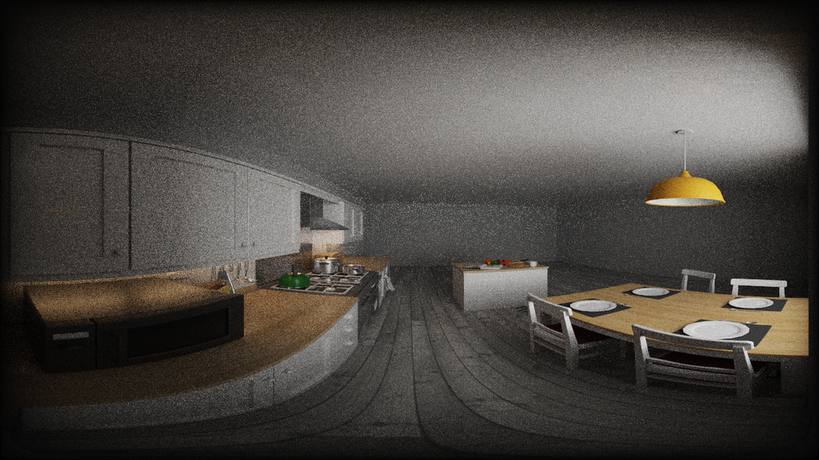} &
    \includegraphics[width=\linewidth]{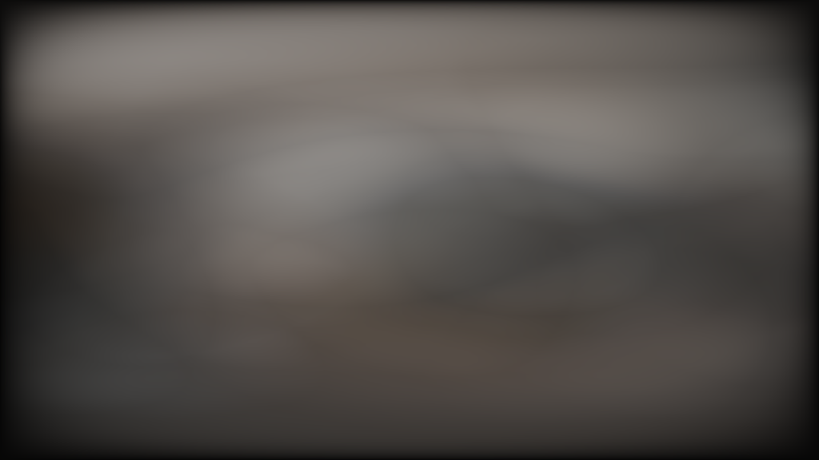} &
    \includegraphics[width=\linewidth]{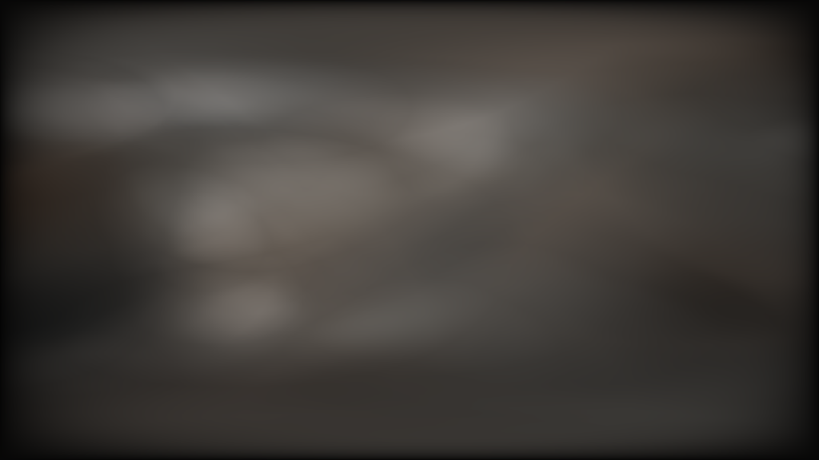} &
    \includegraphics[width=\linewidth]{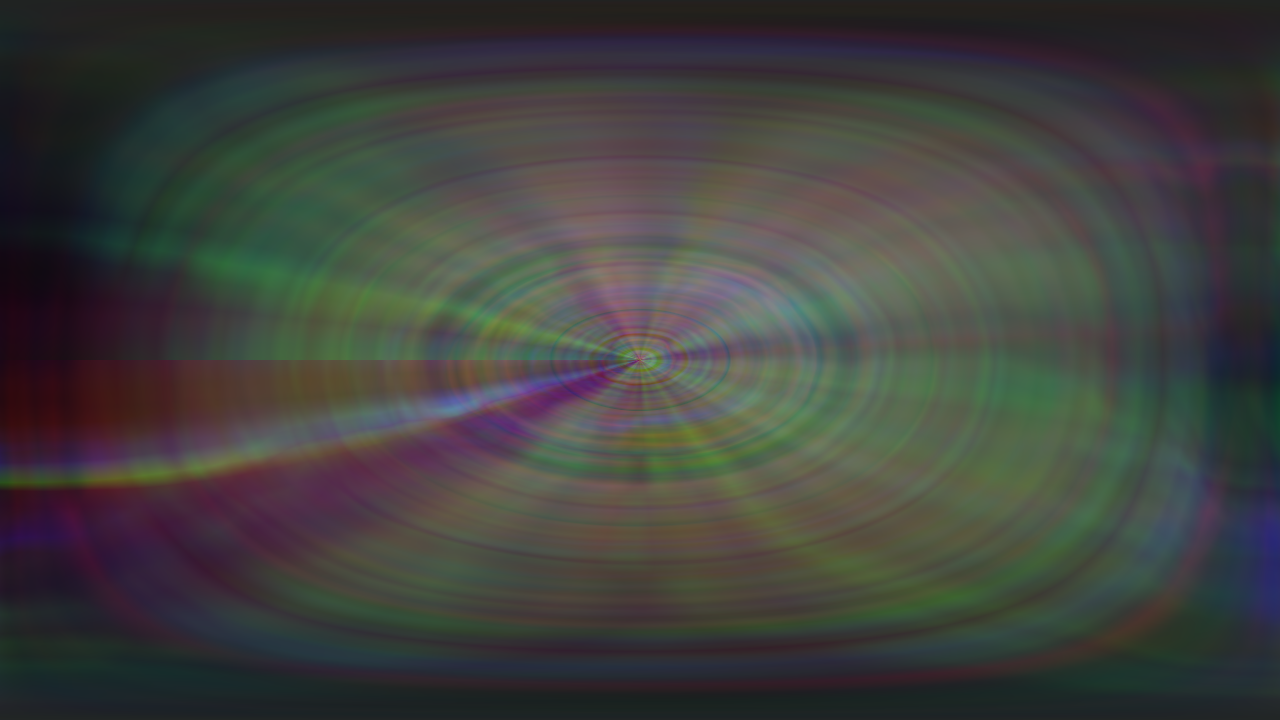} &
    \includegraphics[width=\linewidth]{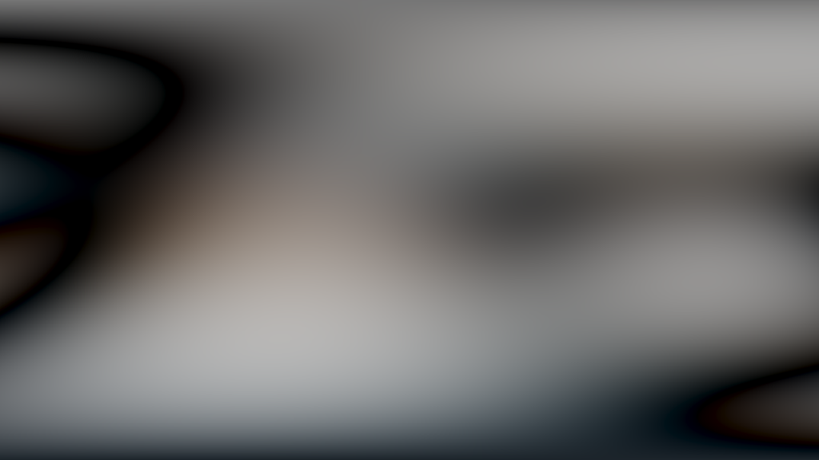} &
    \includegraphics[width=\linewidth]{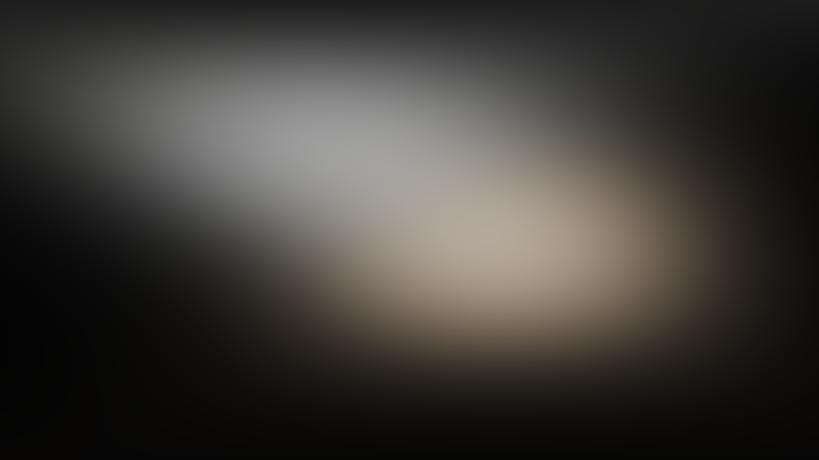} & \\
    & $\mathbb{V}_{rel}(\langle F_{r} \rangle)$: & \small \textbf{34.096} &  \small \textbf{89.90} & \small 44.87 & \small 1548 & \small 118630 & \\
    \\
   \\
\end{tabular}
\caption{Equal-memory footprint comparison between SHs, vMF mixtures, Neural Control Variates \cite{NCV} and two our neural caches: NIRC with world space multi-level hash encodings and screen space latent features (SS NIRC). This figure presents shading integrand values in selected surfaces with different materials in a set of scenes, displayed as tone-mapped latitude-longitude images. NCV, SHs and vMFs serve as control variates for analytical integral computations. NIRC focuses only on incident radiance on surfaces for use in MLMC estimators. SHs generally struggle with ringing and fail to capture high-frequency signals for specular surfaces. Although vMFs can theoretically adapt to any frequency, the stepwise EM algorithm can fail to optimize all lobes efficiently with limited noisy radiance samples. Neural Control Variates can evaluate integrals in closed form, but, as evidenced, the expressive power of NIRC demonstrates higher capacity. The average relative variance of the residual error estimator $\mathbb{V}_{rel}(\langle F_{r} \rangle)$ per each scene and method is presented under the images. The plots on the right show estimated $\mathbb{V}_{rel}(\langle F_{r} \rangle)$ using \emph{Exponential Moving Average} (EMA) per training frame for the both neural models. As illustrated the NIRC requires significantly fewer training frames to achieve even lower $\mathbb{V}_{rel}(\langle F_{r} \rangle)$ than the NCV. }
\label{fig:envmap_comp}
\end{figure*}

\subsection{Cache application}
Our algorithm follows the megakernel path tracing approach.
For each vertex \(x_j\), we generate \(N_{c}^j\) incident vectors \(\omega_{i}\),
where \(N_{c}^j\) denotes the number of samples of the NIRC
at each vertex $x_j$, to estimate the cached radiance term \(L_{c}\).
This process requires collecting all per-surface parameters and incident vectors
\(\omega_{i}\). These values are used for both cache
estimation and for calculating the additional terms in \Cref{eq:L_c} for $\omega_i$.
To estimate the residual error \Cref{eq:L_r}, we need a set of directions
$\omega$ which are uncorrelated to the directions used to compute \Cref{eq:L_c}.

We use independent sampling of $\omega_i$ according to the surface BSDF for both
$L_c$ and $L_r$.
Alternative sampling strategies like path guiding
for residual integral estimation have been explored in other studies \cite{NCV}.
To keep our study focused, we only employ basic importance sampling here.
While we can show an improvement over baseline methods with this approach already,
it would be an interesting future extension to perform direction sampling by
incident radiance here (for instance by rejection sampling with the NIRC).

After the ray tracing pass,
the collected parameters are encoded, aggregated, and
utilized in the NIRC inference. The results from the inference are combined with
the remaining terms in \Cref{eq:L_c} to compute the final rendered image.

For the training of our cache, we follow the algorithm of the original Neural
Radiance Cache \cite{NRC}. A significant modification
is that we use
a separate rendering pass for training.
This decision is based on the following observation: using training paths
derived directly from the main path tracing paths leads to performance issues in
our implementation. Training requires us to trace paths of unbounded length to remain
unbiased (by using Russian roulette).
Since the main rendering paths are considerably shorter in our biased version,
this results in thread divergence.
We found that executing a separate path tracing pass for
training paths not only simplifies the process but also slightly improves
performance. While there will still be thread divergence due to differing path lengths,
given that we only use about 2-3\% the number of training paths as
compared to the number of pixels, this additional pass does not
incur substantial overhead. This method ensures efficient training of the cache
while maintaining the integrity and performance of the main rendering process.

\subsection{Path Termination Heuristics}
\label{sec:heuristics}

The original NRC uses the \emph{Spread Angle Heuristic} (SPH) for path termination, which almost always results in path termination after the first bounce, as shown in \Cref{fig:cache_analysis}. This suggests that we can leverage our cache instantly at the primary bounce without the need for this intermediate bounce since our cache predicts incident radiance directly.

The NRC path termination heuristic is defined as follows:
\begin{align}
a(x_1 \cdots x_n) &= \left( \prod_{i=2}^{n} \frac{\|x_{i-1} - x_i\|}{p(\omega_i | x_{i-1}, \omega) \cos \theta_i} \right)^2, \\
a_0 &= \frac{\|x_0 - x_1\|^2}{4 \pi \cos \theta_1},
\end{align}
where \( p(\omega_i) \) is the BSDF sampling PDF and \( \theta_i \) is the angle between \( \omega_i \) and the surface normal at \( x_i \). Paths are terminated if:
\begin{align}
a(x_1 \cdots x_n) > c \cdot a_0, 
\end{align}
where \( c \) is a hyperparameter equal to 0.01.

This heuristic is not directly applicable in our setting since we want the
ability to terminate on the first directly visible path vertex.
We propose a new criterion inspired by the balance heuristic \cite{MIS}, which we call the \emph{Balanced Termination Heuristic} (BTH).
The BTH leverages the strengths of NIRC while acknowledging its limitations,
particularly with glossy details. It essentially computes the MIS weight of BRDF sampling vs.\ a virtual diffuse sampling scheme with $N_c$ samples.
We calculate the continuation probability \( P_s \) as:

\begin{align}
P_s = \frac{p(\omega_i)}{p(\omega_i) + \frac{N_c}{\pi}},
\end{align}

where \( p(\omega_i) \) is the pdf of the sampled scattering direction, and \( N_c \) is the number of neural samples. The path termination criteria are as follows:

\begin{equation}
\begin{aligned}
i = 1&: \quad \text{stop if } \xi > P_s, \\
i > 1&: \quad \text{stop if } \xi > P_s \text{ or } a(x_1 \cdots x_i) > c \cdot a_0,
\end{aligned}
\end{equation}

where \( \xi\in[0,1) \) is a uniformly distributed random variable. If terminated, the current surface is shaded using \( N_c \) directions sampled according to the surface BRDF, incorporating direct lighting with Monte Carlo integration using Multiple Importance Sampling (MIS) and Next Event Estimation (NEE). Delta reflections automatically continue tracing without termination. For a fair comparison with NRC, we use its heuristic for all surfaces \( x_i \) where \( i > 1 \), ensuring that paths are not terminated later than they would be with NRC.

\section{Neural Visibility Cache}
\label{sec:nvc}
In the previous section, we addressed the computation of incident illumination from other surfaces.
While this approach works for the special case of lighting from an environment map, this type of illumination is important enough to warrant special treatment to improve efficiency.
In particular, for environment map lighting we know that the incident radiance is
the product of visibility $V$ and environment illumination $L_i$:
\( V(x_{i}, x_{i+1})L_{i}(x_{i},\omega_{i}) \).
Since $L_i$ is known from the
scene definition we only have to find $V$. For that purpose, we propose a specialized
\emph{Neural Visibility Cache} (NVC) which is trained to only estimate the visibility term
instead of the final radiance signal. $V$ takes on values between zero and one (potentially
including partial visibility due to transparency or transmittance in volumes), so
we use a sigmoidal activation function on the output neurons of the NVC.
Further details and empirical
observations related to this approach are discussed in \Cref{sec:eval}.

\section{Results}
\subsection{Multi-level Monte Carlo vs. control variates}

To assess the expressiveness of our proposed neural cache, we compare
it to other analytical models, in particular \emph{Spherical Harmonics} (SH)
and mixtures of \emph{von Mises-Fisher} lobes (vMF). These models serve as a basis
in \emph{control variates} (CV) approaches because they can be integrated analytically
and have been used in previous work \cite{pantaleoni2020online}. 
To conduct a meaningful comparison, we allocate each pixel its own set of model parameters in a framebuffer at 720p, ensuring that every pixel holds the same amount of memory supporting the models. We chose 720p because we implemented this experiment in PyTorch \cite{paszke2019pytorch} using Falcor 7 \cite{Falcor} and PyTorch Tiny CUDA NN bindings \cite{tiny-cuda-nn}. This setup facilitates debugging and correct fitting of vMF and SH coefficients but limits the performance. We are planning to publish the source code for reproducing the experiments after the publication.

In our experiments, the vMF mixture model holds 11
lobes per pixel, requiring 7 coefficients per lobe. This model
is similar to the Spherical Gaussian mixture model, which has been previously
used as a basis for CV \cite{pantaleoni2020online}

We initialized the vMF lobes directions using a spherical
Fibonacci lattice for optimal sphere coverage and optimized the lobes parameters
using the stepwise-EM \cite{OnlineEMPG} \cite{pantaleoni2020online}. The batch size for the stepwise EM was set in a range from 15 to 40 samples, depending on the noise level of a scene.

We introduce \emph{Screen Space NIRC} (SS NIRC), a version of NIRC that accepts one latent vector per pixel instead of the multiresolution hash encoding for a fair comparison with the vMFs and SHs models. The neural network additionally stores the network weights in a global buffer with less than 55kB and allocates 68.2 MB for the screen space latent features. The world-space NIRC requires only 20MB for its latent representation and 50kB for the weights respectively. We set both NIRC models to have 6 layers to assess their expressive power at full capacity.

We selected the number of vMF lobes to be 11, SH bands to be 5, and 74 latent
vectors for SS NIRC to ensure a similar memory footprint of approximately 75
floats per pixel. We also limit the number of frames for optimization up to 4000 due to our target application goal of real-time rendering.

We estimate the relative variance of the residual error integral estimator $\mathbb{V}_{rel}(\langle F_{r} \rangle)$ for both control variates (vMFs, SH) and two-level Monte Carlo estimators (NIRC and SS NIRC) per each surface in the frame buffer and average them. We also provide visualizations of the computed integrand values by all four models for a set of selected surfaces. It is important to highlight that SHs and vMFs are trained to memorize not only the incident lighting signal as the NIRC and SS NIRC do but the whole integrand, including the BRDF term, so they can be used as control variates.

As seen in \Cref{fig:envmap_comp}, the neural models achieve lower variance for all scenes, preserving geometrical edges and reconstructing high-frequency illumination details. The standard NIRC with multiresolution hash encoding often reaches even better metrics by consuming less memory, as it effectively redistributes the limited latent features between surfaces by resolving hash collisions using gradient descent.

As shown earlier, analytical models like SH and vMF don’t fully capture the complexity of the scenes and struggle to reconstruct the integrated illumination with reasonable accuracy. This motivated us to conduct further experiments with neural models, specifically implementing a model similar to \emph{Neural Control Variates} (NCV) \cite{NCV}. 
This model is based on a Normalizing Flow using the Piecewise-Quadratic Coupling Transform from Neural Importance Sampling (NIS) \cite{NIS}. We utilized an open-source NIS implementation as the foundation for the warps with the invertible coupling transformations \cite{zunis2021github}, and developed the rest of the components ourselves, as the original works did not release any public code \cite{NIS}\cite{NCV}. Inspired by NCV, our NCV model utilizes one blob encoding for latent variables with 32 bins. The coupling transforms consist of 64 bins and 2 repetitions as in the original work. To ensure fair equal-memory comparison, we use the same encodings and the number of hash levels for the \emph{optional features} (per-surface parameters) as for the NIRC.
Our implementation of NCV optimizes the shape of the control variates using a 6-layer MLP with 64 neurons to predict the coupling matrices, similar to the NIRC. Directions are encoded using the cylindrical transformation. For the sake of simplicity, we instantiate an independent flow for each colour channel. While potentially increasing computation time, it demonstrates accuracy improvements as shown in \cite{NCV}. 

Neural Control Variates also require CV integral optimization, so we employed the NRC \cite{InstantNGP} to predict the CV integral, using the same hash encoding layer for both models, ensuring a fair comparison. Additionally, we conducted experiments using a pre-estimated CV integral instead of employing a unified network for both tasks, but this approach did not yield any significant improvements in the overall performance metrics. Unlike NIRC, NCV needs a larger batch size (57,600 vs. 14,400) due to convergence issues, which is why we perform only 1 gradient descent step per frame for NCV compared to 4 for NIRC.

To compare the performance of NCV and NIRC, we track the relative variance per frame using an exponential moving average ($\alpha = 0.95$).
The results are shown in \cref{fig:envmap_comp} (right).
NCV comes close to the NRC metrics after a thousand frames, while the NIRC
requires only a few hundred.
Additionally, NIRC achieves in 1.1-1.32$\times$ lower relative variance for the
residual error estimator as compared to the NCV.

Our experiments in \Cref{fig:envmap_comp} indicate that the additional accuracy
gained by numerically estimating the integral for MLMC can 
offset the benefits of an analytical solution for the first-level
integral for CVs.
Another advantage is that NIRC does not require the neural network to approximate the BSDF, as is necessary for the NCV, uses less compute, and does not require a CV integral prediction network. While it is not easy to assess the precise compute time for NCV inference, as it can strongly depend on the final optimized CUDA kernels, we estimate that our NCV model requires 18 MLP inferences compared to just one for NIRC alongside the coupling transformation.

\subsection{Evaluation}
\label{sec:eval}

\setlength{\abovecaptionskip}{10pt}
\begin{figure*}[htbp]
\setlength{\abovecaptionskip}{0pt}
    \centering
{\footnotesize
      \begin{tabular*}{\linewidth}{%
    @{}
  p{.02\linewidth}
  @{}
  p{.27\linewidth}
  @{\hspace*{.01\linewidth}}
  p{.700\linewidth}
  @{}}
      \multirow{2}{*}[1.8cm]{\rotatebox[origin=lB]{90}{New Sponza}} &  
      \multirow{1}{*}[2.6cm]{\includegraphics[clip, width=\linewidth]{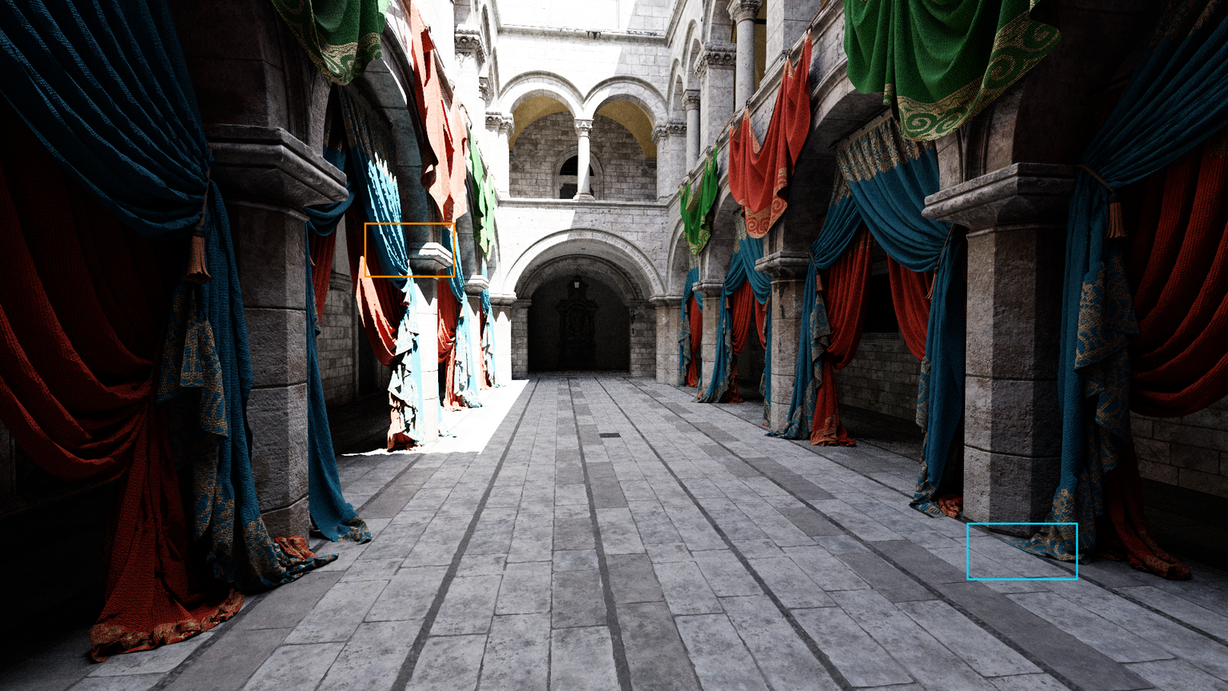}} &
    \begin{tabular*}{\linewidth}[b]{@{}P{.16\linewidth}@{\hspace*{.01\linewidth}}P{.16\linewidth}@{\hspace*{.01\linewidth}}P{.16\linewidth}@{\hspace*{.01\linewidth}}P{.16\linewidth}@{\hspace*{.01\linewidth}}P{.16\linewidth}@{\hspace*{.01\linewidth}}P{.16\linewidth}@{}}
         Reference & PT 2 spp & Ours [Unbiased] & PT 1 spp & NRC [Biased] & Ours [Biased]\\
            \includegraphics[width=\linewidth]{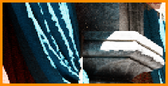}&
        \includegraphics[width=\linewidth]{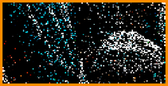} &
        \includegraphics[width=\linewidth]{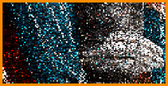} &
        \includegraphics[width=\linewidth]{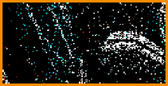} &
        \includegraphics[width=\linewidth]{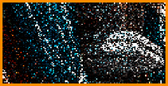} &
        \includegraphics[width=\linewidth]{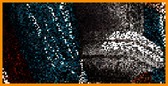} \\
        \includegraphics[width=\linewidth]{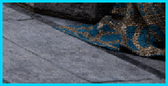}&
        \includegraphics[width=\linewidth]{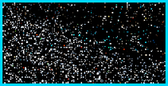} &
        \includegraphics[width=\linewidth]{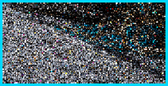} &
        \includegraphics[width=\linewidth]{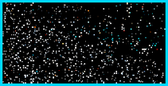} &
        \includegraphics[width=\linewidth]{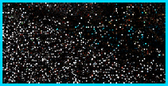} &
        \includegraphics[width=\linewidth]{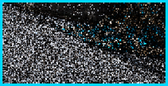} \\
        \reflectbox{F}LIP/MRSE: & 0.34/40.52 & 0.29/\textbf{9.34} & 0.39/47.21 & 0.35/33.03 & \textbf{0.21}/14.87\\ 
         Compute Time: & 87.1ms & 86.7ms & 42.13ms & 35.1ms & 35.2ms 
         
    \end{tabular*} \\
    \multirow{2}{*}[2cm]{\rotatebox[origin=lB]{90}{Country Kitchen}} &   
    \multirow{1}{*}[2.6cm]{\includegraphics[clip, width=\linewidth]{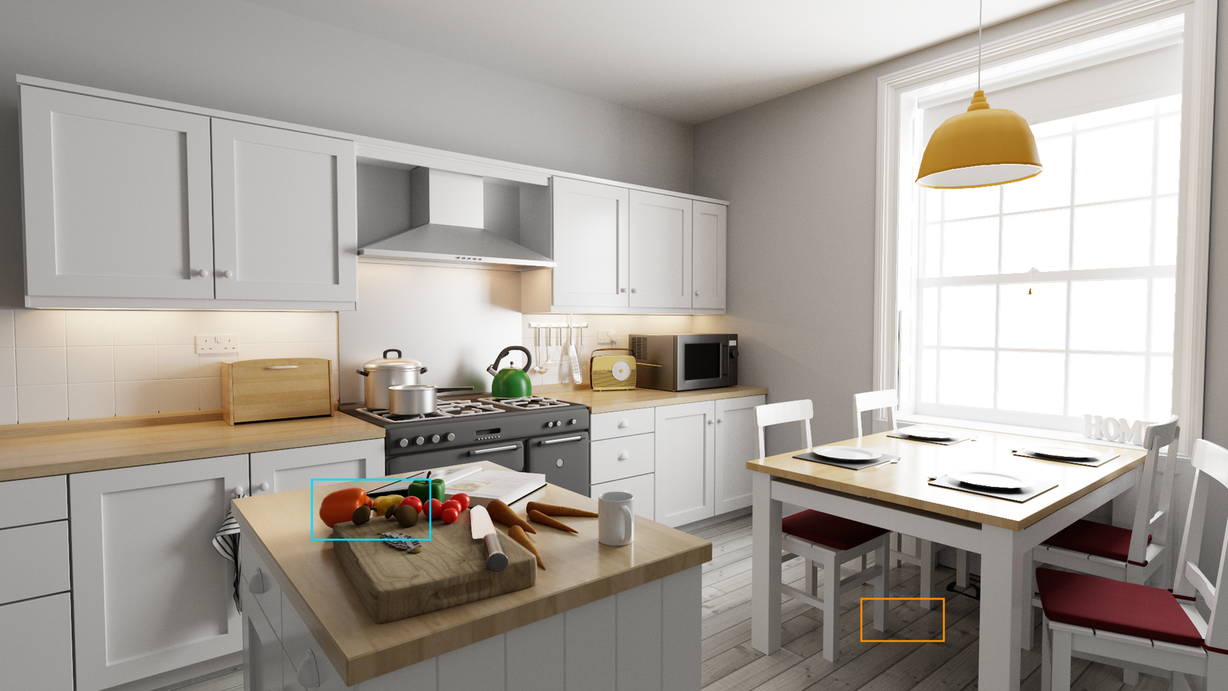}} &
    \begin{tabular*}{\linewidth}[b]{@{}P{.16\linewidth}@{\hspace*{.01\linewidth}}P{.16\linewidth}@{\hspace*{.01\linewidth}}P{.16\linewidth}@{\hspace*{.01\linewidth}}P{.16\linewidth}@{\hspace*{.01\linewidth}}P{.16\linewidth}@{\hspace*{.01\linewidth}}P{.16\linewidth}@{}}
        \includegraphics[width=\linewidth]{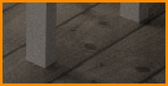}&
        \includegraphics[width=\linewidth]{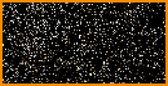} &
        \includegraphics[width=\linewidth]{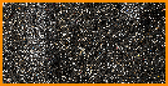} &
        \includegraphics[width=\linewidth]{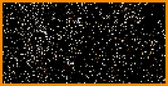} &
        \includegraphics[width=\linewidth]{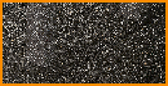} &
        \includegraphics[width=\linewidth]{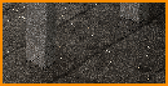} \\
      \includegraphics[width=\linewidth]{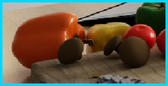}&
        \includegraphics[width=\linewidth]{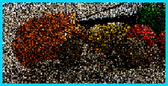} &
        \includegraphics[width=\linewidth]{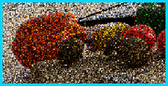} &
        \includegraphics[width=\linewidth]{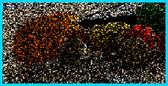} &
        \includegraphics[width=\linewidth]{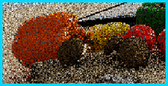} &
        \includegraphics[width=\linewidth]{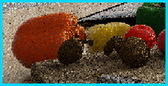} \\
        \reflectbox{F}LIP/MRSE: & 0.43/86.52 & 0.27/19.82 & 0.48/114.5 & 0.27/9.31 & \textbf{0.22}/\textbf{5.61}\\ 
         Compute Time: & 54.2ms & 55.2ms & 24.93ms & 25.9ms & 25.8ms 
         
    \end{tabular*} \\
        \multirow{2}{*}[1.9cm]{\rotatebox[origin=lb]{90}{Bistro Exterior}} &   
            \multirow{1}{*}[2.6cm]{\includegraphics[clip, width=\linewidth]{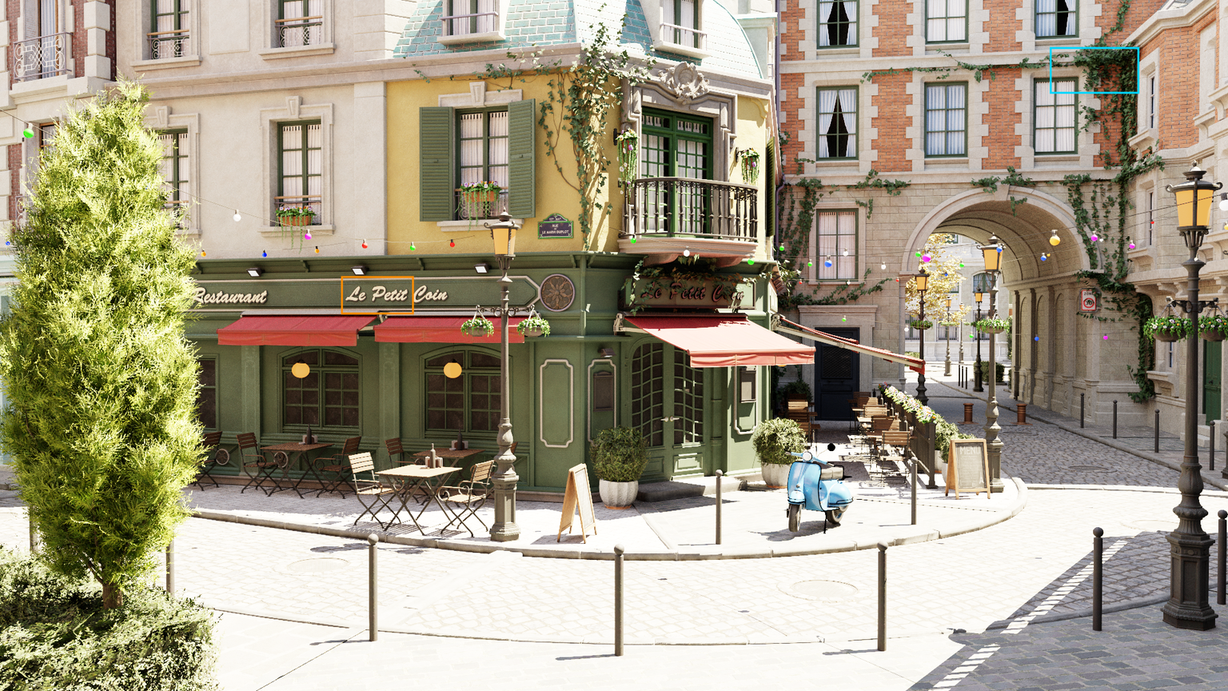}} &
    \begin{tabular*}{\linewidth}[b]{@{}P{.16\linewidth}@{\hspace*{.01\linewidth}}P{.16\linewidth}@{\hspace*{.01\linewidth}}P{.16\linewidth}@{\hspace*{.01\linewidth}}P{.16\linewidth}@{\hspace*{.01\linewidth}}P{.16\linewidth}@{\hspace*{.01\linewidth}}P{.16\linewidth}@{}}
        \includegraphics[width=\linewidth]{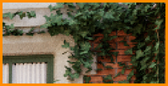}&
        \includegraphics[width=\linewidth]{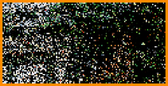} &
        \includegraphics[width=\linewidth]{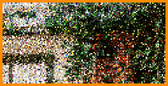} &
        \includegraphics[width=\linewidth]{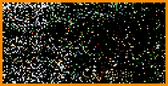} &
        \includegraphics[width=\linewidth]{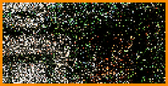} &
        \includegraphics[width=\linewidth]{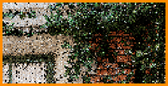} \\
      \includegraphics[width=\linewidth]{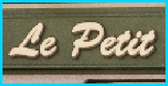}&
        \includegraphics[width=\linewidth]{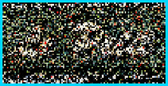} &
        \includegraphics[width=\linewidth]{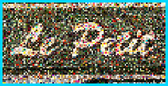} &
        \includegraphics[width=\linewidth]{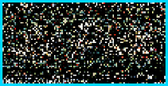} &
        \includegraphics[width=\linewidth]{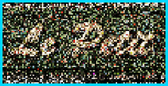} &
        \includegraphics[width=\linewidth]{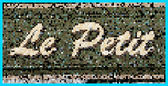} \\
        \reflectbox{F}LIP/MRSE: & 0.5/88.23 & 0.4/\textbf{25.81} & 0.57/121.23 & 0.47/68.23 & \textbf{0.37}/29.32\\ 
         Compute Time: & 83.2ms & 82.1ms & 40.1ms & 37.2ms & 37.7ms 
        
    \end{tabular*} \\
    \multirow{2}{*}[1.85cm]{\rotatebox[origin=lB]{90}{White Room}} &   
        \multirow{1}{*}[2.6cm]{\includegraphics[clip, width=\linewidth]{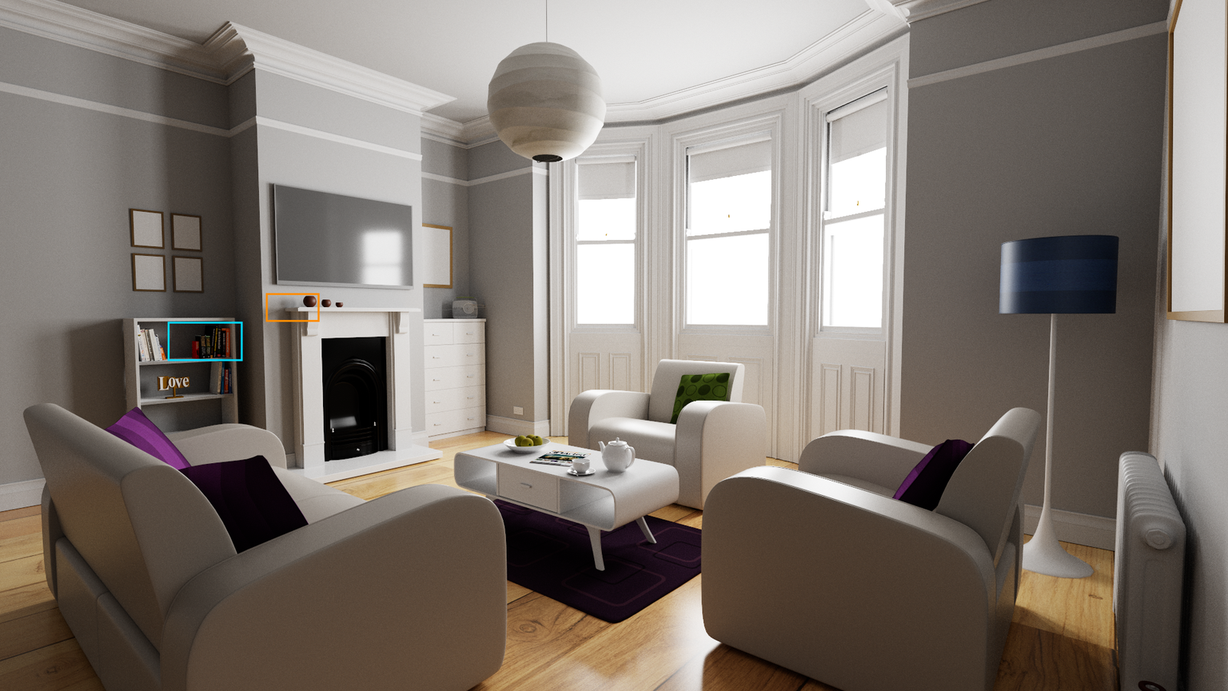}} &
    \begin{tabular*}{\linewidth}[b]{@{}P{.16\linewidth}@{\hspace*{.01\linewidth}}P{.16\linewidth}@{\hspace*{.01\linewidth}}P{.16\linewidth}@{\hspace*{.01\linewidth}}P{.16\linewidth}@{\hspace*{.01\linewidth}}P{.16\linewidth}@{\hspace*{.01\linewidth}}P{.16\linewidth}@{}}
    \includegraphics[width=\linewidth]{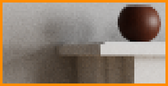}&
        \includegraphics[width=\linewidth]{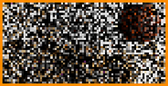} &
        \includegraphics[width=\linewidth]{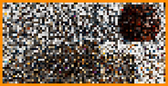} &
        \includegraphics[width=\linewidth]{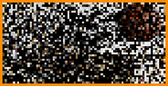} &
        \includegraphics[width=\linewidth]{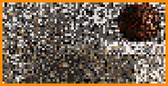} &
        \includegraphics[width=\linewidth]{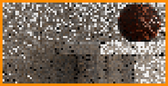} \\
        \includegraphics[width=\linewidth]{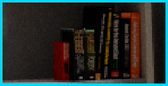}&
        \includegraphics[width=\linewidth]{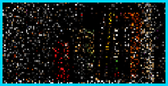} &
        \includegraphics[width=\linewidth]{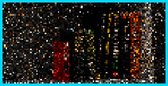} &
        \includegraphics[width=\linewidth]{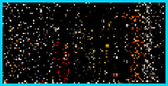} &
        \includegraphics[width=\linewidth]{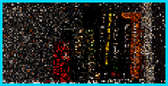} &
        \includegraphics[width=\linewidth]{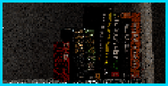} \\
        \reflectbox{F}LIP/MRSE: & 0.32/41.52 & 0.29/13.81 & 0.35/55.75 & 0.26/12.93 & \textbf{0.19}/\textbf{4.42}\\ 
         Compute Time: & 56.2ms & 56.4ms & 24.9ms & 25.1ms & 25.2ms 
        
    \end{tabular*} \\
    \multirow{2}{*}[1.85cm]{\rotatebox[origin=lB]{90}{San Miguel}} &   
    \multirow{1}{*}[2.6cm]{\includegraphics[clip, width=\linewidth]{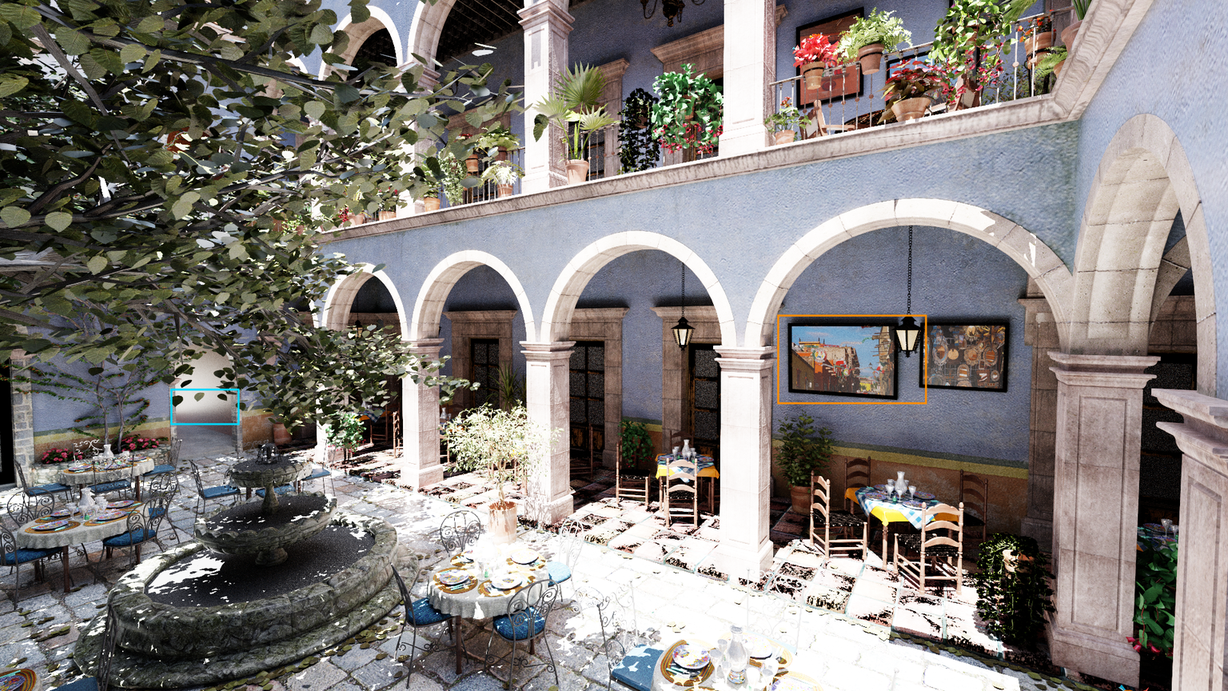}} &
    \begin{tabular*}{\linewidth}[b]{@{}P    {.16\linewidth}@{\hspace*{.01\linewidth}}P{.16\linewidth}@{\hspace*{.01\linewidth}}P{.16\linewidth}@{\hspace*{.01\linewidth}}P{.16\linewidth}@{\hspace*{.01\linewidth}}P{.16\linewidth}@{\hspace*{.01\linewidth}}P{.16\linewidth}@{}}
        \includegraphics[width=\linewidth]{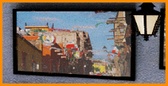}&
        \includegraphics[width=\linewidth]{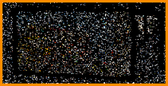} &
        \includegraphics[width=\linewidth]{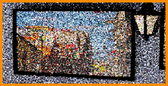} &
        \includegraphics[width=\linewidth]{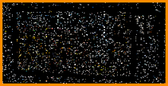} &
        \includegraphics[width=\linewidth]{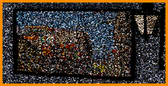} &
        \includegraphics[width=\linewidth]{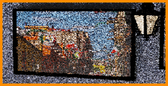} \\
      \includegraphics[width=\linewidth]{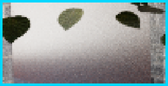}&
        \includegraphics[width=\linewidth]{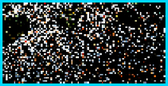} &
        \includegraphics[width=\linewidth]{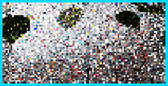} &
        \includegraphics[width=\linewidth]{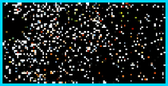} &
        \includegraphics[width=\linewidth]{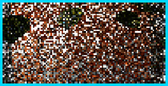} &
        \includegraphics[width=\linewidth]{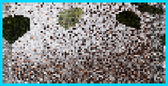} \\
        \reflectbox{F}LIP/MRSE: & 0.53/101.1 & 0.34/\textbf{18.02} & 0.6/119.2 & 0.47/62.3 & \textbf{0.31}/25\\ 
         Compute Time: & 73.2ms & 73.7ms & 35.9ms & 33.2ms & 33.3ms 
        
    \end{tabular*} 
    \end{tabular*}
    }
    \caption{We show different scenes rendered at 1080p on an RTX 3080. The
    experiments highlight the benefits of our Neural Incident Radiance Cache
    (NIRC) in combination with the Multi-Level Monte Carlo (MLMC) approach.
    For an equal-time comparison, the number of neural samples per light
    path is manually adjusted per scene within the range of 15 to 25 for our
    unbiased estimator to match the run time of 2 spp path tracing.
    We compare our biased estimator, which is explicitly
    designed to bypass the computation of the residual error integral, with 1 spp path tracing and the
    Neural Radiance Cache (NRC) algorithm. One key difference is that NIRC
    demands fewer computational resources and does not require casting any rays.
    This allows for 3 to 7 neural samples compared to the single sample of NRC.
    Our results indicate a decrease in the Mean Relative Squared Error (MRSE) of
    path tracing by 1 to 2 orders of magnitude and a 1.5 to 2.5 times reduction
    in \reflectbox{F}LIP relative to conventional path tracing. Moreover, even
    the efficiency comparison between NIRC and NRC based biased estimators leads
    to a 2$\times$ to 3$\times$ reduction in MRSE and a 1.2$\times$ to
    1.5$\times$ decrease in \reflectbox{F}LIP in favor of our method.}
    \label{fig:main_comp}
\end{figure*}
\setlength{\abovecaptionskip}{0pt}

We implemented all components of our NIRC and the required rendering infrastructure for it using CUDA, based on an already released framework for training and inferencing fully-fused neural networks~\cite{NRC}, and Direct3D 12 with the hardware accelerated ray tracing leveraging the Falcor 4.4 engine \cite{Falcor}. 

In order to achieve reasonable performance without a significant cache quality
sacrifice we suggest using neural networks with 4 hidden layers and 64-neuron
width. For the training, we follow the same strategy as was suggested for
NRC~\cite{NRC} but without using the self-training strategy
for our NIRC, because we did not observe any benefits from
it. Also, we perform 4 gradient descent steps per frame using the Adam
optimizer~\cite{Adam} with a learning rate equal to 0.01. ReLU activation
is used as a nonlinear block for the MLP. For the NRC we use just one cache query at a final surface \(x_t\) for each light transport path (render sample) in all experiments. The cache-based Monte Carlo estimator \Cref{eq:L_c} is applied only in combination with the NIRC (see \Cref{fig:paths}).

The main experiments were conducted on an RTX 3080 GPU and an i7-12700k CPU with 32GB of RAM, rendering at 1920$\times$1080 resolution. For the sake of fair judgments, we performed equal-performance comparisons involving computations of Mean Relative Squared Error (MRSE) and the perceptual \reflectbox{F}LIP-metric~\cite{FLIP}. The rendering is performed by a unidirectional path tracing algorithm combined with Next-Event Estimation (NEE) and Multiple Importance Sampling (MIS)~\cite{MIS} for direct lighting estimation. Lights are sampled by a light BVH~\cite{LightSampling}. The path-tracer uses Russian roulette with a termination probability of 0.1.

\paragraph*{Biased and unbiased variants}

We examine two possible ways of using NIRCs to accelerate rendering: %
We achieve an unbiased estimator by combining NIRCs and MLMC.
For this, we estimate the outgoing radiance for the
first three vertices of a path using a predefined number of neural samples
(manually adjusted per scene), as well as full path tracing samples to estimate $L_r$.
We obtain a biased estimator when instead using the Spread Angle Heuristic criteria for the NRC and the Balanced Termination Heuristic for the NIRC. 
Here we compute the shading integral using only the neural approximation of the incident
lighting for truncated paths instead of performing path tracing; this
results in better performance, but introduces bias. The only
difference to the NRC biased estimator is that we can afford to compute the
shading integral with more than 1 cache sample as our cache does not require
casting any rays.
As can be seen in \Cref{fig:main_comp}, our unbiased estimator reduces the MRSE
by a factor of 3 to 5$\times$ and \reflectbox{F}LIP by 10-38\% compared to
path tracing. Our biased estimator reduces MRSE by 3.17 to 20.40$\times$, and
achieves a 35-54\% reduction for the FLIP metric.
Compared to the NRC estimator, our biased algorithm also demonstrates
significant variance reduction in the same computation time:
\reflectbox{F}LIP decreases up to 34\% and MRSE is
reduced by 1.66 up to 2.92$\times$.

\setlength{\abovecaptionskip}{0pt}
\begin{figure}
    \centering
{\footnotesize
\centering
       \includegraphics[clip,trim=0 65 30 0,width=\linewidth]{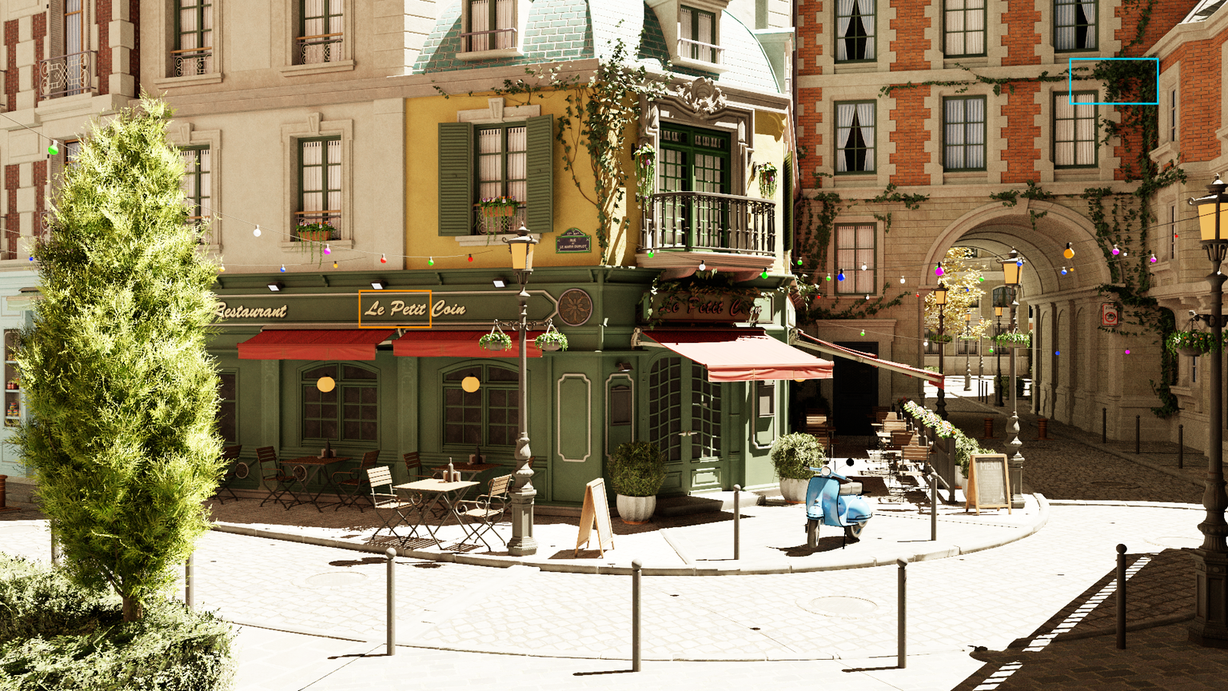}\\
      \begin{tabular}{%
        @{}
        p{.242\linewidth}
        @{\hspace*{.01\linewidth}}
        p{.242\linewidth}
        @{\hspace*{.01\linewidth}}
        p{.242\linewidth}
        @{\hspace*{.01\linewidth}}
        p{.242\linewidth}
        @{}}
        \multicolumn{1}{c}{Reference} & \multicolumn{1}{c}{PT 1spp} & \multicolumn{1}{c}{NRC [Biased]} & \multicolumn{1}{c}{Ours [Biased]} \\
        \includegraphics[clip, width=\linewidth]{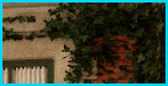} & \includegraphics[clip, width=\linewidth]{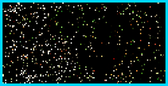} & \includegraphics[clip, width=\linewidth]{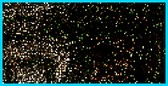} & \includegraphics[clip, width=\linewidth]{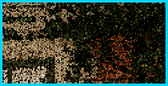}  \\
        \includegraphics[clip, width=\linewidth]{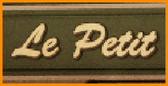} & \includegraphics[clip, width=\linewidth]{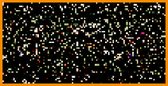} & \includegraphics[clip, width=\linewidth]{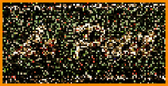} & \includegraphics[clip, width=\linewidth]{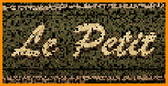}  \\
        \multicolumn{1}{c}{\reflectbox{F}LIP/MRSE:} & \multicolumn{1}{c}{0.55/112.4} & \multicolumn{1}{c}{0.45/82.5} & \multicolumn{1}{c}{\textbf{0.37}/\textbf{68.3}}\\ 
         \multicolumn{1}{c}{Compute Time:} & \multicolumn{1}{c}{39.4ms} & \multicolumn{1}{c}{37.0ms} & \multicolumn{1}{c}{37.4ms}
    \end{tabular}
}
\caption{We present a comparative analysis of the performance between our Neural Incident Radiance Cache (NIRC) and the standard Neural Radiance Cache (NRC) when rendering the Bistro scene with environment lighting disabled on a RTX 3080 and path tracing (PT). It ensures a fair comparison since the standard NRC does not mitigate the variance caused by sky illumination estimation. The performance of both techniques is evaluated based on two metrics: Mean Relative Squared Error (MRSE) and \reflectbox{F}LIP. It is evident from the results that despite its inherent bias, our NIRC estimator consistently outperforms the NRC in the provided scenario.} 
\label{fig:fair_biased}
\end{figure}
\setlength{\abovecaptionskip}{10pt}%
\paragraph*{Environment map lighting}
Our cache includes special support for environment map lighting.
This approach improves our method and thus \Cref{fig:main_comp}
does not compare the pure approximation power of the NIRC and NRC, since NRC does
not decrease the variance caused by direct environmental lighting at the primarily visible surfaces. \Cref{fig:fair_biased} shows a balanced comparison with the NRC. Even when the influence of sky lighting is eliminated,
our biased estimator exhibits outperforms the NRC.

\paragraph*{Distribution of neural samples in a light path}
Our approach can be used for estimating the outgoing radiance at each vertex along a light path.
In our experiments, we limit the NIRC usage to the first three non-specular vertices of a path for the sake of simplicity, although it is obvious that a light path’s variance may be affected by other vertices. This decision is also influenced by the need to predefine the number of vertices that can sample NIRC, which is critical for managing corresponding memory allocations discussed in the following paragraph. Additionally, our experiments do not show substantial improvements when increasing the number of vertices beyond three.
\begin{figure*}[t!]
    \centering
{\footnotesize
      \begin{tabular*}{\linewidth}{%
  @{}
  p{.318\linewidth}
  p{.682\linewidth}
  @{}}
    \includegraphics[clip, width=1.06\linewidth]{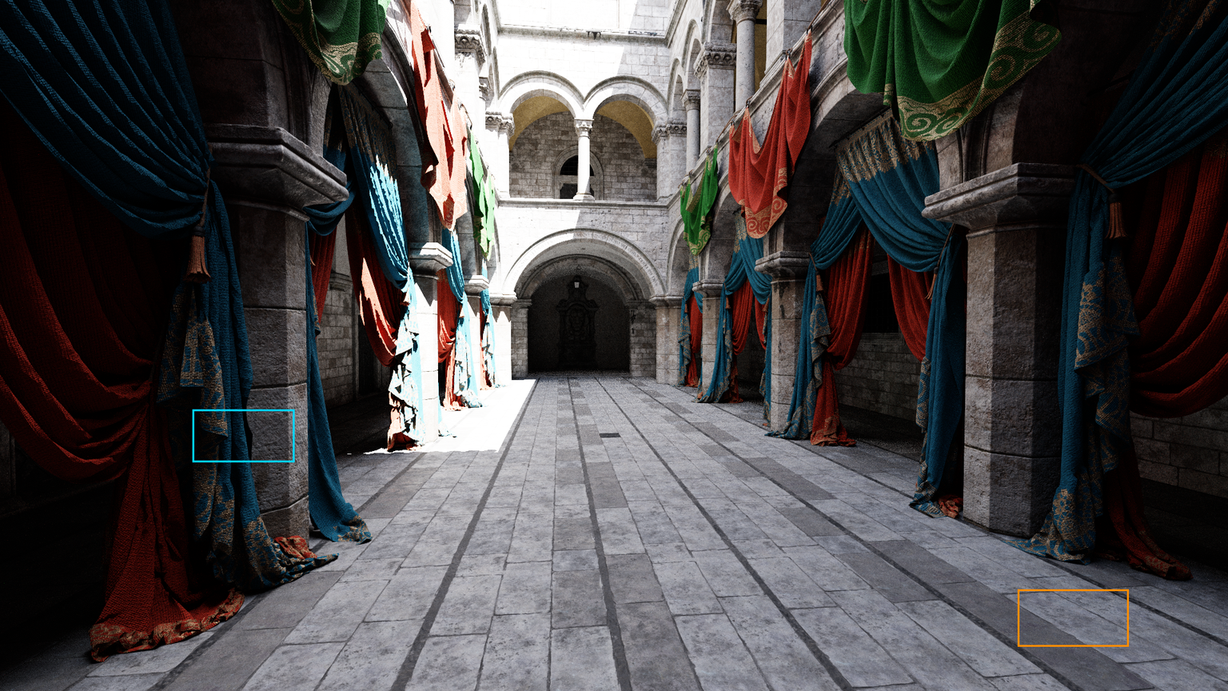} &
    \begin{tabular*}{\linewidth}[b]{@{}P{.18\linewidth}@{\hspace*{.01\linewidth}}P{.18\linewidth}@{\hspace*{.01\linewidth}}P{.18\linewidth}@{\hspace*{.01\linewidth}}P{.18\linewidth}@{\hspace*{.01\linewidth}}P{.18\linewidth}@{}}
         Reference & Path Tracing 2 spp & Our {\scriptsize \(N_{c}^1=25, N_{c}^2=0\)}& Our {\scriptsize \( N_{c}^1=20, N_{c}^2=5\)}& Our {\scriptsize \(N_{c}^1=15, N_{c}^2=10\)}
        \\
        \includegraphics[width=\linewidth]{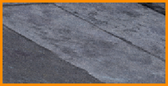}&
        \includegraphics[width=\linewidth]{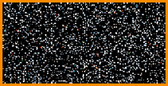} &
        \includegraphics[width=\linewidth]{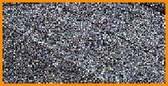} &
        \includegraphics[width=\linewidth]{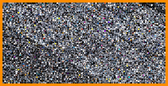} &
        \includegraphics[width=\linewidth]{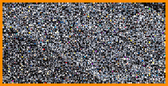}\\
        \reflectbox{F}LIP/MRSE: & 0.45/76.05 & \textbf{0.33}/\textbf{11.81} & 0.34/12.42 & 0.34/13.66 \\ 
        \includegraphics[width=\linewidth]{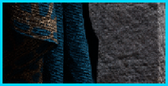}&
        \includegraphics[width=\linewidth]{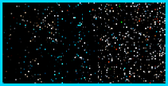} &
        \includegraphics[width=\linewidth]{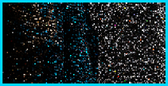} &
        \includegraphics[width=\linewidth]{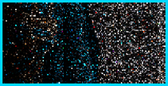} &
        \includegraphics[width=\linewidth]{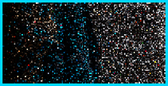} \\
        \reflectbox{F}LIP/MRSE: & 0.30/75.26 & \textbf{0.28}/26.49 & 0.30/22.61 & 0.31/\textbf{21.49} \\ 
         Compute Time: & 108.8ms & 104.9ms & 109.8ms & 107.6ms
    \end{tabular*}\\\\
        \includegraphics[clip, width=1.06\linewidth]{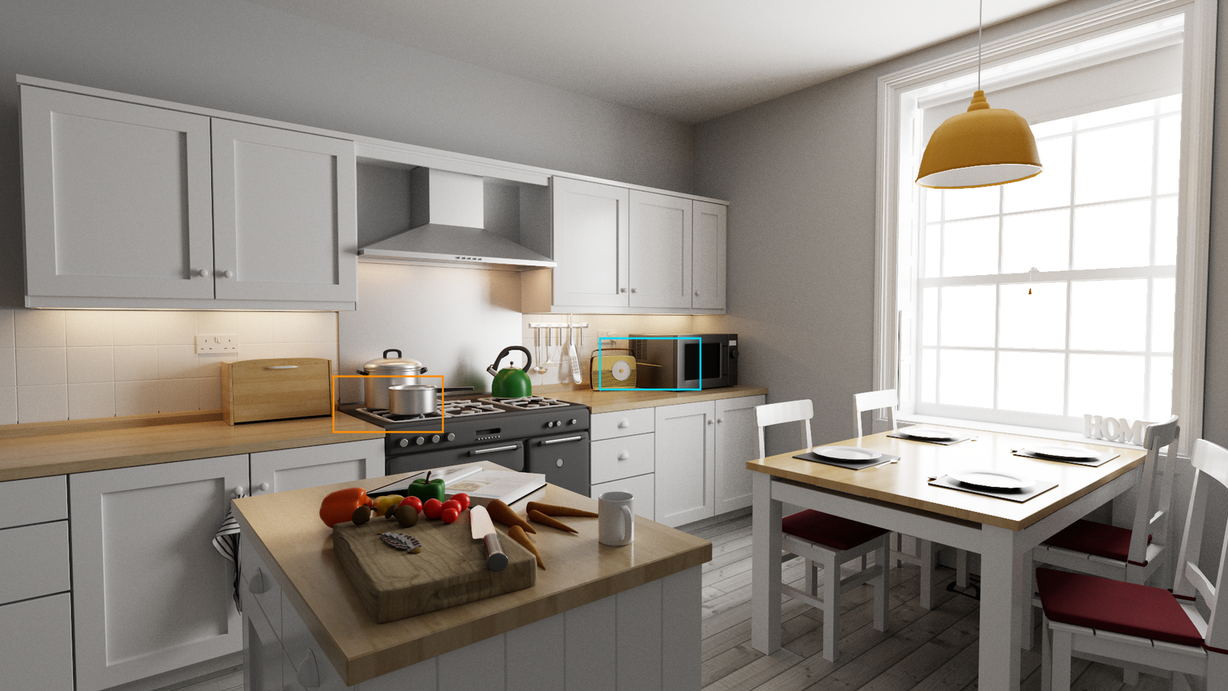} &
    \begin{tabular*}{\linewidth}[b]{@{}P    {.18\linewidth}@{\hspace*{.01\linewidth}}P{.18\linewidth}@{\hspace*{.01\linewidth}}P{.18\linewidth}@{\hspace*{.01\linewidth}}P{.18\linewidth}@{\hspace*{.01\linewidth}}P{.18\linewidth}@{}}
         Reference & Path Tracing 2 spp & Our {\scriptsize \(N_{c}^1=15, N_{c}^2=0\)}& Our {\scriptsize \( N_{c}^1=10, N_{c}^2=5\)}& Our {\scriptsize \(N_{c}^1=5, N_{c}^2=10\)}
        \\
        \includegraphics[width=\linewidth]{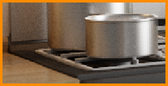}&
        \includegraphics[width=\linewidth]{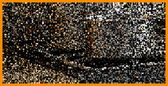} &
        \includegraphics[width=\linewidth]{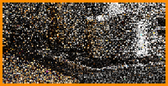} &
        \includegraphics[width=\linewidth]{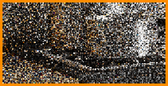} &
        \includegraphics[width=\linewidth]{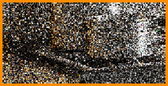}\\
        \reflectbox{F}LIP/MRSE: & 0.37/42.64 & 0.36/20.61 & \textbf{0.34}/\textbf{17.69} & 0.37/19.26 \\ 
        \includegraphics[width=\linewidth]{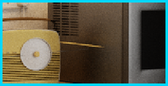}&
        \includegraphics[width=\linewidth]{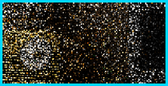} &
        \includegraphics[width=\linewidth]{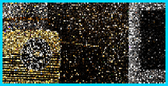} &
        \includegraphics[width=\linewidth]{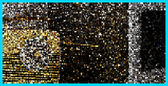} &
        \includegraphics[width=\linewidth]{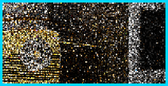} \\
        \reflectbox{F}LIP/MRSE: & 0.41/47.03 & 0.39/32.92 & 0.37/\textbf{28.16} & \textbf{0.36}/35.01 \\ 
         Compute Time: & 50.2ms & 51.1ms & 55.3ms & 54.3ms
    \end{tabular*}
    \end{tabular*}
    }
    \caption{We explore the effect of redistribution of the number of neural samples between the first and second vertex of a light path, denoted as \(N_{c}^1\) and \(N_{c}^2\), respectively. The experiments show the Sponza and Country Kitchen scenes rendered with up to 5 vertices in a light path without Russian roulette in order to match the render time. We observe that the optimal allocation of neural samples between the first and second vertex is highly scene-dependent, influenced by, for example, material properties and illumination.}
    \label{fig:samples_redistribution}
\end{figure*}
\setlength{\abovecaptionskip}{10pt}
\Cref{fig:samples_redistribution} shows experiments on the distribution of the neural samples within a light path.
Ideally, the amount of variance introduced at each vertex should be taken into account to determine the number of neural samples; we believe this is an interesting aspect for future work.

\paragraph*{Memory consumption}
The memory consumption for NIRC inference is mainly determined by the number of
surfaces where the NIRC is queried and the number of requests per surface. Our
implementation includes several buffers: the \emph{surface parameter buffer}, which
requires 9 floats (36 bytes per surface), the \emph{encoded parameter buffer},
utilizing 39 half floats or 78 bytes per surface and an additional 4 bytes for
each direction, where a neural sample is requested, are stored in the
\emph{directional request buffer}. The unit direction vector is mapped to an
octahedron, stored as 32 bits as in
previous works \cite{Cigolle2014Vector, directions}. In our experiments,
the NIRC operates on up to 3 non-specular vertices per light path, with a total of
25 neural samples. This results in memory usage of 223.9 MB for the surface
parameter buffer, 448.2 MB for the encoded parameter buffer, and 207 MB for the
directional request buffer for 1080p resolution. We discuss potential memory
optimization strategies in \Cref{sec:discussion}.

\begin{figure}
    \centering
    \includegraphics[width=.5\textwidth]{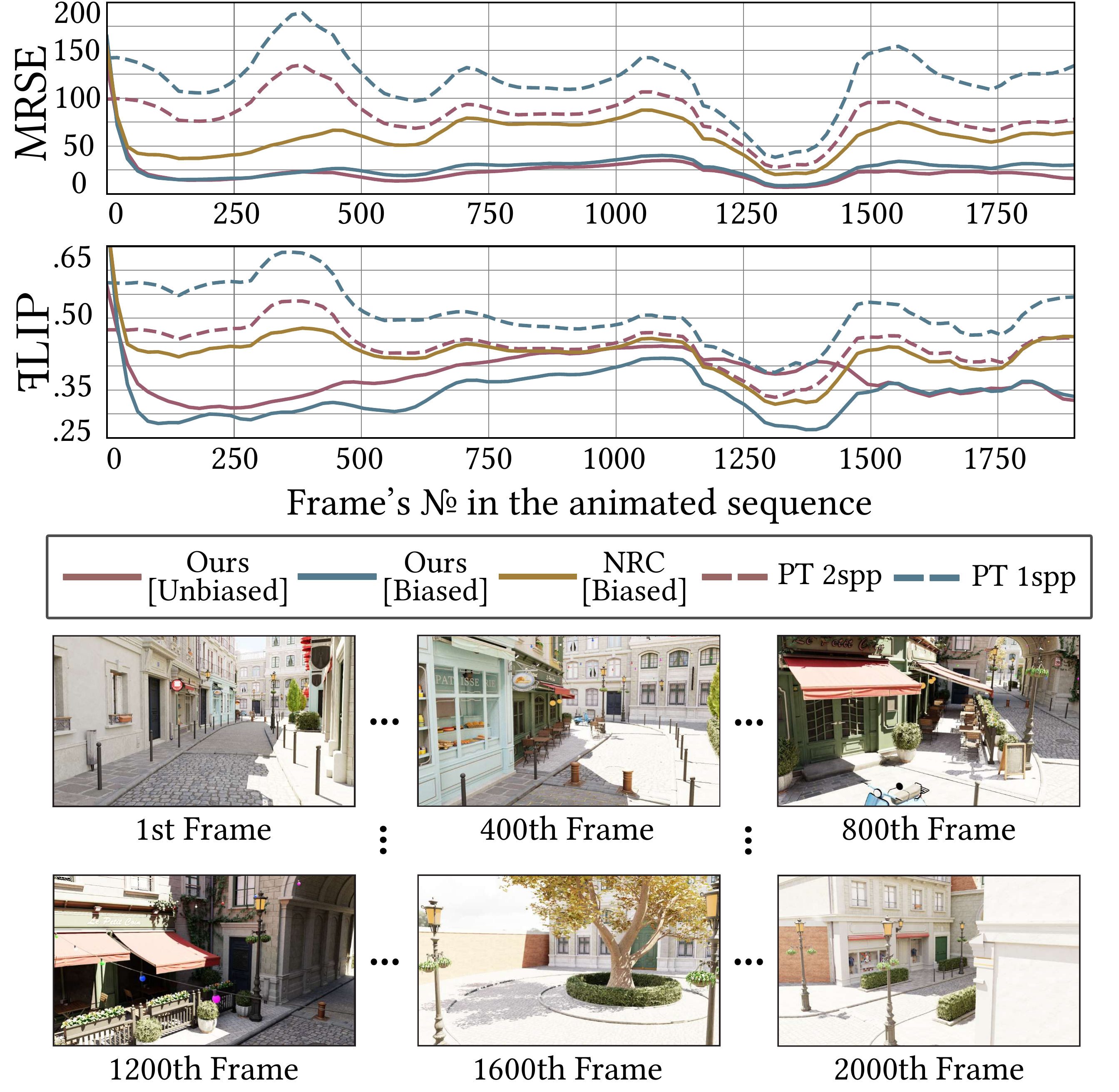}
    \caption{To illustrate the dynamic adaptation of the
    \emph{Neural Incident Radiance Cache} (NIRC) in a changing environment,
    we conducted experiments using the animated Bistro Exterior scene
    with moving objects. The figure demonstrates a quick
    reduction in both Mean Related Squared Error (MSE) and \(\reflectbox{F}\)LIP
    metrics, reaching levels better than the conventional Monte Carlo
    estimator within a span of 10-20 frames. The
    compute time for path tracing (PT) with one sample per pixel (spp) and
    the biased estimators based on NIRC and NRC, as well as PT with two spp and
    our unbiased estimator, are comparable, respectively.}
    \label{fig:dynamic}
\end{figure}

\paragraph*{Dynamic scenes} 
Our method demonstrates good performance even in non-static scenarios such as
the Bistro Exterior scene with dynamic objects and a moving camera (\Cref{fig:dynamic}). 
We compare our real-time rendering using online trainable NIRCs in
combination with the MLMC estimator to ground truth images computed using an
offline renderer with 1000 samples per pixel (spp) for each frame. 
As visible in the plots, our method starts to outperform the standard Monte Carlo approach with the same compute budget after approximately 20 frames.

Note that our model is trained in an online fashion and is defined in world
space. This means that with each new frame, the model can improve the
quality of subsequent renders and show good generalization rather than an
overfit to generated paths from previous frames.
We encourage readers to also watch the accompanying videos in the supplementary materials. %

\setlength{\abovecaptionskip}{0pt}
\setlength{\abovecaptionskip}{10pt}

\begin{figure}
    \centering
    \includegraphics[width=.5\textwidth]{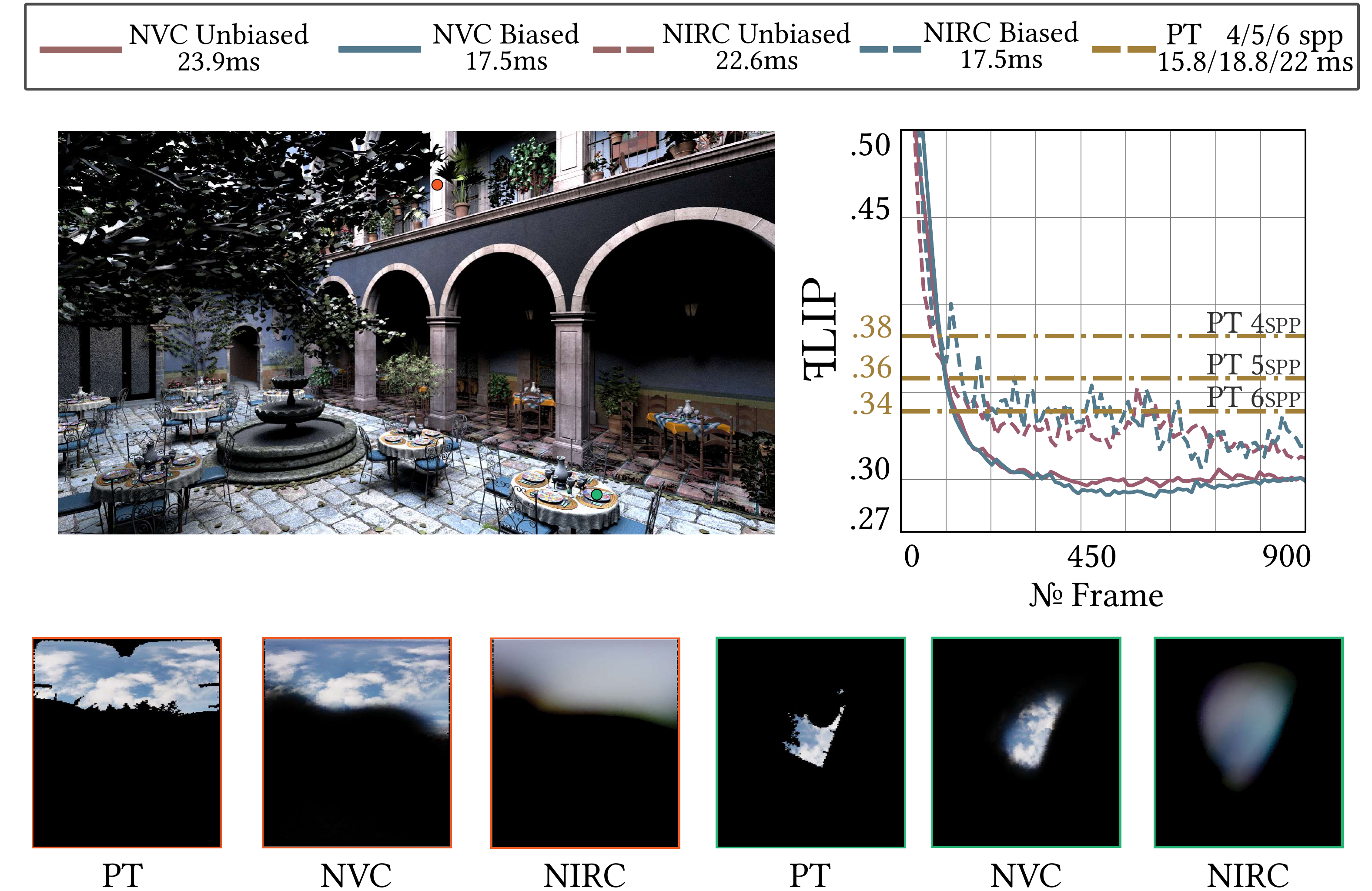}

    \caption{We assess the effectiveness of the Neural Incident Radiance Cache
    (NIRC) and Neural Visibility Cache (NVC) to decrease the variance of the
    direct illumination estimator utilizing our MLMC estimator with $N_{c} = 7$
    in the San Miguel scene. Despite using more memory
    bandwidth, the NVC exhibits better reconstruction quality and training
    stability than the NIRC. Since the baseline path tracing estimator is very fast
    in this setting, we limited the MLPs to 2 layers for both NIRC and NVC to achieve
    comparable timings.}
    \label{fig:nvc}
\end{figure}

\paragraph*{Neural Visibility Cache (NVC)}
Our results demonstrate that inferencing the visibility
term instead of the final radiance value (\Cref{fig:nvc}) outperforms
basic path tracing as well as the generic NIRC. The benefits are particularly apparent when comparing
the reconstructed hemispherical maps of incoming radiance for shading points.
Focusing on learning the visibility term only also significantly simplifies the training process, and as demonstrated by the corresponding plot, NVCs converge more rapidly and provide more stable training compared to NIRCs. 
However, it is important to note that the NVC requires higher bandwidth due to
the required sampling of the environment map; this is also illustrated in
\Cref{fig:nvc}.%

\subsection{Cache Analysis}

To assess the effectiveness of the NIRC in comparison to the NRC, we
investigated their influence on bias, variance, path length, and
render time.

While the NIRC allows us to achieve lower image errors within a limited sample
budget by invoking it multiple times and saving bounces (\Cref{fig:performance_plot}), 
it remains to be shown
that this cache can approximate the incident radiance precisely enough to
achieve converged renders. Moreover, our new heuristic may
negatively impact the final render by causing most paths to terminate at the
first vertex. This saves performance but can introduce bias. Therefore, we aim
to answer several questions:\\
\begin{itemize}
    \item How do different caches (the NIRC vs. the NRC) influence the bias and variance of the final estimator?
    \item Which cache produces more accurate renders for real-time rendering within limited compute time?
    \item How does the decision to stop earlier or later impact the bias vs.\ variance trade-off and render time?
\end{itemize}

To investigate these questions, we conducted a comprehensive analysis on a set of scenes using an RTX 4080, 32GB RAM, and an i7-12700K. For the NIRC, we only considered the task of capturing indirect light, excluding NCV and direct lighting estimation from an environment map for a fair comparison. We trained the caches for 2000 frames with 4 epochs to ensure their convergence.

\begin{figure*}[t!]
    \centering
    \footnotesize
    {\footnotesize
        \begin{tabular*}{\linewidth}{@{}>{\centering\arraybackslash}p{.02\linewidth}@{\hspace*{.01\linewidth}}>{\centering\arraybackslash}p{.15\linewidth}@{\hspace*{.01\linewidth}}>{\centering\arraybackslash}p{.15\linewidth}@{\hspace*{.01\linewidth}}>{\centering\arraybackslash}p{.15\linewidth}@{\hspace*{.01\linewidth}}>{\centering\arraybackslash}p{.15\linewidth}@{\hspace*{.01\linewidth}}|@{\hspace*{.01\linewidth}}>{\centering\arraybackslash}p{.15\linewidth}@{\hspace*{.01\linewidth}}>{\centering\arraybackslash}p{.15\linewidth}@{}}
            & \multicolumn{4}{c}{\textbf{\small Heuristics Based Comparison}} & \multicolumn{2}{c}{\textbf{\small Approx. Equal Bias Comparison}}  \\
            \hline
            \addlinespace[3pt]
            & Reference & NIRC Ours BTH  & NIRC Ours + SPH & NRC SPH & NIRC Adaptive & NRC Adaptive \\
            &  & 1 neural sample &  \cite{NRC} &  1 neural sample & 1-5 neural samples & \cite{NRC} \\
            \addlinespace[1pt]
            \multirow{2}{*}[1.5cm]{\rotatebox[origin=lB]{90}{Bistro Exterior}}  &
            \includegraphics[width=\linewidth]{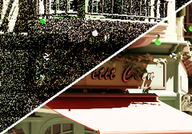}&
            \includegraphics[width=\linewidth]{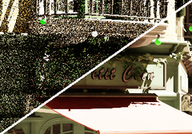} &
            \includegraphics[width=\linewidth]{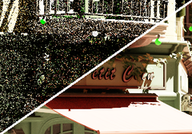} &
            \includegraphics[width=\linewidth]{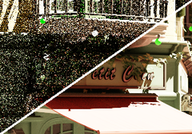} &
            \includegraphics[width=\linewidth]{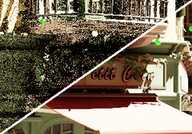} &
            \includegraphics[width=\linewidth]{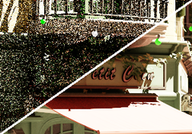} \\
        &1pp \reflectbox{F}LIP/MRSE/Time:&\textbf{0.468}/36.84/\textbf{15.6ms} & 0.517/52.80/19.82ms & 0.480/\textbf{27.53}/18.79ms& \textbf{0.392}/\textbf{16.35}/20.41ms & 0.468/25.89/19.98ms \\
        &1spp rBias$^2$ / rVar: & 1.19e-02/\textbf{1.95} & \textbf{9.81e-04}/2.87 & 3.12e-03/2.49 & 2.87e-03/\textbf{1.69} & 3.08e-03/2.41 \\
        &Average Path Length: &  \textbf{1.22} & 1.77 & 1.78 & \textbf{1.19} & 1.74 \\
                    \addlinespace[1pt]
            \multirow{2}{*}[1.5cm]{\rotatebox[origin=lB]{90}{Glossy Sponza}} &
            \includegraphics[width=\linewidth]{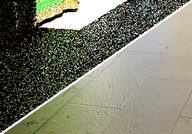}&
            \includegraphics[width=\linewidth]{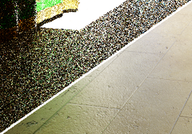} &
            \includegraphics[width=\linewidth]{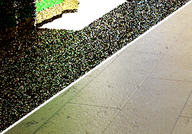} &
            \includegraphics[width=\linewidth]{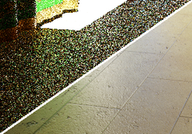} &
            \includegraphics[width=\linewidth]{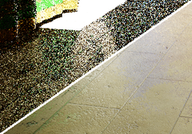} &
            \includegraphics[width=\linewidth]{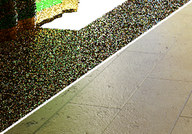} \\
       
        &1pp \reflectbox{F}LIP/MRSE/Time:& \textbf{0.278}/6.69/\textbf{14.48ms} & 0.337/10.98/17.16ms & 0.307/\textbf{4.75}/17.06ms& \textbf{0.249}/\textbf{4.46}/18.11ms & 0.302/4.65/18.14ms \\
        &rBias$^2$ / rVar: & 9.21e-03/\textbf{2.38} & \textbf{5.41e-03}/5.68 & 6.40e-03/2.63 & 6.35e-03/\textbf{1.61} & 6.34e-03/2.57 \\
        &Average Path Length: &  \textbf{1.38} & 1.94 & 1.94 & \textbf{1.40} & 1.92 \\
        \addlinespace[1pt]
            \multirow{2}{*}[1.5cm]{\rotatebox[origin=lB]{90}{Country Kitchen}} &
            \includegraphics[width=\linewidth]{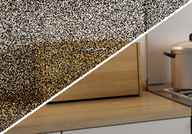}&
            \includegraphics[width=\linewidth]{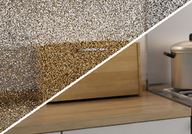} &
            \includegraphics[width=\linewidth]{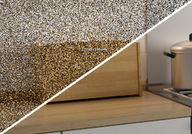} &
            \includegraphics[width=\linewidth]{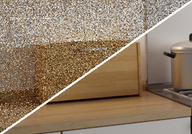} &
            \includegraphics[width=\linewidth]{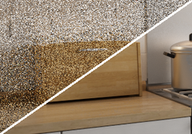} &
            \includegraphics[width=\linewidth]{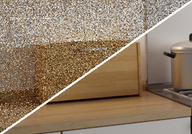} \\
        &1pp \reflectbox{F}LIP/MRSE/Time:& 0.241/1.96/\textbf{14.76ms} & 0.283/2.96/17.08ms & \textbf{0.236}/\textbf{1.17}/16.61ms& \textbf{0.213}/\textbf{0.88}/17.99ms & 0.235/1.16/17.73ms \\
        &rBias$^2$ / rVar: & 1.86e-03/0.85 & \textbf{4.37e-04}/1.28 & 1.49e-03/\textbf{0.73} & 1.11e-03/\textbf{0.69} & 1.47e-03/0.73 \\
        &Average Path Length: &  \textbf{1.43} & 1.98 & 1.99 & \textbf{1.24} & 1.99 \\
        \addlinespace[1pt]
            \multirow{2}{*}[1.5cm]{\rotatebox[origin=lB]{90}{The White Room}} &
            \includegraphics[width=\linewidth]{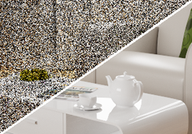}&
            \includegraphics[width=\linewidth]{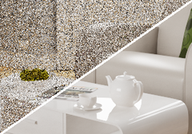} &
            \includegraphics[width=\linewidth]{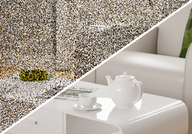} &
            \includegraphics[width=\linewidth]{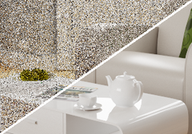} &
            \includegraphics[width=\linewidth]{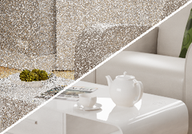} &
            \includegraphics[width=\linewidth]{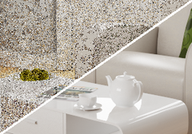} \\
        &1spp \reflectbox{F}LIP/MRSE/Time:& 0.308/6.81/\textbf{14.45ms} & 0.351/8.83/17.52ms & \textbf{0.293}/\textbf{4.11}/16.43ms& 0.269/\textbf{0.54}/17.04ms & \textbf{0.201}/1.98/17.60ms \\
        &1spp rBias$^2$ / rVar: & 1.50e-03/0.67 & \textbf{7.73e-04}/0.89 & 3.19e-03/\textbf{0.57} & 1.54e-03/0.50 & 2.90e-03/\textbf{0.33} \\
        &Average Path Length: &  \textbf{1.40} & 1.96 & 1.99 & \textbf{1.05} & 1.56 \\
        \end{tabular*}
    }
    \caption{Comparison of the \emph{Neural Incident Radiance Cache} (NIRC) and the \emph{Neural Radiance Cache} (NRC) using different heuristics and adaptive methods rendered on an RTX 4080 on a set of scenes at 1080p with biased estimators. The left part of the figure provides a heuristics-based comparison, showing our NIRC with the \emph{Balanced Termination Heuristic} (BTH) and the \emph{Spread Angle Heuristic} (SPH) against NRC SPH, each using 1 neural sample. The NIRC achieves the lowest bias in all scenes with SPH and even when the paths are terminated at the 1st visible surface in Bistro and The White Room. The right comparison provides a fair representation of the performance of the algorithms with closely similar bias based on the per-pixel adaptive path termination policy. The figure shows that our NIRC method reduces the \emph{Mean Relative Squared Error} (MRSE) and \reflectbox{F}LIP metrics in scenes with a high influence of indirect lighting on the final variance while terminating the paths earlier and using limited computational resources more efficiently than NRC. }
    \label{fig:cache_analysis}
\end{figure*}
As shown in \Cref{fig:cache_analysis}, relying on the \emph{Spread Angle Heuristic} (SPH) for NIRC results in a lower \emph{Relative Squared Bias} (rBias\(^2\)) of the final estimator by up to 4.12$\times$ compared with the NRC. However, this approach increases noise due to the need to integrate over the cache and estimate direct lighting, sometimes resulting in higher image-based error within a limited sample budget with just one sample per pixel. Conversely, our \emph{Balanced Termination Heuristic} (BTH) demonstrates less noisy results for real-time rendering purposes and reduces the number of light bounces required for estimating indirect lighting by up to 3.54$\times$, though this comes at the cost of increased bias in our estimator.

Our analysis shows that while invoking the NIRC at the same location as the NRC can lead to superior bias reduction, it degrades other metrics because we need to stochastically estimate the outgoing radiance from there, incorporating indirect lighting based on random samples of the cache along with direct lighting estimation. %
To gain theoretical insight, we explore an adaptive strategy for applying NIRC and NRC caches
during rendering to analyze overall performance and a range of characteristics.

The experiment is set up as follows:
We estimate the relative bias per pixel and decide whether to
query the cache at the first visible path vertex based on this.
If the relative bias of the converged
render terminating the path at the cache compared to the ground truth is less than $\varepsilon$, the cache is used.
By iterating over a set of \(\varepsilon\) values, specifically \{0.05, 0.1, 0.15,
0.20, 0.25, 0.30, 0.35,...,1.0\}, we collected a comprehensive set of metrics:
MRSE, \reflectbox{F}LIP, compute time for one-sample renders, (rBias\(^2\)),
relative variance (rVar), and the average index of a vertex at which the cache
is invoked (this is either the first or the second path vertex).
We calculated at least 5,000 samples per pixel up to 50,000 to
achieve fully converged ground truth renders and error maps. This was done for
both the NIRC and the NRC. We excluded NIRC usage at \(v_{1}\) for delta
scattering rays and the NRC if the roughness is less than 0.0625 due to extreme
glossiness. The NRC authors suggest a similar heuristic to invoke the cache only
at the first \textit{"non-specular vertex"} \cite{NRC} as motivation for the
SPH. For \(v_i\) where \(i > 1\), if the error pass fails or the first
interaction is "too glossy", we always follow the Spread Angle Heuristic for
both cases. The algorithm uses the screen space error estimation approach for
maximum precision, understanding that this approach limits the ability to guide
the termination policy for the second and further bounces. However, it remains
the most accurate method for the initial termination decision.

\setlength{\tabcolsep}{0.1em} %
\begin{figure*}[t!]
\footnotesize
\centering
    \begin{tabular*}{\linewidth}{%
        @{}
        >{\raggedleft}P{.02\linewidth}
        P{.11\linewidth}
        P{.11\linewidth}
        P{.11\linewidth}
        P{.11\linewidth}
        @{\hspace*{.025\linewidth}} |
        >{\raggedleft}P{.02\linewidth}
        P{.11\linewidth}
        P{.11\linewidth}
        P{.11\linewidth}
        P{.11\linewidth}
        @{}}
        & Bistro Exterior & The White Room & Sponza Specular & Country Kitchen & & Bistro Exterior & The White Room & Sponza Specular & Country Kitchen \\
        \multirow{2}{*}[2.1cm]{\rotatebox[origin=lB]{90}{rBias\(^2\)}} &
        \includeinkscape[width=\linewidth]{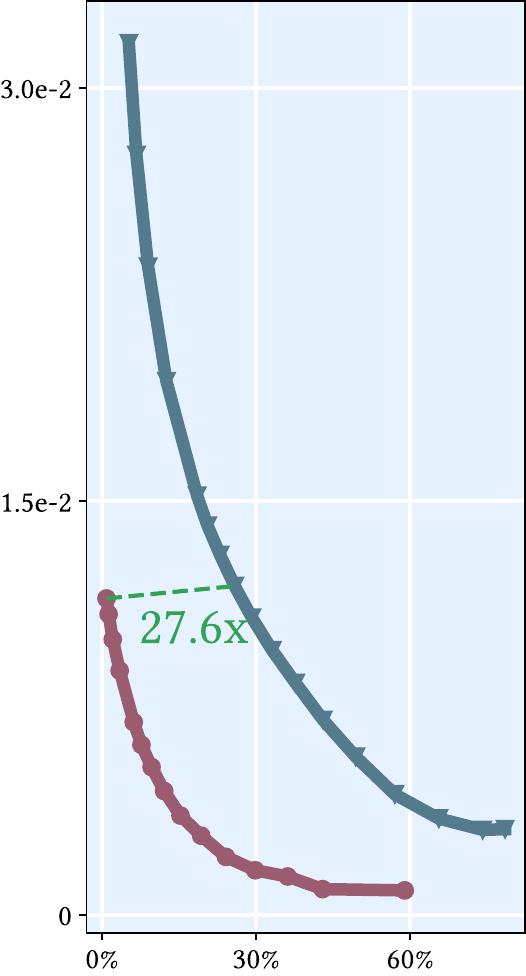} &
        \includeinkscape[width=\linewidth]{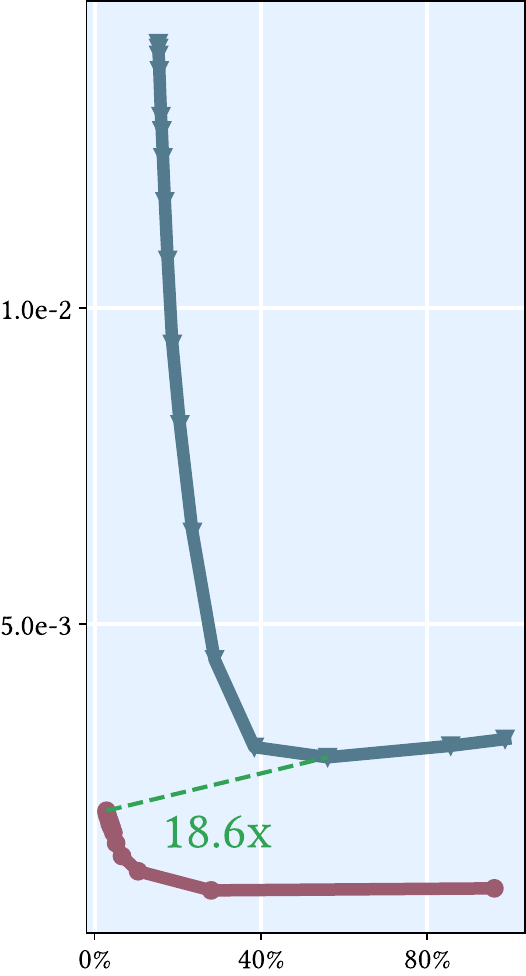} &
        \includeinkscape[width=\linewidth]{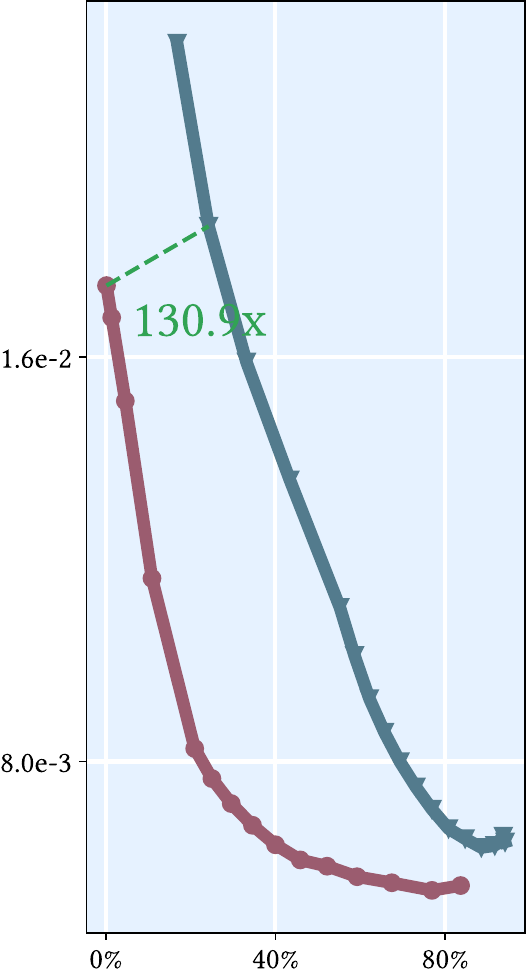} &
        \includeinkscape[width=\linewidth]{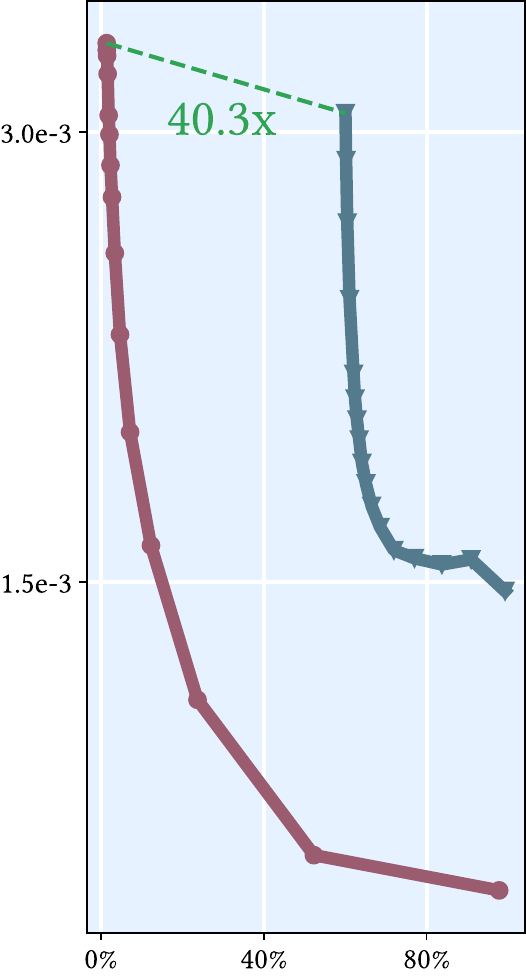} &
        \multirow{2}{*}[2.1cm]{\rotatebox[origin=lB]{90}{\reflectbox{F}LIP}} &
        \includeinkscape[width=\linewidth]{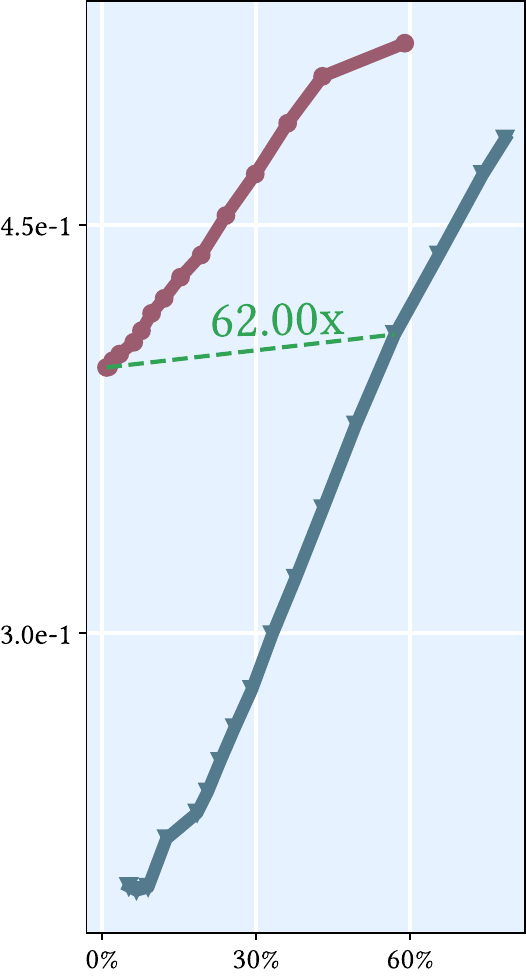} &
        \includeinkscape[width=\linewidth]{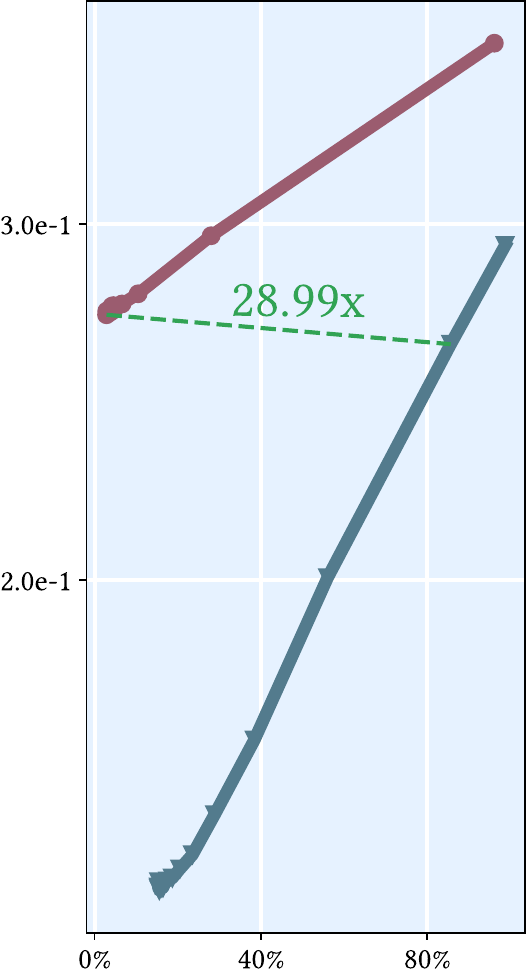} &
        \includeinkscape[width=\linewidth]{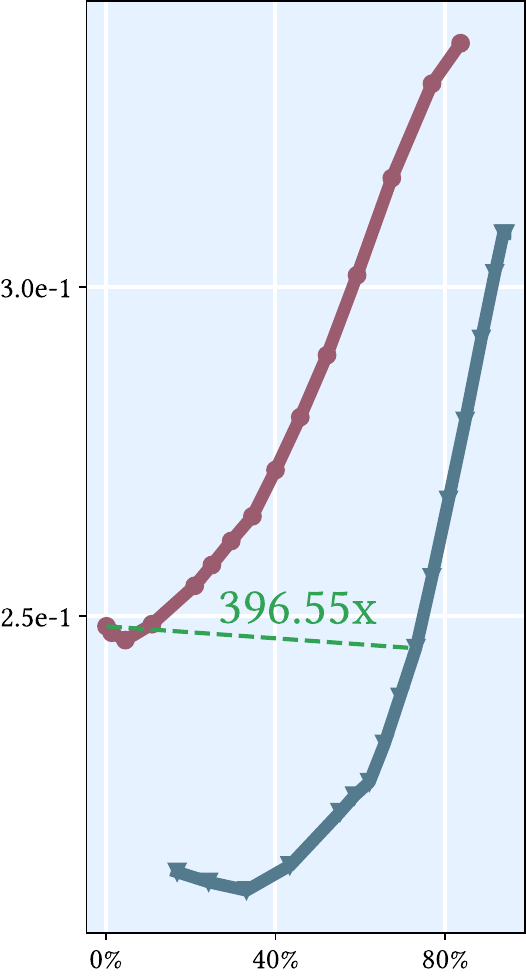} &
        \includeinkscape[width=\linewidth]{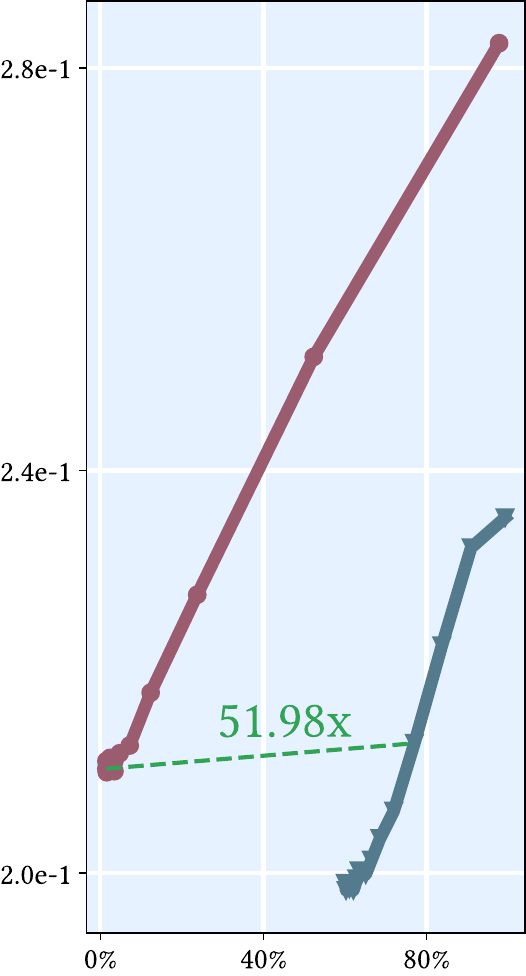} \\
        & IR Bounces & IR Bounces & IR Bounces & IR Bounces &
        
        & IR Bounces & IR Bounces & IR Bounces & IR Bounces \\ 
        &&& & \\ 
        \multirow{2}{*}[2.1cm]{\rotatebox[origin=lB]{90}{rVar}} &
        \includeinkscape[width=\linewidth]{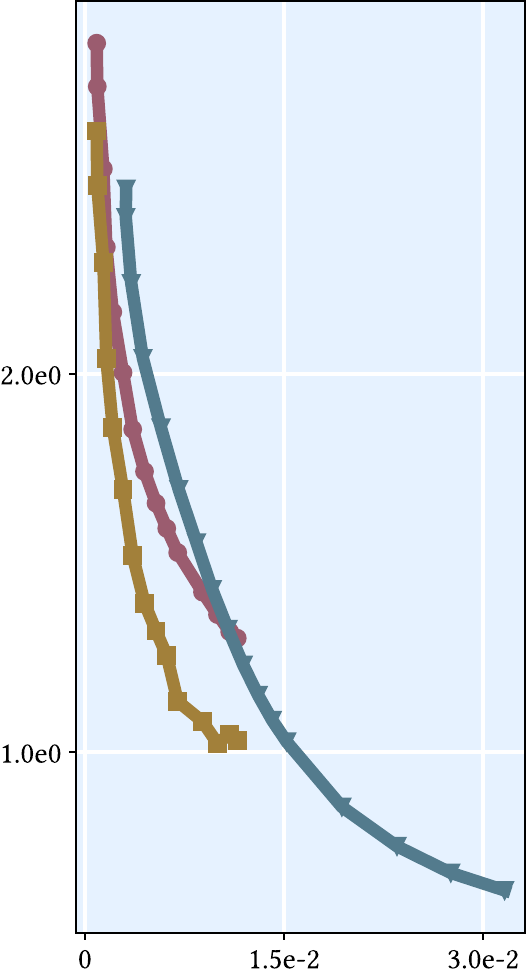} &
        \includeinkscape[width=\linewidth]{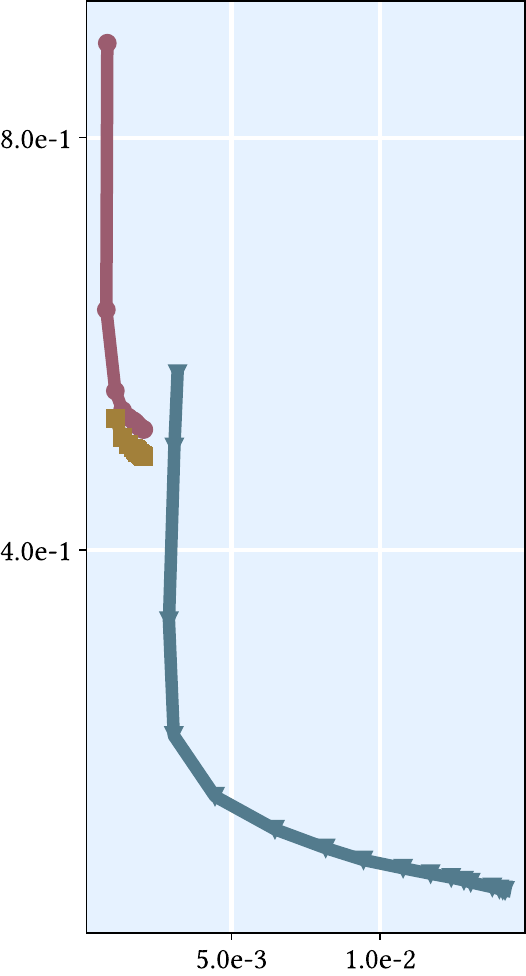} &
        \includeinkscape[width=\linewidth]{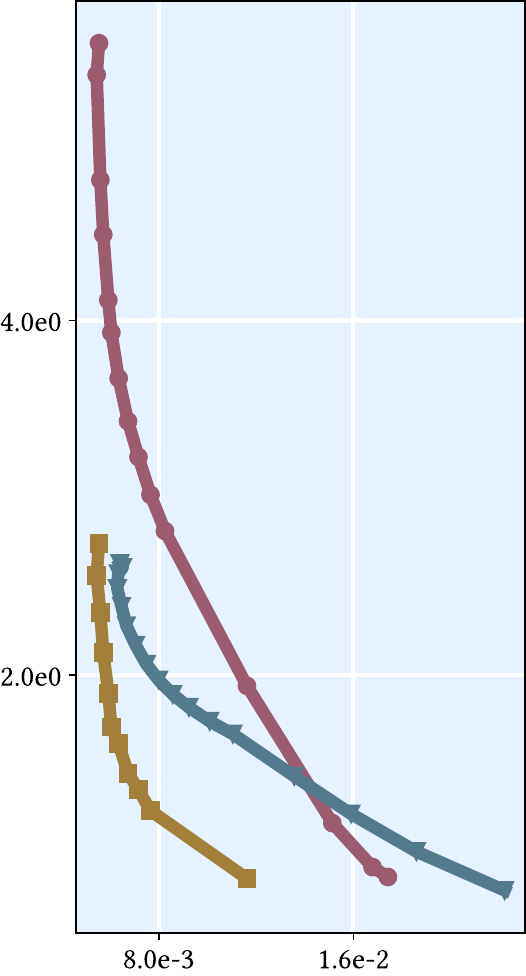} &
        \includeinkscape[width=\linewidth]{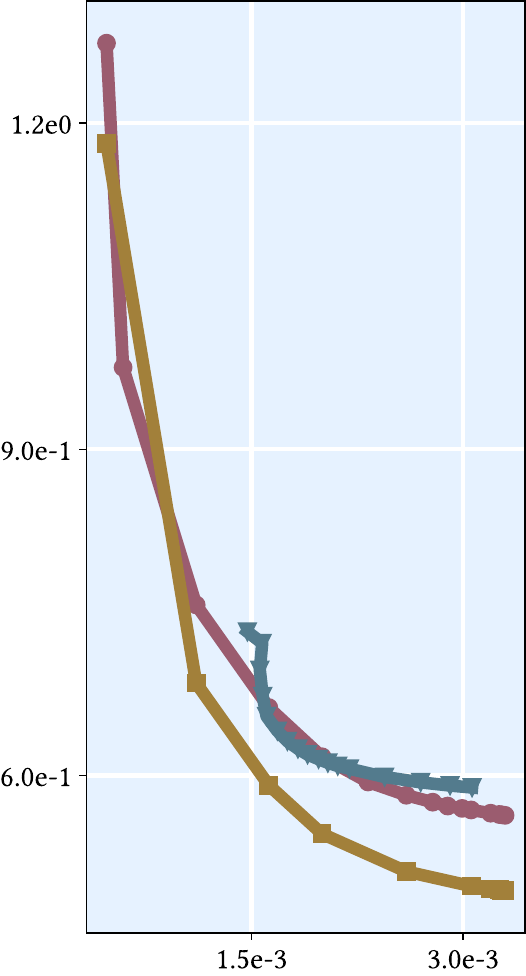} &
        \multirow{2}{*}[2.1cm]{\rotatebox[origin=lB]{90}{MRSE}} &
        \includeinkscape[width=\linewidth]{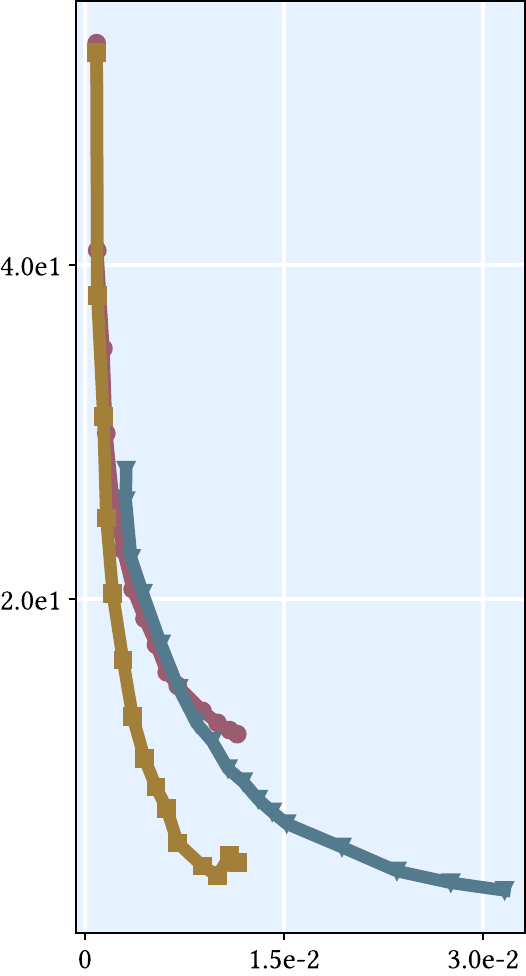} &
        \includeinkscape[width=\linewidth]{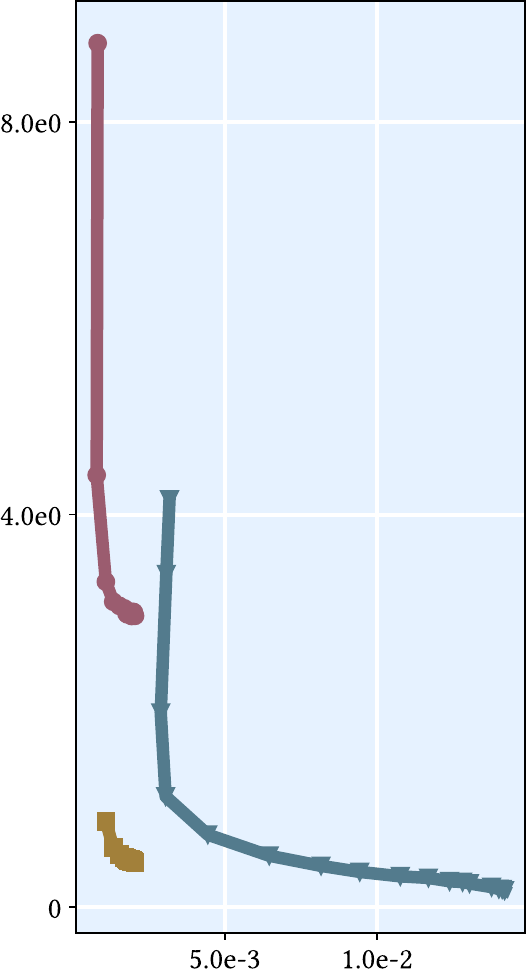} &
        \includeinkscape[width=\linewidth]{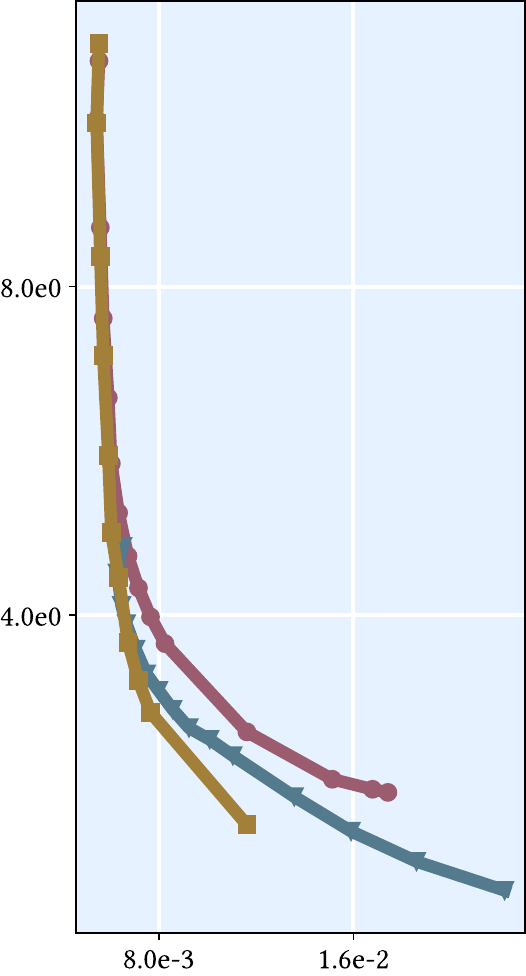} &
        \includeinkscape[width=\linewidth]{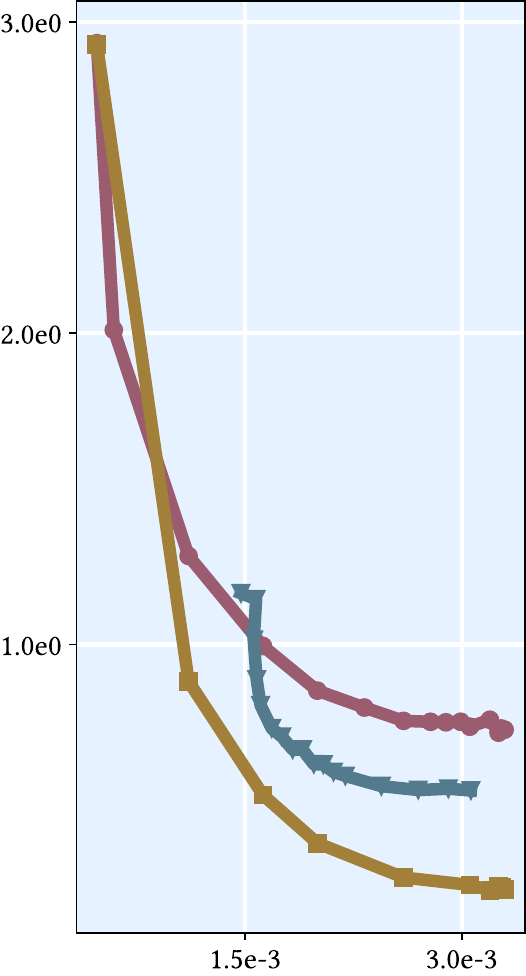} \\
        & rBias\(^2\) & rBias\(^2\) & rBias\(^2\) & rBias\(^2\) & & rBias\(^2\) & rBias\(^2\) & rBias\(^2\) & rBias\(^2\)
        \end{tabular*}
        \includeinkscape[width=0.4\linewidth]{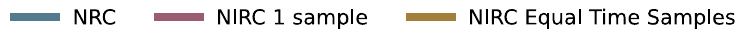}
    \caption{ For each scene, we analyze the performance of the \emph{Neural Incident Radiance Cache} (NIRC) and the \emph{Neural Radiance Cache} (NRC), highlighting the significant reduction in \emph{Indirect Radiance} (IR) bounces and improvements in \emph{Mean Relative Squared Error} (MRSE) and \reflectbox{F}LIP based on the per-pixel adaptive path termination policy. The results show that in most scenes the NIRC curves (in red and brown) are consistently below the NRC curves (in red), indicating superior performance. Additionally, we include a comparison of the NIRC with equal compute time to NRC (brown) that maximizes the number of allocated neural samples per path for the NIRC, demonstrating the accuracy gains and variance reduction achieved by our algorithm. In particular, plots show a significant reduction in IR Bounces over the NRC for achieving closely similar bias and \reflectbox{F}LIP, as highlighted by the green dashed lines in extreme cases with up to 396x difference. These improvements suggest that NIRC achieves lower error and variance at a similar computational cost to NRC.}
    \label{fig:metrics_comparison}
\end{figure*}
As shown in \Cref{fig:metrics_comparison}, we introduced a new metric, \% \emph{Indirect Radiance} (IR) Bounces, representing the averaged index of a vertex at which the path was invoked minus one. Our results show that we can achieve up to 130.5$\times$ fewer IR bounces for estimating indirect illumination while maintaining a similar bias in the final estimator as for the NRC. Furthermore, we consistently achieved lower bias for any  \% IR bounces and lower variance in all cases where NRC can achieve the same bias. In the other cases, NRC is simply not able to provide a precise approximation high enough to have the same bias level as NIRC without delaying further the path termination decision. Moreover, it can be seen that a set of experiments leads to a significant saving in the number of IR bounces up to 28.99-396$\times$ compared to the NRC depending on the scene when we consider equal-error comparisons based on \reflectbox{F}LIP with just 1 spp. We may want to emphasize that these extreme numbers are produced when NIRC leads to allocating IR bounces only to less than 0.2\% of pixels while the NRC needs to trace further in more than 10\% of pixels to achieve the same \reflectbox{F}LIP metric or the rBias\(^2\) of the final estimator.

Despite not exploiting the main architecture feature of the NIRC and allocating
only 1 neural request per path for the NRC as well, we focused on more
fundamental metrics regardless of the final compute time. We were also
interested in applied cases for real-time applications. This is why we
introduced the NIRC Equal Time Samples plot in
\Cref{fig:metrics_comparison}. We suggested maximizing the number of allocated
neural samples for the NIRC per pixel where the cache is invoked at the
\(v_{1}\) until we match the compute time of the corresponding render with the
closest rBias\(^2\) based on the NRC (within a 1ms epsilon). By having more
neural samples, we managed to achieve FLIP: 1.05-1.14$\times$ decrease, MRSE:
0.93-6.67$\times$ improvement, Variance: 1.11-1.82$\times$ improvement over the
constant 1 neural sample. This improvement lets us achieve superior MRSE and
rVar compared to the NRCs almost in all cases with equal bias. So, even considering the
need to trace direct light rays and the associated costs, efficient adaptive
usage combined with cheaper neural samples dominate over these costs.

This adaptive error-based algorithm requires a lot of precomputations and cannot
be used in real-time applications. We discuss it to demonstrate potential theoretically; in
reality, a classical hash table based data structure could be used to estimate
the error gradually. Alternatively, very cheap Bayesian uncertainty estimation
\cite{durasov2024zigzag} does not require any additional infrastructure and
could be theoretically used to estimate the bias.

\setlength{\abovecaptionskip}{10pt}

\section{Discussion and Future Work}
\label{sec:discussion}
\paragraph*{Examining the quality of NIRC}
The decision to store incident radiance in our cache, as opposed to outgoing
radiance in Neural Radiance Caching (NRC), is an inherent challenge for the
model as it effectively has to learn the geometrical structure of a scene. As we discussed earlier
this can lead to missed local reflection effects and impact performance with highly specular surfaces. When comparing converged renders using the \emph{biased} estimators based on  NRC and NIRC, the
latter tends to produce renders of lower accuracy (the unbiased estimator does
not share these difficulties).
A venue for future research might be to enhance the quality of NIRC in the angular domain or to explore ways to combine  NIRCs and NRCs to mitigate their limitations and achieve a better balance between performance and quality of biased renders.

\begin{figure}
    \centering
    \includegraphics[width=.48\textwidth]{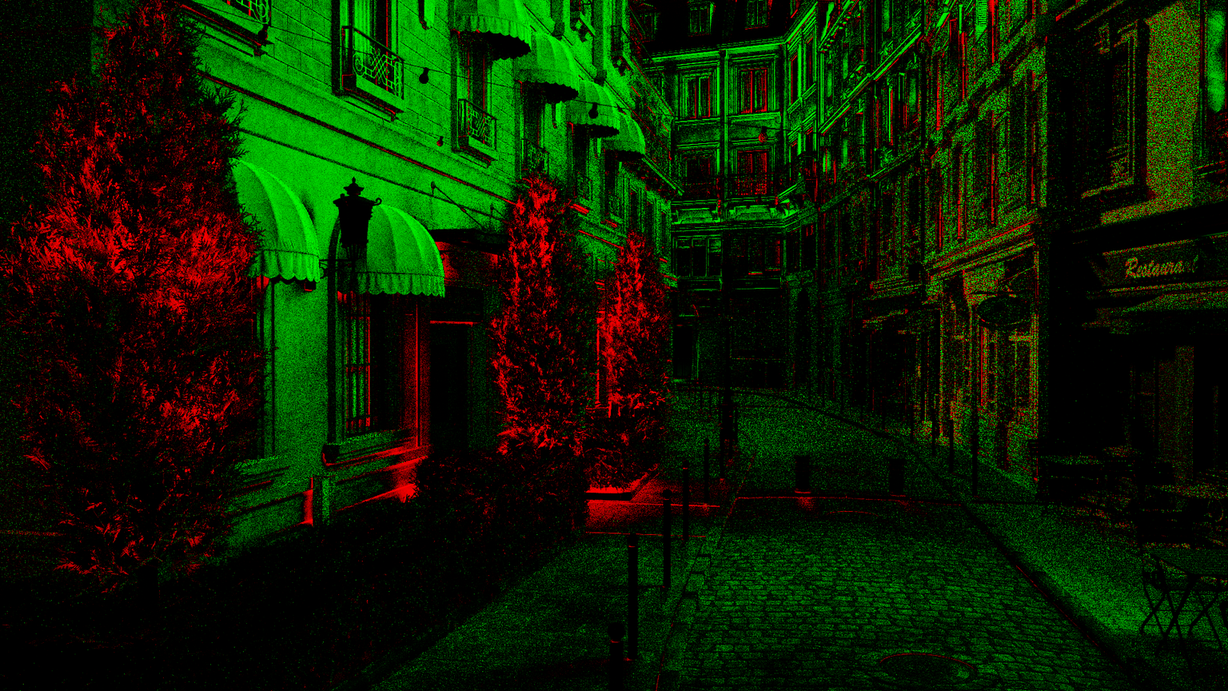}

    \caption{This image shows a tone-mapped visualization of signed
    squared differences between the standard deviation of the original Monte
    Carlo and our multi-level Monte Carlo estimator for
    direct illumination only. Negative outcomes (red) indicate that our approach
    increases variance and positive outcomes (green) signify variance reduction.
    Our method works best in smooth regions and can fail near complex geometric occlusion
    such as in trees and bushes.}
    \label{fig:failure_sd}
\end{figure}

\paragraph*{MLMC Adaptivity}
MLMC removes the bias from the estimator but can increase variance. This is evident from \Cref{fig:failure_sd} where NIRC struggles to accurately represent incident light patterns for complex geometry like trees and bushes. 
We believe it is worthwhile to explore ways to improve MLMC's adaptivity. Müller et al.~\cite{NIS} have demonstrated the advantages of using a separate neural network to calculate the probability of relying on a trained neural guiding model instead of a regular sampling routine. Similar strategies might be used to restrict the use of MLMC to configurations where the cache quality is high.

\paragraph*{Quality of caches}
Undesirable results may emerge, particularly in regions with high variance of radiance estimates, as the training of neural networks can become unstable due to noisy gradients. We believe that the use of techniques such as ReSTIR~\cite{ReSTIR} might lead to better cache quality and reduce the occurrence of undesirable patterns in the render. 

\paragraph*{Dynamic adaptivity}
Although our algorithm typically performs well in dynamic scenes, the neural network might reach a local minimum and cease further adaptation.  
This can also be observed even when stochastic gradient descent~\cite{SGD} is used without adaptive estimation of the gradient's momentum. We believe this deserves further work in the future.

\setlength{\abovecaptionskip}{0pt}
\begin{figure*}[t!]
    \centering
{\footnotesize
      \begin{tabular*}{\linewidth}{%
  @{}
  p{.32667\linewidth}
  @{\hspace*{.01\linewidth}}
  p{.32667\linewidth}
  @{\hspace*{.01\linewidth}}
  p{.32667\linewidth}
  @{}}
    \includegraphics[clip, width=\linewidth]{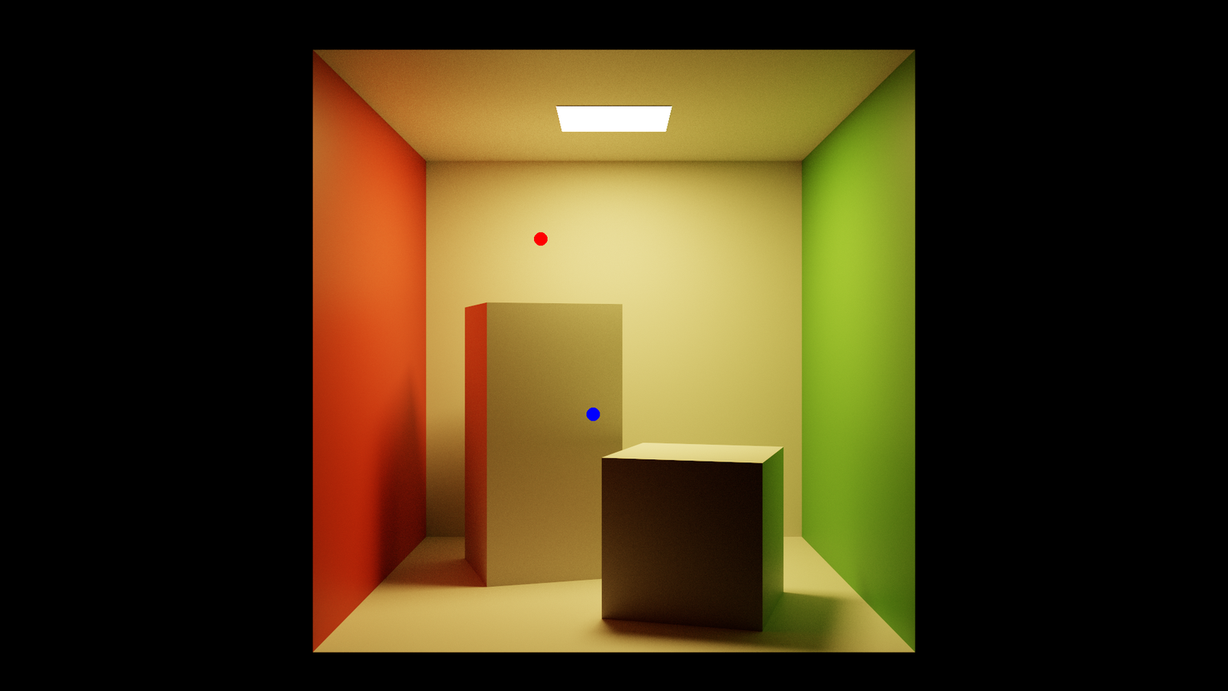} &
    \includegraphics[clip, width=\linewidth]{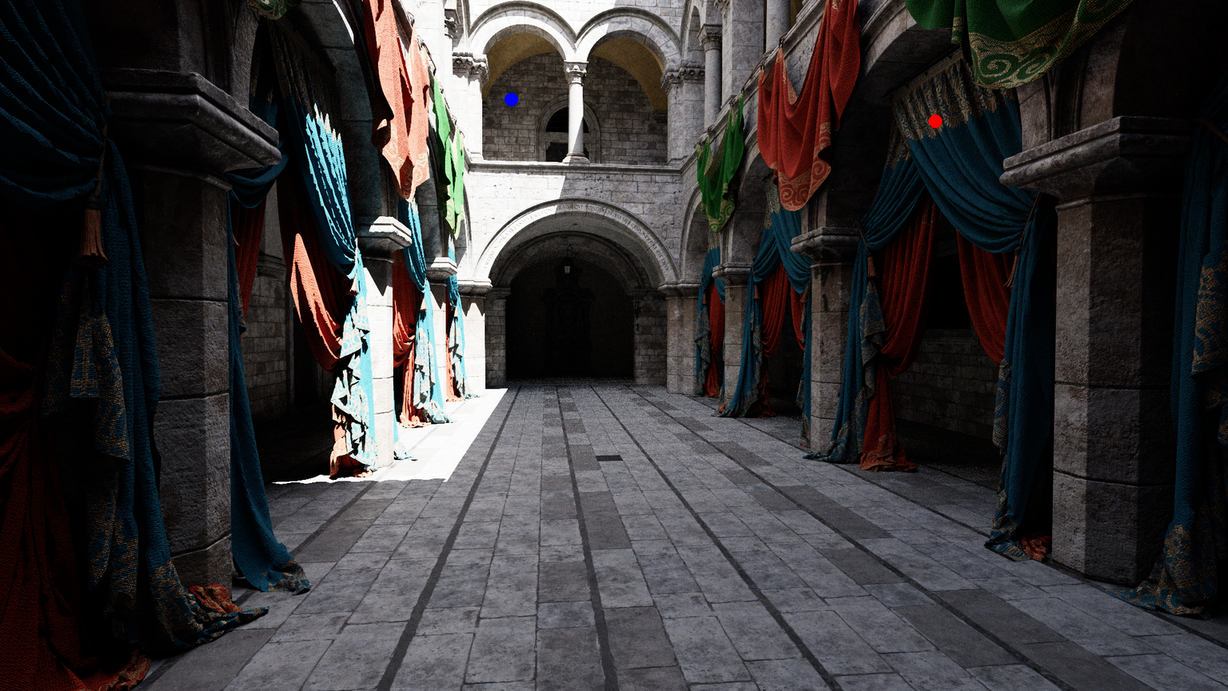} &
    \includegraphics[clip, width=\linewidth]{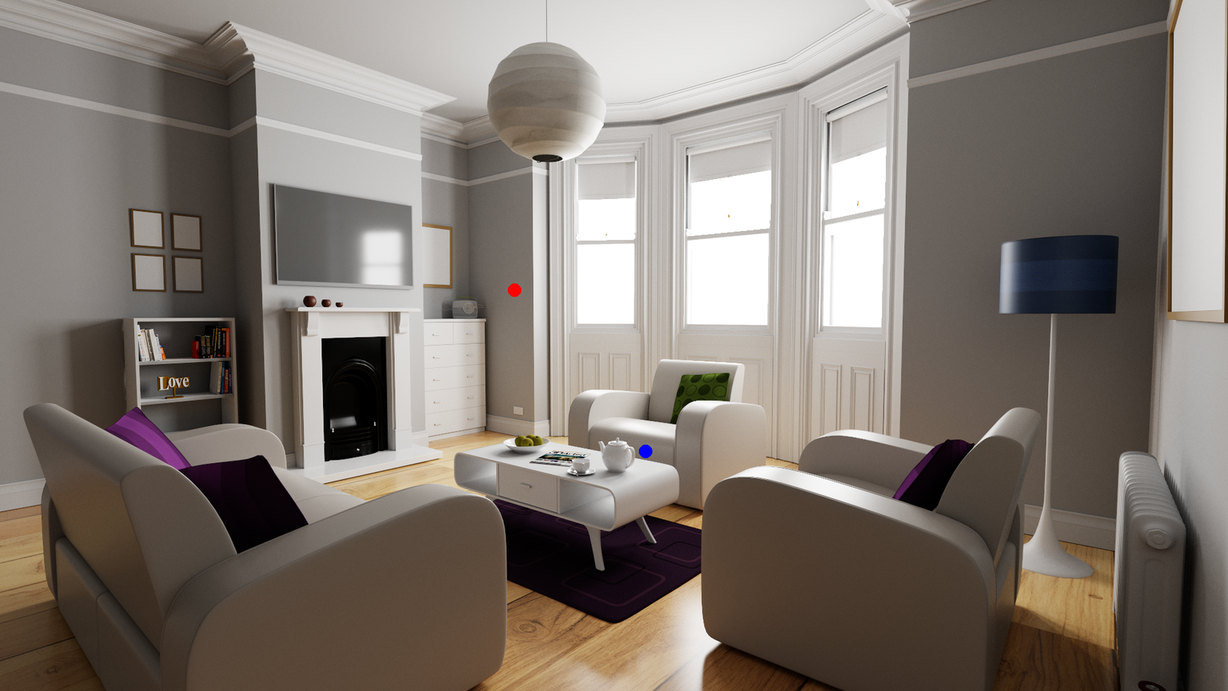} \\
    
    \begin{tabular*}{\linewidth}[b]{@{}P{.24\linewidth}@{\hspace*{.0133\linewidth}}P{.24\linewidth}@{\hspace*{.0133\linewidth}}P{.24\linewidth}@{\hspace*{.0133\linewidth}}P{.24\linewidth}@{}}
        \includegraphics[width=\linewidth]{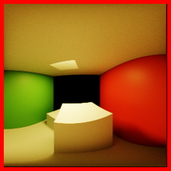} &
        \includegraphics[width=\linewidth]{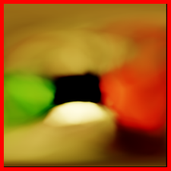}&
        \includegraphics[width=\linewidth]{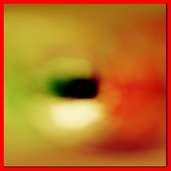} &
        \includegraphics[width=\linewidth]{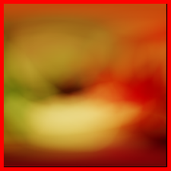} \\
        MSE: & 0.14 & 0.27 & 0.56 \\ 
        \includegraphics[width=\linewidth]{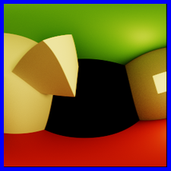} &
        \includegraphics[width=\linewidth]{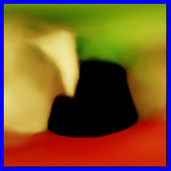}&
        \includegraphics[width=\linewidth]{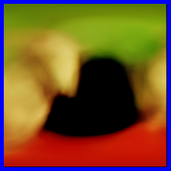} &
        \includegraphics[width=\linewidth]{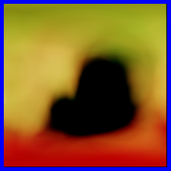} \\
        MSE: & 0.13 & 0.21 & 0.23 \\ 
        Reference & Ours, \(D = 6\) & Ours, \(D = 4\) & Ours, \(D = 2\) \\
         &  1.64ms & 1.24ms & 0.84ms
    \end{tabular*}

    &

    \begin{tabular*}{\linewidth}[b]{@{}P{.24\linewidth}@{\hspace*{.0133\linewidth}}P{.24\linewidth}@{\hspace*{.0133\linewidth}}P{.24\linewidth}@{\hspace*{.0133\linewidth}}P{.24\linewidth}@{}}
        \includegraphics[width=\linewidth]{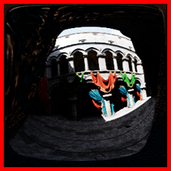} &
        \includegraphics[width=\linewidth]{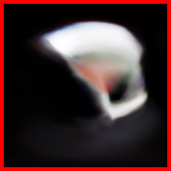}&
        \includegraphics[width=\linewidth]{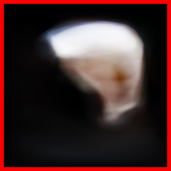} &
        \includegraphics[width=\linewidth]{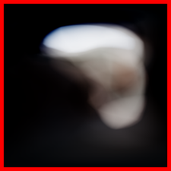} \\
        MSE: & 1.52 & 1.83 & 2.15 \\ 
        \includegraphics[width=\linewidth]{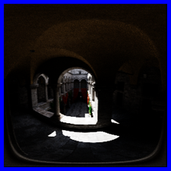} &
        \includegraphics[width=\linewidth]{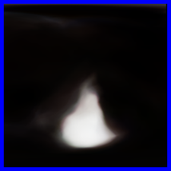}&
        \includegraphics[width=\linewidth]{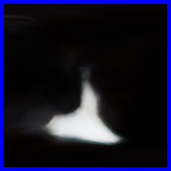} &
        \includegraphics[width=\linewidth]{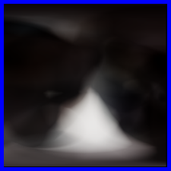} \\
        MSE: & 0.94 & 0.98 & 1.02 \\ 
        Reference & Ours, \(D = 6\) & Ours, \(D = 4\) & Ours, \(D = 2\) \\
         &  1.64ms & 1.24ms & 0.84ms
    \end{tabular*}
    &
    \begin{tabular*}{\linewidth}[b]{@{}P{.24\linewidth}@{\hspace*{.0133\linewidth}}P{.24\linewidth}@{\hspace*{.0133\linewidth}}P{.24\linewidth}@{\hspace*{.0133\linewidth}}P{.24\linewidth}@{}}
        \includegraphics[width=\linewidth]{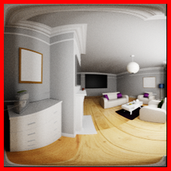} &
        \includegraphics[width=\linewidth]{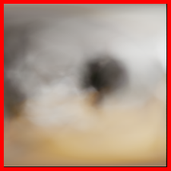}&
        \includegraphics[width=\linewidth]{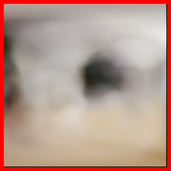} &
        \includegraphics[width=\linewidth]{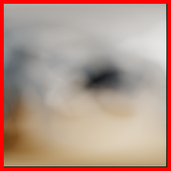} \\
        MSE: & 0.77 & 0.79 & 0.78 \\ 
        \includegraphics[width=\linewidth]{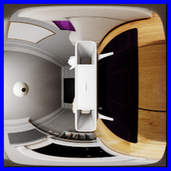} &
        \includegraphics[width=\linewidth]{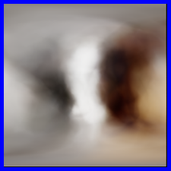}&
        \includegraphics[width=\linewidth]{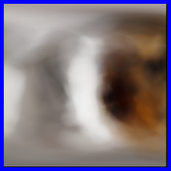} &
        \includegraphics[width=\linewidth]{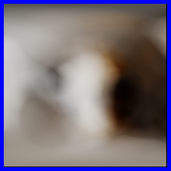} \\
        MSE: & 0.97 & 0.97 & 1.13 \\ 
        Reference & Ours, \(D = 6\) & Ours, \(D = 4\) & Ours, \(D = 2\) \\
         &  1.64ms & 1.24ms & 0.84ms
    \end{tabular*}
    \end{tabular*}
    }
    \caption{In this figure, we examine the impact of varying the number of
    hidden layers \(D\) within our \emph{Neural Incident Radiance Cache} (NIRC)
    on the precision of the reconstructed radiance signal for multiple scenes,
    arranged from left to right: Cornel Box, New Sponza, and White Room. The HDR
    cache is showcased for a certain set of points using octahedral projection,
    and we calculate the \emph{Mean Squared Error} (MSE) between it and the
    corresponding ground truth reference renders. For better visualization, we
    convert HDR values to LDR using the ACES tone mapper.
    Our cache does not store lighting information from emissive surfaces.
    Additionally, we present the computation time for each additional neural
    sample per neural architecture, aimed to illustrate scalability and the
    balance between performance and quality. As observed, a greater number of
    hidden layers typically result in a substantial improvement in the
    reconstruction capabilities of the NIRC, although this is not consistently
    the case across all scenarios.}
    \label{fig:cache_quality}
\end{figure*}
\setlength{\abovecaptionskip}{10pt}
\paragraph*{More than two-levels Monte Carlo}
In our work, we have studied a two-level Monte Carlo method. 
The experiments in \Cref{fig:cache_quality} indicate that often 2-layer MLPs reproduce the original signals quite well. Therefore, it might be beneficial to use such MLPs as a first level, and further MLPs, possibly with varying reconstruction qualities, in higher levels of MLMC.

\paragraph*{Negative values}
The rendering equation does not yield negative values, ensuring that our NIRC model consistently predicts positive values, making it well-suited for our applications. However, when considering the residual error integral in our two-level Monte Carlo estimator, we encounter the possibility of negative values. Unlike other works \cite{NCV, NIS}, we do not employ path guiding, as it is considered orthogonal research. This absence of path guiding might result in slower convergence due to sign-based variance. While techniques like the positivisation method \cite{Owen2000SafeAE} could potentially mitigate this issue by addressing the negative values, implementing such techniques would necessitate learning distributions, fitting Control Variates or path resampling. The latter would involve tracing multiple paths from a vertex where we invoke the cache, complicating the approach and degrading the performance of the whole algorithm. Therefore, we acknowledge this as an important area for future research.

\paragraph*{Challenges in Optimization}
Our experiments showed that using L2 differences as a loss function does not consistently lead to the best outcomes.
Taking into account tone mapping, or even perceptual metrics, might be beneficial, however, non-linear functions lead to biased gradients due to the interference of noisy radiance estimators.
Previous work~\cite{DiffRender_CV,DiffRender_ReSTIR} proposes to decrease the variance of estimators, which in turn reduces the bias of the gradients. We believe that this is an interesting direction for research.%

\paragraph*{In Place Execution}
Our inference pipeline temporarily stores all neural network requests in
buffers, a topic we discuss in detail in \Cref{sec:eval} concerning memory
consumption. This approach not only requires substantial memory but also
increases the consumption of memory bandwidth, which may lead to performance
degradation. Keeping CUDA and Direct3D in sync within our implementation adds an
overhead of 2-3 milliseconds. Implementing encoding evaluations and neural
network inference directly in the path tracing kernel might eliminate this
overhead and reduce the overall implementation complexity and memory
consumption. 

\section{Conclusion}
In this paper, we presented a two-level Monte Carlo estimator for real-time
global illumination rendering.
We train and evaluate tiny neural networks on the fly to approximate incident radiance,
as the first level of the estimator. A second level is used to create an unbiased estimator
in the context of multi-level Monte Carlo.
We also present a biased variant which can reduce the residual error further in the same rendering time.
Both variants produce lower errors in equal time comparisons in most experiments than previously published work on
neural radiance caching \cite{NRC}. In addition, our experiments revealed that combining the MLMC framework with NIRC can offset integration costs while achieving superior variance reduction of the residual error estimator compared to Control Variates.
Our key insights to achieving this are a data representation that stores incident radiance
(avoiding a costly ray trace to determine visibility when querying the cache),
a carefully selected set of inputs to the neural network (producing more accurate output
to support this new storage), and a tightly coupled implementation that avoids round trips
to global memory where possible. Our comprehensive cache analysis further demonstrated that we can significantly reduce the number of rays traced for indirect illumination by leveraging our cache, leading to more efficient rendering without significantly sacrificing quality.

We found that the allocation of sample counts for path vertices along a
transport path influences the final variance significantly and should be
investigated in more detail. This concept aligns with principles previously explored in classical light transport simulations, as demonstrated in the previous works \cite{ears} \cite{adjointsplittingjo}.

We only experimented with dynamic scene content but did not specifically
optimize the cache to quickly react to drastic changes.
For instance, distributing the training workload more into regions of the cache that
have low quality might improve the results.
Utilizing hierarchical caches has the potential to more quickly distribute information
about light sources in the scenes, and could provide interesting interplay with the multi-level
Monte Carlo paradigm.
We believe that research in this area promises
to be applicable for real-time and offline rendering.

\bibliographystyle{ACM-Reference-Format}
\bibliography{sample-base}

\end{document}

%% file: system2.pdf_tex
\begingroup%
  \makeatletter%
  \providecommand\color[2][]{%
    \errmessage{(Inkscape) Color is used for the text in Inkscape, but the package 'color.sty' is not loaded}%
    \renewcommand\color[2][]{}%
  }%
  \providecommand\transparent[1]{%
    \errmessage{(Inkscape) Transparency is used (non-zero) for the text in Inkscape, but the package 'transparent.sty' is not loaded}%
    \renewcommand\transparent[1]{}%
  }%
  \providecommand\rotatebox[2]{#2}%
  \newcommand*\fsize{\dimexpr\f@size pt\relax}%
  \newcommand*\lineheight[1]{\fontsize{\fsize}{#1\fsize}\selectfont}%
  \ifx\svgwidth\undefined%
    \setlength{\unitlength}{571.5bp}%
    \ifx\svgscale\undefined%
      \relax%
    \else%
      \setlength{\unitlength}{\unitlength * \real{\svgscale}}%
    \fi%
  \else%
    \setlength{\unitlength}{\svgwidth}%
  \fi%
  \global\let\svgwidth\undefined%
  \global\let\svgscale\undefined%
  \makeatother%
  \begin{picture}(1,0.61023622)%
    \lineheight{1}%
    \setlength\tabcolsep{0pt}%
    \put(0,0){\includegraphics[width=\unitlength,page=1]{system2.pdf}}%
    \put(0.92827034,0.56657677){\color[rgb]{0.30980392,0.25098039,0.25098039}\makebox(0,0)[lt]{\lineheight{1.25}\smash{\begin{tabular}[t]{l}MLP\end{tabular}}}}%
    \put(0,0){\includegraphics[width=\unitlength,page=2]{system2.pdf}}%
    \put(0.88914829,0.50583845){\color[rgb]{0.30980392,0.25098039,0.25098039}\makebox(0,0)[lt]{\lineheight{1.25}\smash{\begin{tabular}[t]{l}64\end{tabular}}}}%
    \put(0.55283858,0.35098675){\color[rgb]{0.30980392,0.25098039,0.25098039}\makebox(0,0)[lt]{\lineheight{1.25}\smash{\begin{tabular}[t]{l}64\end{tabular}}}}%
    \put(0.53353675,0.07987782){\color[rgb]{0.30980392,0.25098039,0.25098039}\makebox(0,0)[lt]{\lineheight{1.25}\smash{\begin{tabular}[t]{l}3+8+4\end{tabular}}}}%
    \put(0.65929003,0.22317966){\color[rgb]{0.30980392,0.25098039,0.25098039}\makebox(0,0)[lt]{\lineheight{1.25}\smash{\begin{tabular}[t]{l}24\end{tabular}}}}%
    \put(0.61519816,0.49305236){\color[rgb]{0.30980392,0.25098039,0.25098039}\makebox(0,0)[lt]{\lineheight{1.25}\smash{\begin{tabular}[t]{l}25\end{tabular}}}}%
    \put(0.16482546,0.24776627){\color[rgb]{0.30980392,0.25098039,0.25098039}\makebox(0,0)[lt]{\lineheight{1.25}\smash{\begin{tabular}[t]{l}$x_i$\end{tabular}}}}%
    \put(0.03195144,0.37571903){\color[rgb]{0.30980392,0.25098039,0.25098039}\makebox(0,0)[lt]{\lineheight{1.25}\smash{\begin{tabular}[t]{l}$\omega_{i,k}$\end{tabular}}}}%
    \put(0.11703543,0.4485563){\color[rgb]{0.30980392,0.25098039,0.25098039}\makebox(0,0)[lt]{\lineheight{1.25}\smash{\begin{tabular}[t]{l}$\omega_{i,k}$\end{tabular}}}}%
    \put(0.19675984,0.39901562){\color[rgb]{0.30980392,0.25098039,0.25098039}\makebox(0,0)[lt]{\lineheight{1.25}\smash{\begin{tabular}[t]{l}$\omega_{i,k}$\end{tabular}}}}%
    \put(0.32557913,0.31568228){\color[rgb]{0.30980392,0.25098039,0.25098039}\makebox(0,0)[lt]{\lineheight{1.25}\smash{\begin{tabular}[t]{l}$\omega_{i,k}$\end{tabular}}}}%
    \put(-0.0016273,0.20803661){\color[rgb]{0.30980392,0.25098039,0.25098039}\makebox(0,0)[lt]{\lineheight{1.25}\smash{\begin{tabular}[t]{l}$\bar{x}$\end{tabular}}}}%
    \put(-0.0016273,0.56063766){\color[rgb]{0.30980392,0.25098039,0.25098039}\makebox(0,0)[lt]{\lineheight{1.25}\smash{\begin{tabular}[t]{l}$\bar{\omega}$\end{tabular}}}}%
    \put(-0.0016273,0.14056155){\color[rgb]{0.30980392,0.25098039,0.25098039}\makebox(0,0)[lt]{\lineheight{1.25}\smash{\begin{tabular}[t]{l}$\bar{\phi}$\end{tabular}}}}%
    \put(0.0381378,0.18069409){\color[rgb]{0.30980392,0.25098039,0.25098039}\makebox(0,0)[lt]{\lineheight{1.25}\smash{\begin{tabular}[t]{l}position\end{tabular}}}}%
    \put(0.03916142,0.10682008){\color[rgb]{0.30980392,0.25098039,0.25098039}\makebox(0,0)[lt]{\lineheight{1.25}\smash{\begin{tabular}[t]{l}albedo, normal,\end{tabular}}}}%
    \put(0.03916142,0.07401168){\color[rgb]{0.30980392,0.25098039,0.25098039}\makebox(0,0)[lt]{\lineheight{1.25}\smash{\begin{tabular}[t]{l}roughness\end{tabular}}}}%
    \put(0,0){\includegraphics[width=\unitlength,page=3]{system2.pdf}}%
    \put(0.8898832,0.38651824){\color[rgb]{0.30980392,0.25098039,0.25098039}\makebox(0,0)[lt]{\lineheight{1.25}\smash{\begin{tabular}[t]{l}64\end{tabular}}}}%
    \put(0,0){\includegraphics[width=\unitlength,page=4]{system2.pdf}}%
    \put(0.8898832,0.21919541){\color[rgb]{0.30980392,0.25098039,0.25098039}\makebox(0,0)[lt]{\lineheight{1.25}\smash{\begin{tabular}[t]{l}64\end{tabular}}}}%
    \put(0,0){\includegraphics[width=\unitlength,page=5]{system2.pdf}}%
    \put(0.89339108,0.27793556){\makebox(0,0)[lt]{\lineheight{1.25}\smash{\begin{tabular}[t]{l}...\end{tabular}}}}%
    \put(0.6217874,0.22674134){\makebox(0,0)[lt]{\lineheight{1.25}\smash{\begin{tabular}[t]{l}...\end{tabular}}}}%
    \put(0.8558189,0.10105892){\makebox(0,0)[lt]{\lineheight{1.25}\smash{\begin{tabular}[t]{l}Output\end{tabular}}}}%
    \put(0,0){\includegraphics[width=\unitlength,page=6]{system2.pdf}}%
    \put(0.7998084,0.00790932){\makebox(0,0)[lt]{\lineheight{1.25}\smash{\begin{tabular}[t]{l}(3) inference\end{tabular}}}}%
    \put(0.51122743,0.01006155){\makebox(0,0)[lt]{\lineheight{1.25}\smash{\begin{tabular}[t]{l}(2) encode\end{tabular}}}}%
    \put(0.13244882,0.01006155){\makebox(0,0)[lt]{\lineheight{1.25}\smash{\begin{tabular}[t]{l}(1) gather\end{tabular}}}}%
    \put(0.40850328,0.12993031){\makebox(0,0)[lt]{\lineheight{1.25}\smash{\begin{tabular}[t]{l}identity/one blob\end{tabular}}}}%
    \put(0.40831955,0.28680958){\makebox(0,0)[lt]{\lineheight{1.25}\smash{\begin{tabular}[t]{l}hashed grid\end{tabular}}}}%
    \put(0.40921194,0.55512717){\makebox(0,0)[lt]{\lineheight{1.25}\smash{\begin{tabular}[t]{l}spherical harmonics\end{tabular}}}}%
    \put(0,0){\includegraphics[width=\unitlength,page=7]{system2.pdf}}%
    \put(0.40901745,0.40354318){\color[rgb]{0.02745098,0.02745098,0.02745098}\makebox(0,0)[lt]{\lineheight{1.25}\smash{\begin{tabular}[t]{l}neural input vector\end{tabular}}}}%
    \put(0,0){\includegraphics[width=\unitlength,page=8]{system2.pdf}}%
  \end{picture}%
\endgroup%